\documentclass[12pt,twoside,a4paper]{article}
\usepackage{amsmath,amssymb,latexsym,theorem,natbib,epsfig,color,subfigure}
\usepackage{multirow,graphicx,array,rotating}  
\usepackage{epstopdf,url,afterpage}
\usepackage[normalem]{ulem}
\newfont{\msa}{msam10 scaled\magstep1}
\newfont{\ssmsa}{msam9}

\def\crps{\mathop{\hbox{\rm CRPS}}}

\def\bs{\mathop{\hbox {\rm BS}}}
\def\bss{\mathop{\hbox{\rm BSS}}}
\def\qs{{\mathrm {QS}}}
\def\qss{{\mathrm {QSS}}}

\numberwithin{equation}{section}

\title{Statistical post-processing of dual-resolution ensemble forecasts}

\author{{\sc S\'andor Baran$^{1}$},  {\sc Martin Leutbecher$^{2}$}, {\sc Marianna Szab\'o$^{1}$}\\ and {\sc Zied Ben Bouall\`egue$^{2}$} \vspace*{0.5cm}\\
         $^1$Faculty of Informatics, University of Debrecen\\
         Kassai \'ut 26, H-4028 Debrecen, Hungary \\
         $^2$European Centre for Medium-Range Weather Forecasts\\ 
         Shinfield Park, Reading, RG2 9AX, United Kingdom
        }

        \date{}

\begin{document}
\pagestyle{myheadings}

\maketitle

\begin{abstract}
The computational cost as well as the probabilistic skill of ensemble forecasts depends on the spatial resolution of the numerical weather prediction model and the ensemble size. Periodically, e.g.\ when more computational resources become available, it is appropriate to reassess the balance between resolution and ensemble size. Recently, it has been proposed to investigate this balance in the context of dual-resolution ensembles, which use members with two different resolutions to make probabilistic forecasts. This study investigates whether statistical post-processing of such dual-resolution ensemble forecasts changes the conclusions regarding the optimal dual-resolution configuration.

Medium-range dual-resolution ensemble forecasts of 2-metre temperature have been calibrated using ensemble model output statistics. The forecasts are produced with ECMWF's Integrated Forecast System and have horizontal resolutions between 18\,km and 45\,km. The ensemble sizes range from 8 to 254 members. The forecasts are verified with SYNOP station data. Results show that score differences between various single and dual-resolution configurations are strongly reduced by statistical post-processing. Therefore, the benefit of some dual-resolution configurations over single resolution configurations appears to be less pronounced than for raw forecasts. Moreover, the ranking of the ensemble configurations can be affected by the statistical post-processing.

\end{abstract}

\section{Introduction}
\label{sec:sec1}

Ensemble forecast systems evolve as more computational resources become available. Defining how to upgrade a forecast systems requires among other things decisions regarding the spatial resolution of the forecast model as well as the ensemble size. The required computational resources strongly depend on these decisions. Thus, it is necessary to find an appropriate compromise between increasing spatial resolution and increasing ensemble size. Recently, \cite{lbb} looked at the question of the most skillful ensemble configuration for given computational resources in a {\em dual-resolution\/} setting, where $k$ lower-resolution members and $m$ higher-resolution members were combined. For 2-metre temperature forecasts in the medium-range, they found that dual-resolution ensembles with about equal number of lower- and higher-resolution members provided more skillful predictions than alternative single resolution configurations with either only lower-resolution or only higher-resolution members. Their study looked at forecasts obtained from raw model output by simply pooling together all members. 

However, raw ensemble forecasts tend to be underdispersive and can be subject to systematic bias. These deficiencies result in a lack of calibration and they have been documented for several different EPSs \citep[e.g.][]{bhtp}. Any lack of calibration calls for some form of statistical post-processing \citep{buizza18}. 
In the last decade, various methods of statistical calibration have been developed \citep[for comparison see e.g.][]{sk10,rs12,wfk}. Methods like Bayesian model averaging \citep[BMA;][]{rgbp} and ensemble model output statistics \citep[EMOS;][]{grwg} provide full predictive distributions. Once the predictive distribution is given, its functionals (e.g. median or mean) can easily be calculated and considered as point forecasts.

The BMA predictive distribution of a future weather quantity is a weighted mixture of probability distributions corresponding to the individual ensemble members with weights determined by the predictive performance of the members during the training period. BMA models for various weather quantities differ in the distribution of the mixture components: for temperature or pressure a normal mixture is suggested \citep{rgbp}, wind speed can be modeled using gamma \citep{sgr10} or truncated normal distribution \citep{bar}, whereas for precipitation accumulation a discrete-continuous gamma model was developed \citep{srgf}. 

Here we concentrate on the essentially simpler EMOS, or non-homogeneous regression, approach where the predictive distribution is given by a single parametric probability density function (PDF) with parameters depending on the ensemble. Similar to the BMA, different weather quantities require different distributions. Temperature and pressure can again be modeled by normal distributions \citep{grwg}, wind speed requires non-negative and skewed distributions such as truncated normal \citep{tg}, generalized extreme value \citep{lt}, log-normal \citep{bl15} or their mixture \citep{bl16}, whereas censored generalized extreme value \citep{sch} and censored shifted gamma \citep{schham, bn} distributions and their various combinations \citep{bl18} provide good models for precipitation accumulation.

In this paper, we examine whether dual-resolution ensembles investigated by \citet{lbb} are still preferable to single resolution ensembles after statistical post-processing.
Dual-resolution ensembles fall in the category of multi-model ensembles, where each contributing model has specific error characteristics. Previous studies on post-processing of multi-model forecasts have focused on the optimization of the weights attributed to each contributing model \citep{doblasreyes2005,casanova2009,delsole2013,raynaud2015}.  Here, following an EMOS approach, the target is to provide as a forecast a full probability distribution optimized in terms of probabilistic skill. 

The forecast and verification data used in this study are described in Section~\ref{sec:sec2}. Then, the EMOS calibration and verification methodologies are presented in Section~\ref{sec:sec3}. The results and conclusions follow in Sections~\ref{sec:sec4} and \ref{sec:sec5}, respectively.

\section{Forecast and observation data}
\label{sec:sec2}

This study applies EMOS calibration to the same ensemble forecasts examined by \cite{lbb}. Global medium-range forecasts with ECMWF's Integrated Forecast System (IFS) with three horizontal resolutions are examined. The lower resolution ensembles have four to five times more members than the highest resolution ensemble
\begin{itemize}
\item 50 members at TCo639 (grid resolution $\sim$18km),
\item 200 members at TCo399 (grid resolution $\sim$29km),
\item 254 members at TCo255 (grid resolution $\sim$50km).
\end{itemize}
The 50-member TCo639 ensemble is the operational ECMWF medium-range ensemble while the lower resolution ensembles are generated with the same model version as the operational ensemble at the time (cycle 41r2). The perturbation methodology for the initial conditions and the stochastic representation of model uncertainties are identical in the three ensembles \cite[see][ for further details]{lbb}. The period of investigation is boreal summer 2016. The ensemble forecasts are initialized once daily between 1 June and 31 August 2016.

The cost ratio between a TCo639 forecast and a TCo399 forecast is about 4:1 and the cost ratio between a TCo639 forecast and a TCo255 forecast is around 16:1. When constructing different dual-resolution configurations, one TCo639 member can be traded against 4 TCo399 members or against 16 TCo255 members. Similar to \citet{lbb}, various configurations of higher- and lower-resolution ensemble forecasts are studied. We will consider TCo399-TCo639 dual-resolution ensembles and TCo255-TCo639 ensembles as well as two scenarios corresponding to different assumptions on available HPC resources. The {\em large supercomputer\/} (LHPC) scenario assumes the availability of current HPC resources of the ECMWF, whereas the {\em small supercomputer\/} (SHPC) scenario assumes 1/6th of these resources, for an overview see Table \ref{tab:cases}. Note that TCo399-TCo639 and TCo255-TCo639 combinations are based on different ensemble sizes of TCo639 members, so in fact one has two different LHPC scenarios. The reason is completely technical, as the cost of 50 TCo639 members is equivalent to the cost of 800 TCo255 members, however, with the current GRIB settings the largest possible ensemble size to be handled is 255.

\begin{table}[t]
  \begin{center}
  \begin{tabular}{cc|cc|cc|cc}\hline
  \multicolumn{4}{c|}{TCo399 - TCo639}&\multicolumn{4}{c}{TCo255 - TCo639} \\   
  \hline
  \multicolumn{2}{c|}{Large HPC}&\multicolumn{2}{c|}{Small HPC}& 
  \multicolumn{2}{c|}{Large HPC}&\multicolumn{2}{c}{Small HPC} \\ \hline
  $M_L$&$M_H$&$M_L$&$M_H$&$M_L$&$M_H$&$M_L$&$M_H$ \\ \hline
  0&50&0&8&0&16&0&8\\
  40&40&8&6&16&15&16&7\\
  120&20&16&4&32&14&32&6\\
  160&10&24&2&64&12&64&4\\
  200&0&28&1&128&8&128&0\\
  &&32&0&254&0&&\\ \hline
  \end{tabular}
  \caption{Investigated dual-resolution mixtures.}\label{tab:cases}
 \end{center}
\end{table}

The focus is on forecasts of 2-metre temperature. Post-processing and verification of the ensemble forecasts rely on measurements from surface synoptic observation (SYNOP) stations.  Measurements are reported from various locations around the globe with low observation densities in tropical and sub-tropical regions and high densities in Europe and the North-American continent. The number of observations available vary from day to day. A subset of about 4500 stations with full availability over the verification period is used in this study.

Observations are compared to forecasts at the nearest grid point of the native forecast grid.
To account for systematic representativity errors related to the coarse description of the orography in the model, an orographic correction is applied to the raw temperature forecasts. For each forecast, the correction corresponds to an adjustment linear with the height difference between station and model representation ($\Delta T = -0.0065$\,K\,m$^{-1}\,\Delta z$).

\begin{figure}[t]
\epsfig{file=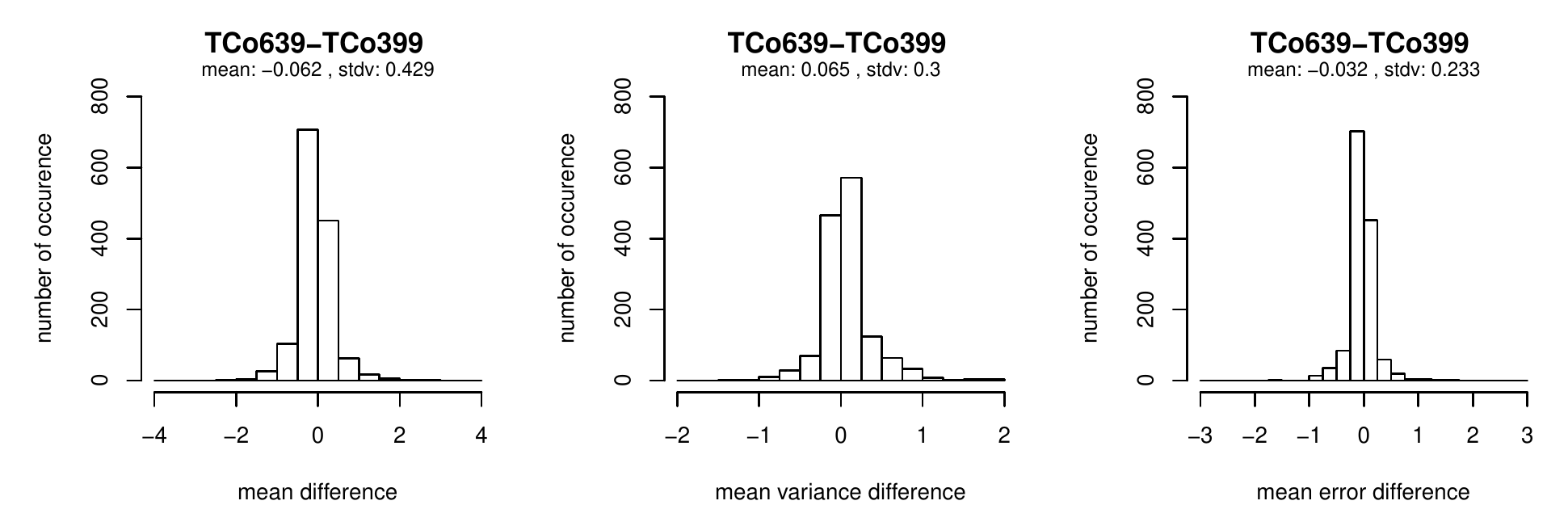, width=.99\textwidth} \\
\epsfig{file=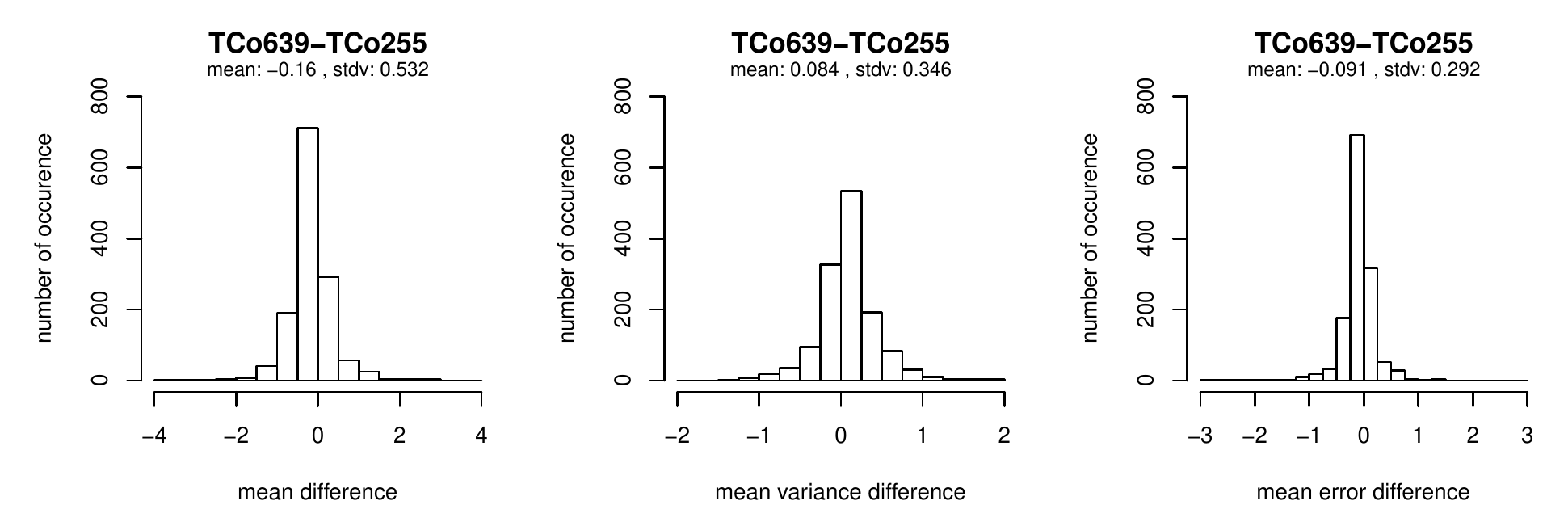, width=.99\textwidth} \\
\caption{Stationwise differences between high resolution (TCo639) and low resolution (TCo399, top panels, and TCo255, bottom panels) ensemble characteristics at day 5 over Europe ($\sim$ 1300 stations): mean differences (left, unit: K), mean ensemble variance differences (middle, unit: K$^2$), and mean root mean squared error differences (right, unit: K). Statistics are computed for ensembles with 50 members each.}
\label{fig:stat}
\end{figure}

A basic comparison of the statistics of lower and higher resolution forecasts is shown in Figure~\ref{fig:stat}.
The distributions of mean differences at the station level are plotted: differences in terms of mean ensemble forecast, mean ensemble variance, and mean forecast error, as measured by the root mean squared error of the ensemble mean. The comparison is performed considering ensembles of the same size (50) in order to focus on the impact of the model resolution and not on the impact of the ensemble size. The statistics shown in Figure~\ref{fig:stat} are valid for forecasts at day 5, for stations located in Europe. 

Ensembles with different model resolutions have on average very small mean differences, but they can exhibit large mean differences (up to 2\,K) at certain locations (Figure~\ref{fig:stat}, left).
The ensemble forecasts at high resolution (TCo639) have on average  a larger mean variance (Figure~\ref{fig:stat}, middle) and a smaller mean error  (Figure~\ref{fig:stat}, right) at the station level than ensemble forecasts at a lower resolution. Here again, the station-to-station variability can be large. Comparing the top and bottom panels of Figure~\ref{fig:stat} shows that the differences in terms of variance and error are slightly larger when the difference of the underlying model resolutions increases.

\section{Ensemble model output statistics}
\label{sec:sec3}

As mentioned earlier, the normal distribution provides a good model for temperature, however, one has to appropriately link the predictive PDF with the ensemble members. 

\subsection{Model and parameter estimation}
\label{subs:subs3.1}
The EMOS approach suggested by \citet{grwg} considers the mean and variance of the predictive PDF to be affine functions of the ensemble forecasts and the ensemble variance, respectively. If \ $f_1,f_2, \ldots , f_K$ \ denote the ensemble forecasts for temperature for a given location, time and lead time, the associated predictive distribution of temperature will be
\begin{equation}
  \label{eq:emos_non_exch}
{\mathcal N}\big(a+b_1f_1 +\cdots + b_Kf_K, c+d S^2 \big), \qquad \text{with} \qquad S^2:= \frac 1{K-1}\sum_{k=1}^K \big(f_k-\overline f\big)^2,
\end{equation}
where \ $\overline f$\  denotes the ensemble mean.

In a multi-model context with exchangeable ensemble groups, \citet{gneiting14} suggests to use the same coefficients within a given group \citep[see also][]{frg2010}. Hence, if we have \ $M$ \ ensemble members divided into \ $K$ \ exchangeable groups, where the \ $k$th \ group contains \ $M_k\geq 1$ \ ensemble members ($\sum_{k=1}^KM_k=M$) and has mean  \ $\overline f_k$, \ the EMOS predictive distribution will be
\begin{equation}
  \label{eq:emos_exch}
{\mathcal N}\big(a+b_1\overline f_1 +\cdots + b_K\overline f_K, c+d S^2 \big),
\end{equation}
with \ $S^2$ \ denoting again the ensemble variance.

According to the optimal score estimation approach of \citet{grjasa}, model parameters \ $a,b_1,\ldots ,b_K$ \ and \ $c,d$ \ are determined by optimizing the mean value of a proper scoring rule as a function of the parameters over suitably chosen training data. Scoring rules measure the forecast skill by numerical values assigned to pairs of forecasts and observations, and for predictive distributions the most popular ones are the logarithmic score, that is, the negative logarithm of the predictive PDF evaluated at the verifying observation \citep{grjasa} and the continuous ranked probability score \citep[CRPS;][]{grjasa,wilks}. Given a (predictive) cumulative distribution function (CDF) \ $F(y)$ \ and real value (observation) \ $x$ \ the CRPS is defined as
\begin{equation}
\label{eq:CRPS}
\crps\big(F,x\big):=\int_{-\infty}^{\infty}
\big(F(y)-{\mathbb I}_{\{y \geq x\}}\big )^2{\mathrm d}y
={\mathsf E}|X-x|-\frac 12
{\mathsf E}|X-X'|, 
\end{equation}
where \ ${\mathbb I}_H$ \ denotes the indicator of a set \ $H$, \ whereas \ $X$ \ and \ $X'$ \ are independent random variables with CDF \ $F$ \ and finite first moment. The right-hand side of \eqref{eq:CRPS} shows that the CRPS has the same unit as the observation, moreover, for normal distribution it has a simple closed form \citep{grwg}. One should also mention that both CRPS and logarithmic score are negatively oriented, that is the smaller the better, and optimization with respect to the latter provides the maximum-likelihood estimates of the model parameters. 

The choice of the training data is important for statistical post-processing. For estimating the EMOS model parameters a rolling training period is applied and the estimates are obtained using ensemble forecasts and corresponding validating observations for the preceding \ $n$ \ calendar days. Given a training period length there are two traditional approaches for selecting the training data \citep{tg}. In the global (regional) approach, parameters are estimated using all available forecast cases from the training period resulting in a single universal set of parameters across the entire ensemble domain. It requires quite short training periods, but usually it is unsuitable for large and heterogeneous observation domains. For local parameter estimation, one has distinct parameter estimates for the different stations obtained only using training data of the given station. To avoid numerical stability problems local EMOS requires much longer training periods \citep[for optimal training period length for different weather quantities see e.g.][]{hemri14}, but if the training data is large enough, it will usually outperform the regional approach. To combine the advantages of local and regional EMOS, \citet{lb17} introduced two semi-local methods where the training data for a given station is augmented with data from stations with similar characteristics. The choice of similar stations is based either on suitably
defined distance functions or on clustering. Here we focus on the clustering based semi-local estimation, where the observation sites are grouped into clusters using $k$-means clustering of feature vectors depending both on the station climatology and the forecast errors of the raw ensemble during the training period. A regional parameter estimation is then performed within each cluster.  With the help of this method one can get reliable parameter estimates even for short training periods and the obtained models may outperform the local EMOS approach \citep{lb17}. Hence, Section \ref{sec:sec4} focuses mainly on results corresponding to semi-local EMOS post-processing.

\subsection{Verification scores}
  \label{subs:subs3.2}

The fundamental aim of probabilistic forecasting is to access the maximal sharpness of the forecast distribution subject to calibration \citep{gbr}, where the former refers to the concentration of the predictive distribution and the latter to the statistical consistency between the predictive distributions and the validating observations. A standard tool of quantifying the predictive performance of probabilistic forecasts both in terms of calibration and sharpness is the mean CRPS over all forecast cases. 

Besides the CRPS one can also consider Brier scores \citep[BS;][Section 8.4.2]{wilks} for the dichotomous event that the observed temperature  \ $x$ \ exceeds a given threshold \ $y$. \  For a predictive CDF \ $F(y)$ \ the Brier score is defined as
\begin{equation*}
 \bs \big(F,x;y\big):= \big (F(y)-{\mathbb I}_{\{y \geq x\}}\big )^2,
\end{equation*}
\citep[see e.g.][]{gr11}, and note that the CRPS is the integral of the BS over all possible thresholds. In the results provided in Section \ref{sec:sec4} we consider  as thresholds the \ $5$, $10$, $\ldots$, $90$, and $95$ \ percentiles of the corresponding station climatology for the verification period.

Further, let \ $q_{\tau} (F)$ \ denote the \ $\tau$-quantile \ ($0\leq \tau\leq 1$) \ of a CDF \ $F(y)$, \  that is
\begin{equation*}
q_\tau(F) := F^{-1}(\tau):= \inf\{y:F(y)\geq\tau\},
\end{equation*}
and consider the loss function
\begin{equation*}
\rho_\tau(x) := 
\begin{cases}
	\tau|x|,  & \text{if  $x \geq 0$,} \\
		(1-\tau)|x|, & \text{if  $x < 0$.}
        \end{cases}
\end{equation*}
Then for a given value \ $x$ \ the quantile score \citep[QS; see e.g.][]{bf} is defined as
\begin{equation*}
\qs _{\tau}(F,x):=\rho_\tau\big(x -q_{\tau} (F) \big).
\end{equation*}
In the present study QS values corresponding to the \ $2$, $5$, $10$, $20$, $50$, $80$, $90$, $95$, and $98$ \ percentiles of the predictive distribution and the raw ensemble are considered. Note that to evaluate CRPS, BS and QS of the raw ensemble one has to replace the predictive CDF by the empirical one.

The improvement in BS and QS with respect to a reference predictive distribution \ $F_{ref}$ \ can be measured with the help of the Brier skill score (BSS) and the quantile skill score (QSS) defined as
\begin{equation*}
\bss \big(F,F_{ref},x;y\big):=1-\frac{\bs \big(F,x;y\big)}{\bs \big(F_{ref},x;y\big)}
\qquad \text{and} \qquad
\qss_{\tau} \big(F,F_{ref},x\big):=1-\frac{\qs_{\tau} \big(F,x\big)}{\qs_{\tau} \big(F_{ref},x\big)},
\end{equation*}
respectively \citep{ft12}. Obviously, in contrast to the BS and QS, the corresponding skill scores are positively oriented, that is the larger the better.

Finally, point forecasts such as ensemble/EMOS medians and means are evaluated using mean absolute errors (MAEs) and root mean squared errors (RMSEs), where the former is optimal for the median and the latter for the mean \citep{gneiting11}, although for the normal EMOS model these quantities coincide. As mean of \ $\qs_{50}$, \ the quantile score of the median forecast,  over all forecast cases is exactly half of the MAE, in what follows we will report only the RMSE values.

As suggested by \citet{gr11}, to assess the statistical significance of the differences between the verification scores we make use of the Diebold-Mariano \citep[DM;][]{dm95} test of equal predictive performance, as it allows to account for the temporal dependencies in the forecast errors. For more details about the DM test see e.g. \citet{bl18}. Further, confidence intervals for mean score values and mean score differences are obtained with the help of $2000$ block bootstrap samples  using the stationary bootstrap scheme with mean block length according to \citet{pl94}.

\section{Results}
  \label{sec:sec4}
\subsection{Preliminaries}

\begin{figure}[t!]
\epsfig{file=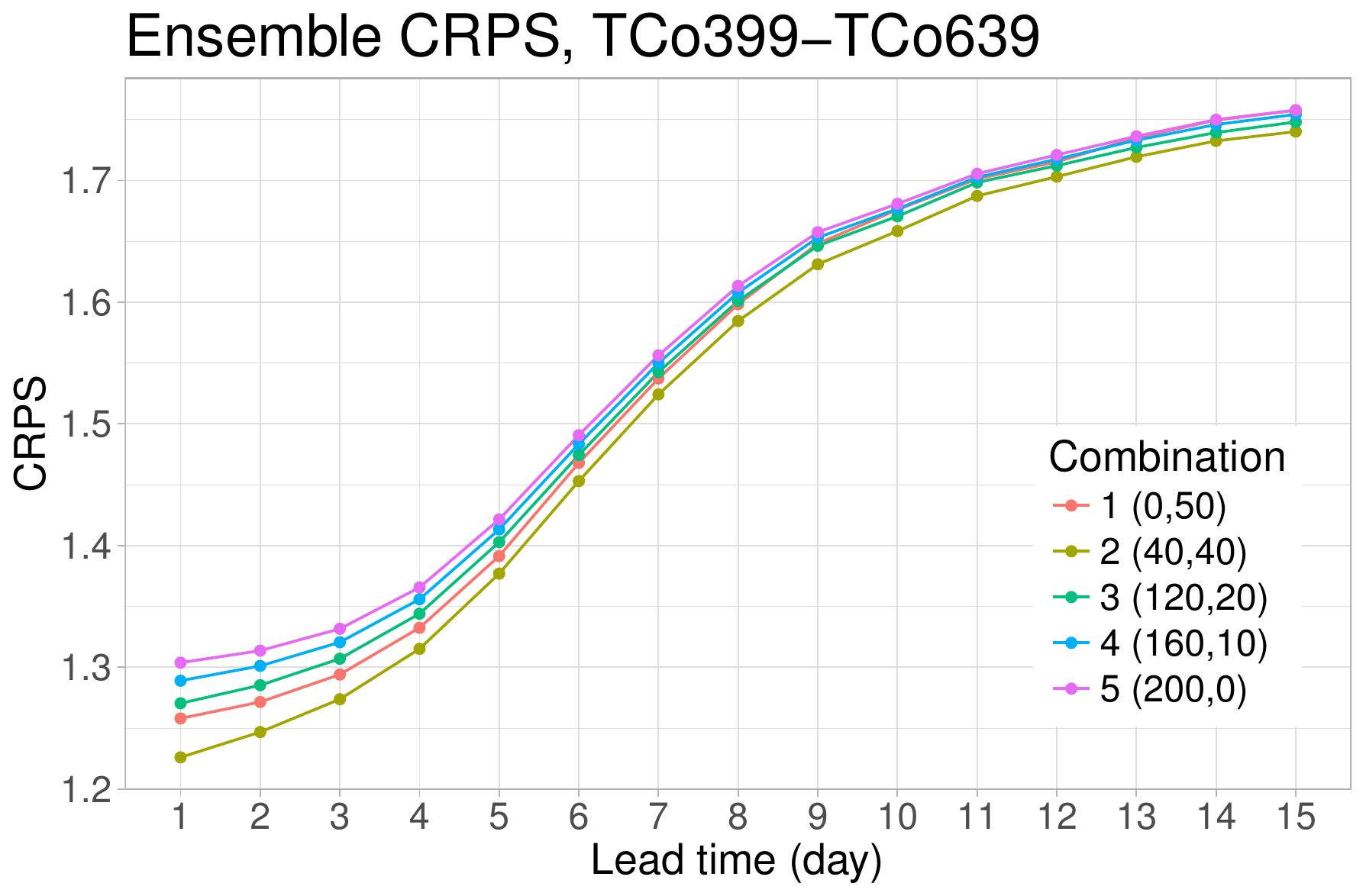, width=.49\textwidth} \
\epsfig{file=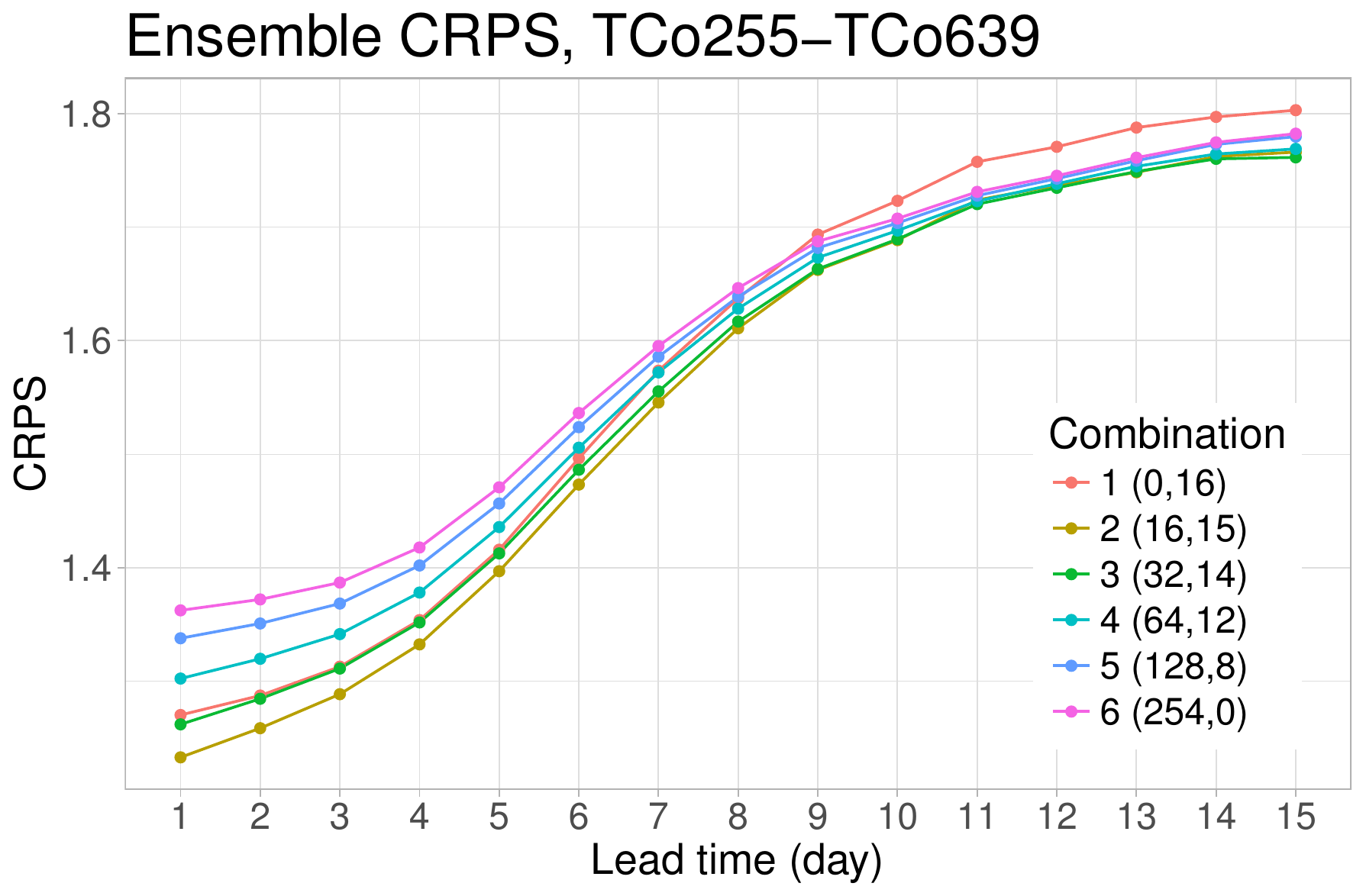, width=.49\textwidth}

\medskip
\epsfig{file=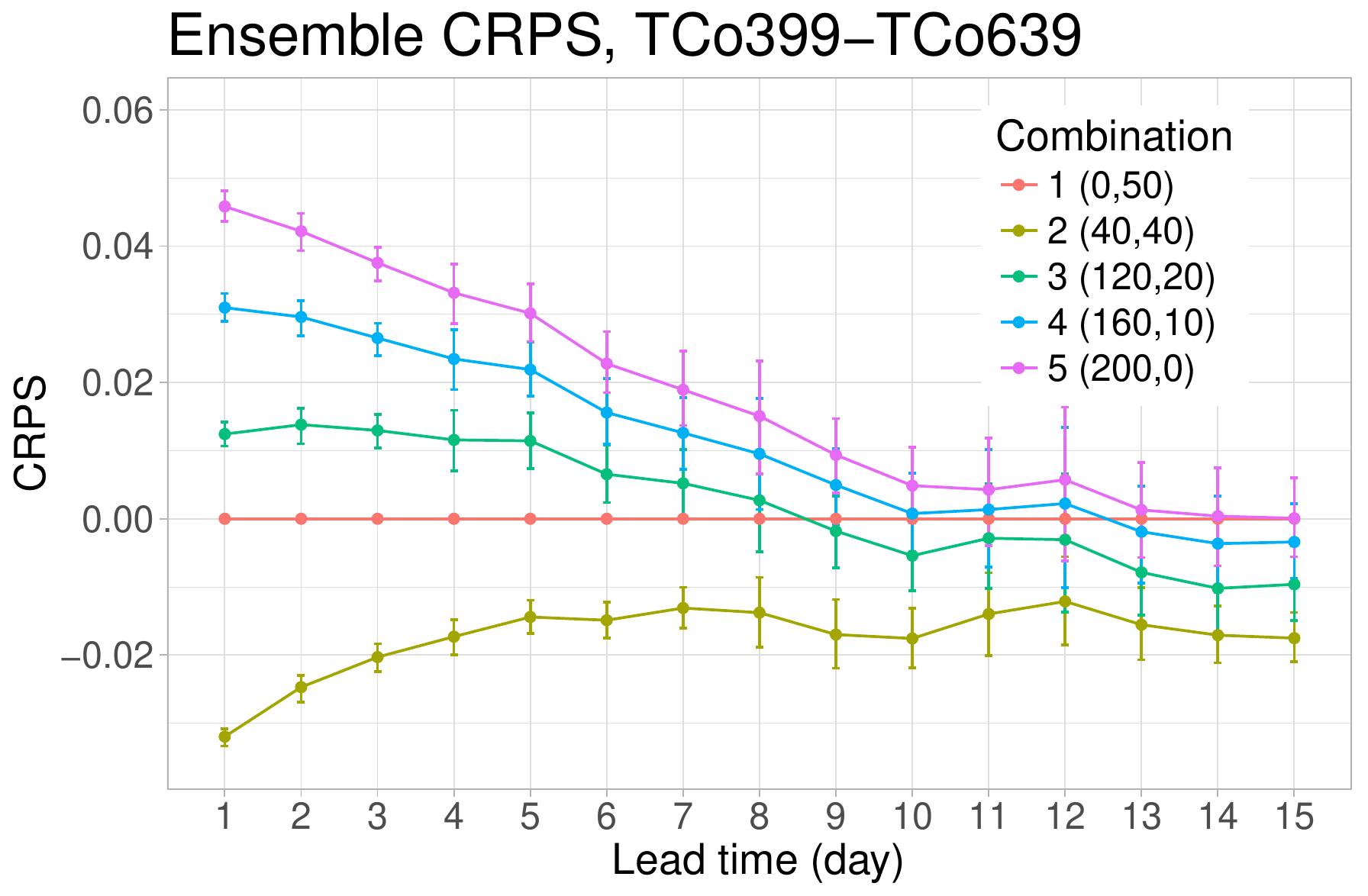, width=.49\textwidth} \
\epsfig{file=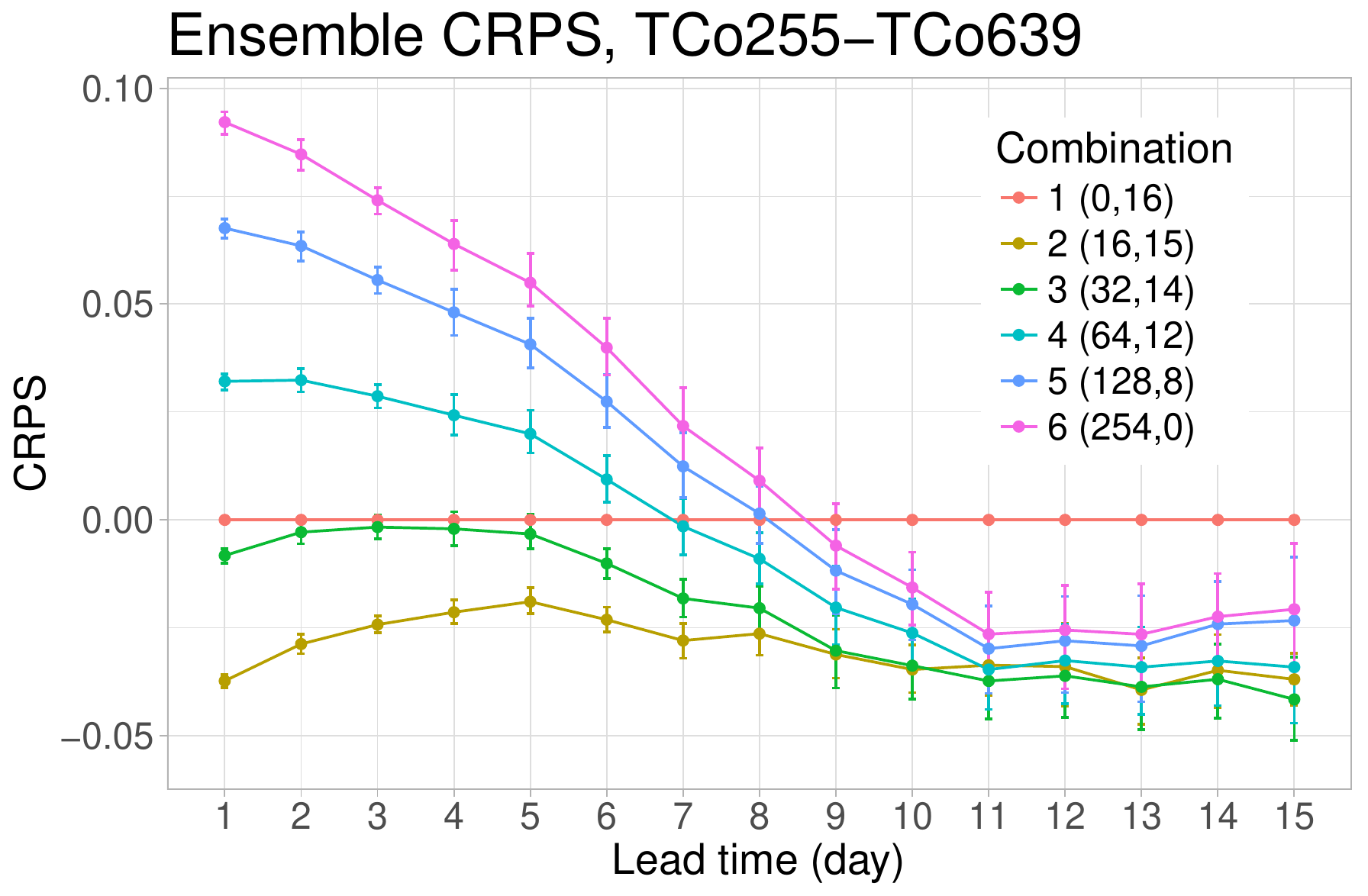, width=.49\textwidth} 
\caption{Mean CRPS values (the lower the better) of global dual-resolution ensemble forecasts for 2m temperature ({\em top}) and the difference in mean CRPS (the lower the better) from the reference pure high resolution ensemble ({\em bottom}) with $95\,\%$ confidence intervals, LHPC scenario.}
\label{fig:crpsRawL}
\end{figure}

As mentioned in the Introduction, EMOS calibration is applied to different dual-resolution ensemble forecasts for 2-metre temperature. In what follows, let \ $f_{H,1},f_{H,2},\ldots ,f_{H,M_H}$ \ and \ $f_{L,1},f_{L,2},\ldots ,f_{L,M_L}$ \ denote the higher and lower resolution ensemble members, respectively, and denote by \ $\overline f_H$ \ and \ $\overline f_L$ \ the corresponding ensemble means. As ensemble members of a given resolution can be considered exchangeable, model \eqref{eq:emos_exch} reduces to
\begin{equation}
  \label{eq:emos_dual}
{\mathcal N}\big(a+b_H\overline f_H + b_L\overline f_L, c+d S^2 \big),
\end{equation}
with \ $b_L=0$ \ for pure higher resolution \ ($M_L=0$) \  and \ $b_H=0$ \ for pure lower resolution \ ($M_H=0$) \ configurations and  \ $S^2$ \ being the variance of the pooled ensemble defined by \eqref{eq:emos_non_exch}. Following the ideas of \citet{grwg}, model parameters are estimated by minimizing the mean CRPS of the predictive distributions and validating observations corresponding to the forecast cases of the training data.

For the case of TCo399 - TCo639 combinations, stationwise statistical tests of equality of the variances and distributions of the two full-size ensembles \ ($M_H=50, \ M_L=200$) \ suggest to treat the variances of the component ensembles separately by considering the predictive distribution 
\begin{equation}
  \label{eq:emos_dual_var}
{\mathcal N}\big(a+b_H\overline f_H + b_L\overline f_L, c+d_H S_H^2 + d_L S_L^2 \big),
\end{equation}
with \ $S_H^2$ \ and \ $S_L^2$ \ denoting the variances of the high- and low resolution components, respectively. However, at least for the TCo639 - TCo399 combination, the verification scores of the more complex model \eqref{eq:emos_dual_var} do not differ significantly from those of model \eqref{eq:emos_dual}, so all our calibration results are obtained using the latter.

\begin{figure}[t!]
\epsfig{file=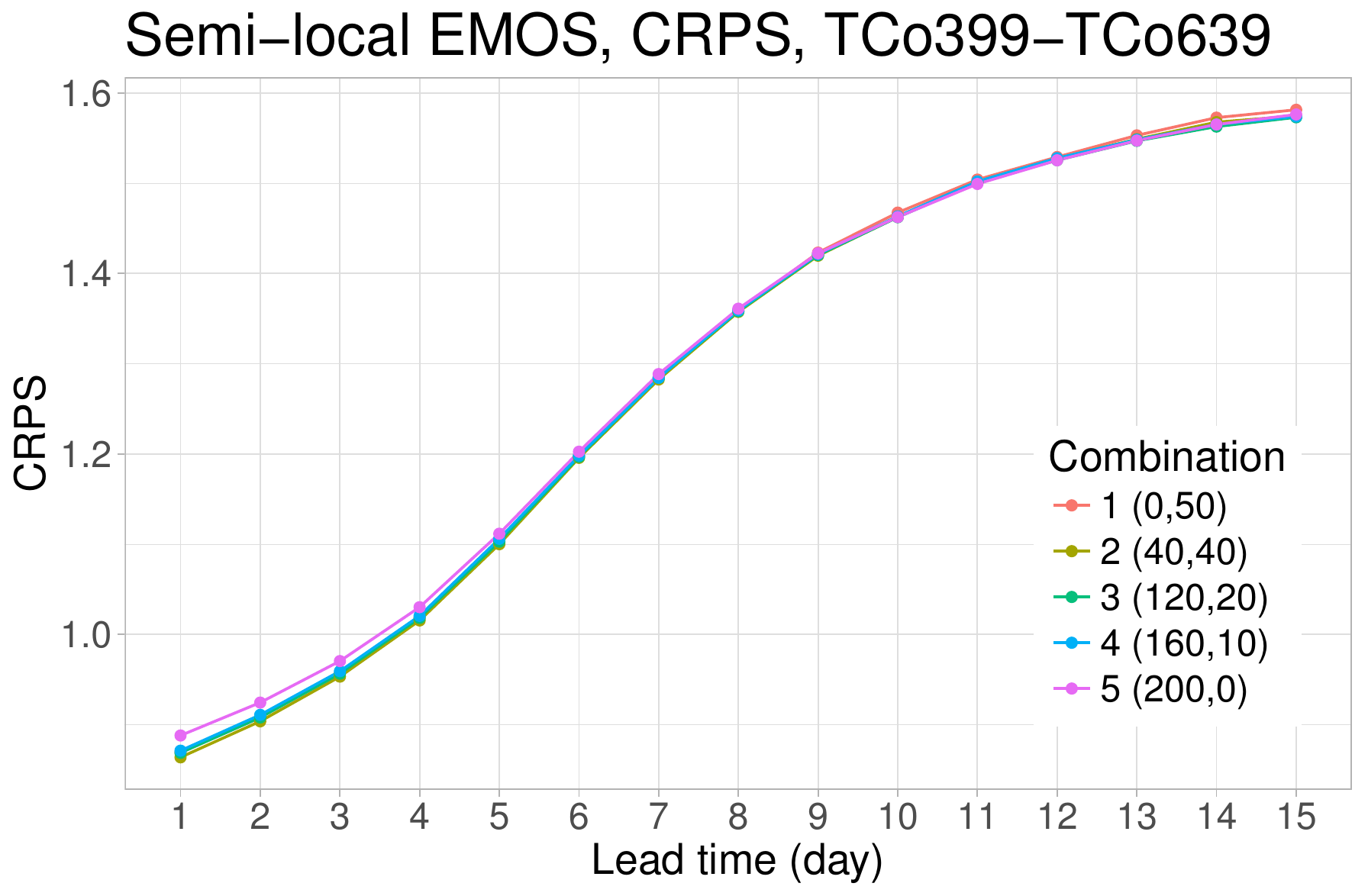, width=.49\textwidth} \
\epsfig{file=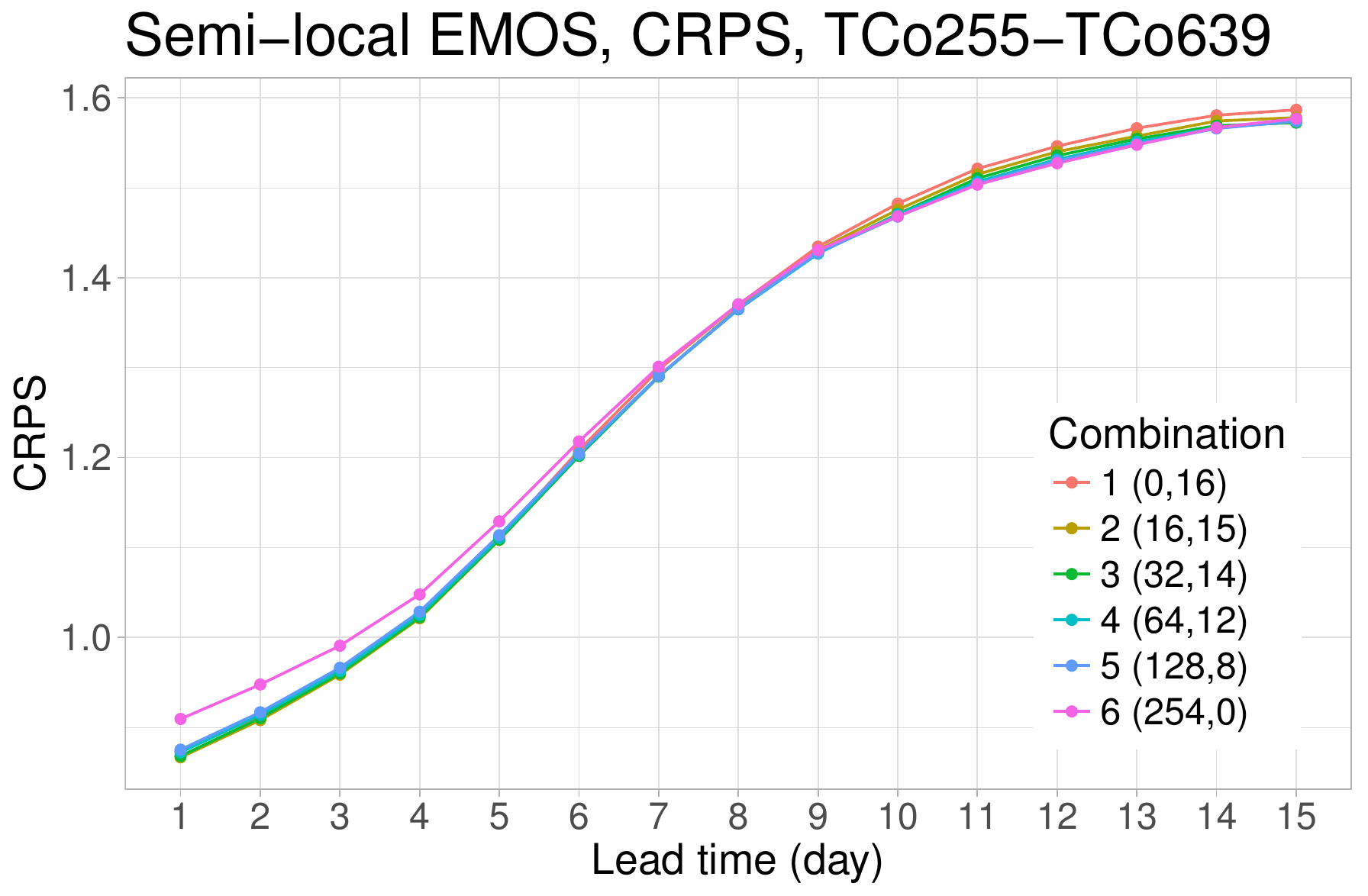, width=.49\textwidth} 
\caption{Mean CRPS values (the lower the better) of semi-local EMOS post-processed global dual-resolution ensemble forecasts for 2m temperature, LHPC scenario.}
\label{fig:crpsEMOSL}
\end{figure}

As the data set at hand covers only the boreal summer 2016 (92 calendar days), one has to consider relatively short training period length that still allow reliable parameter estimation to leave enough data for model verification. As a trade-off between the two requirements, calibration is performed using a rolling 30-day training period. This means that verification scores are calculated for ensemble forecasts initialized between 1 July and 31 August 2016 and the corresponding validating observations. Obviously, the forecast periods are shifted by 1--15 days according to the lead times of the ensemble predictions. 

As global EMOS is not justifiable for global data, local and semi-local approaches to parameter estimation are considered. Similar to \citet{lb17}, the $k$-means clustering of stations is based on $24$-dimensional feature vectors consisting of $12$ equidistant quantiles of the climatological CDF and $12$ equidistant quantiles of the empirical CDF of forecast errors of the ensemble mean over the training period. $200$ clusters are considered which yields a similar mean station number per cluster as in \citet{lb17}.  Local EMOS estimates 4--5 parameters from 30 forecast-observation pairs while  semi-local EMOS estimates the same parameters from about 600 forecast-observation pairs. Therefore, the latter should be able to constrain the parameters much better.

In order to highlight the differences between local and semi-local approaches,  a very short 10-day training period is also investigated.

To be fully consistent with the results for the raw ensemble, EMOS predictive distributions are obtained using the orographically corrected ensemble forecasts. Local EMOS does not require such a preliminary bias correction but, for semi-local EMOS, the use of this correction yields improved skill in terms of verification scores because local variability within a cluster can still be represented through the orographic correction.

\subsection{Calibration of mixtures for large supercomputer}
  \label{subs:subs4.1}

\begin{figure}[t!]

\epsfig{file=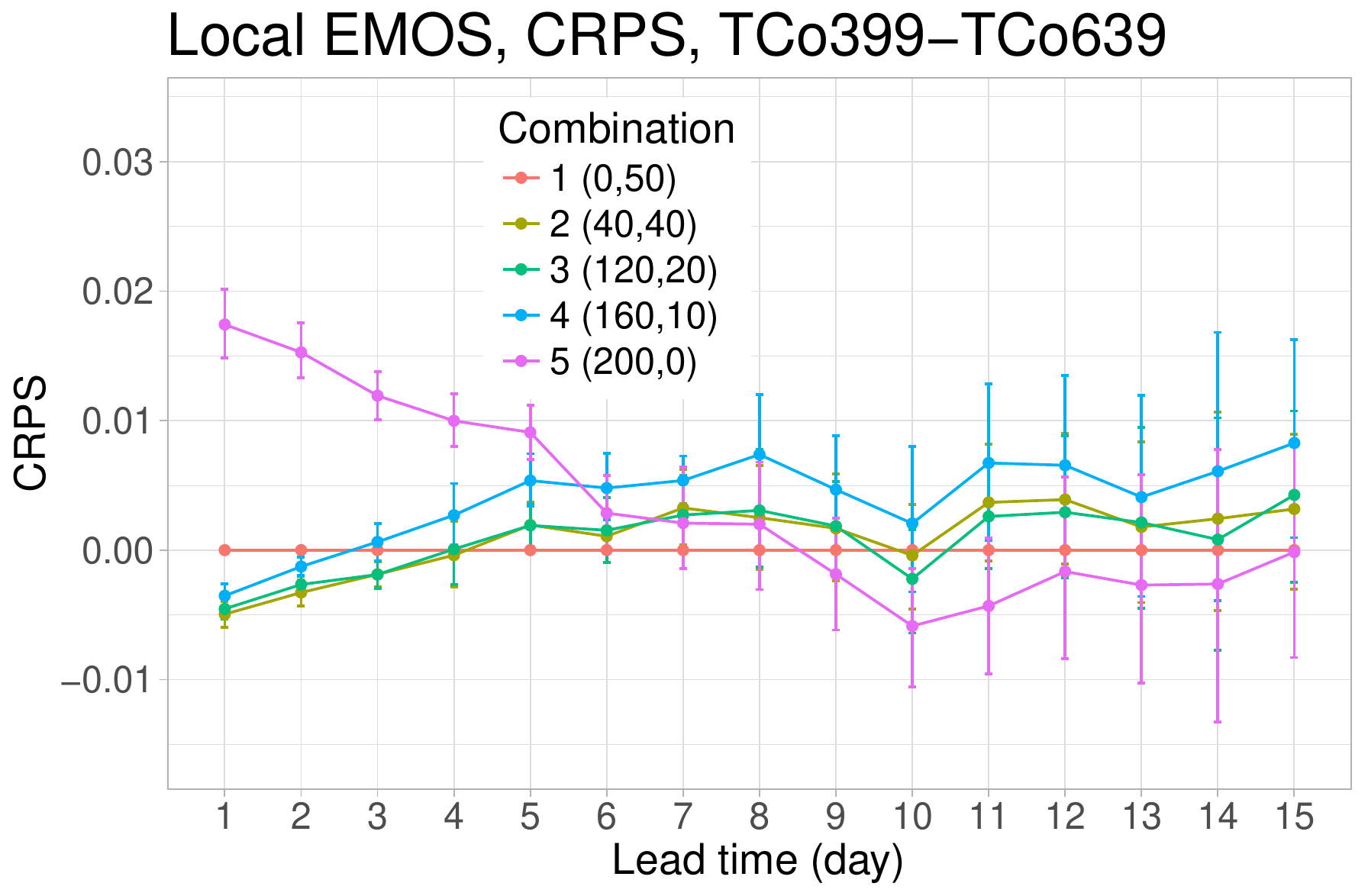, width=.49\textwidth} \
\epsfig{file=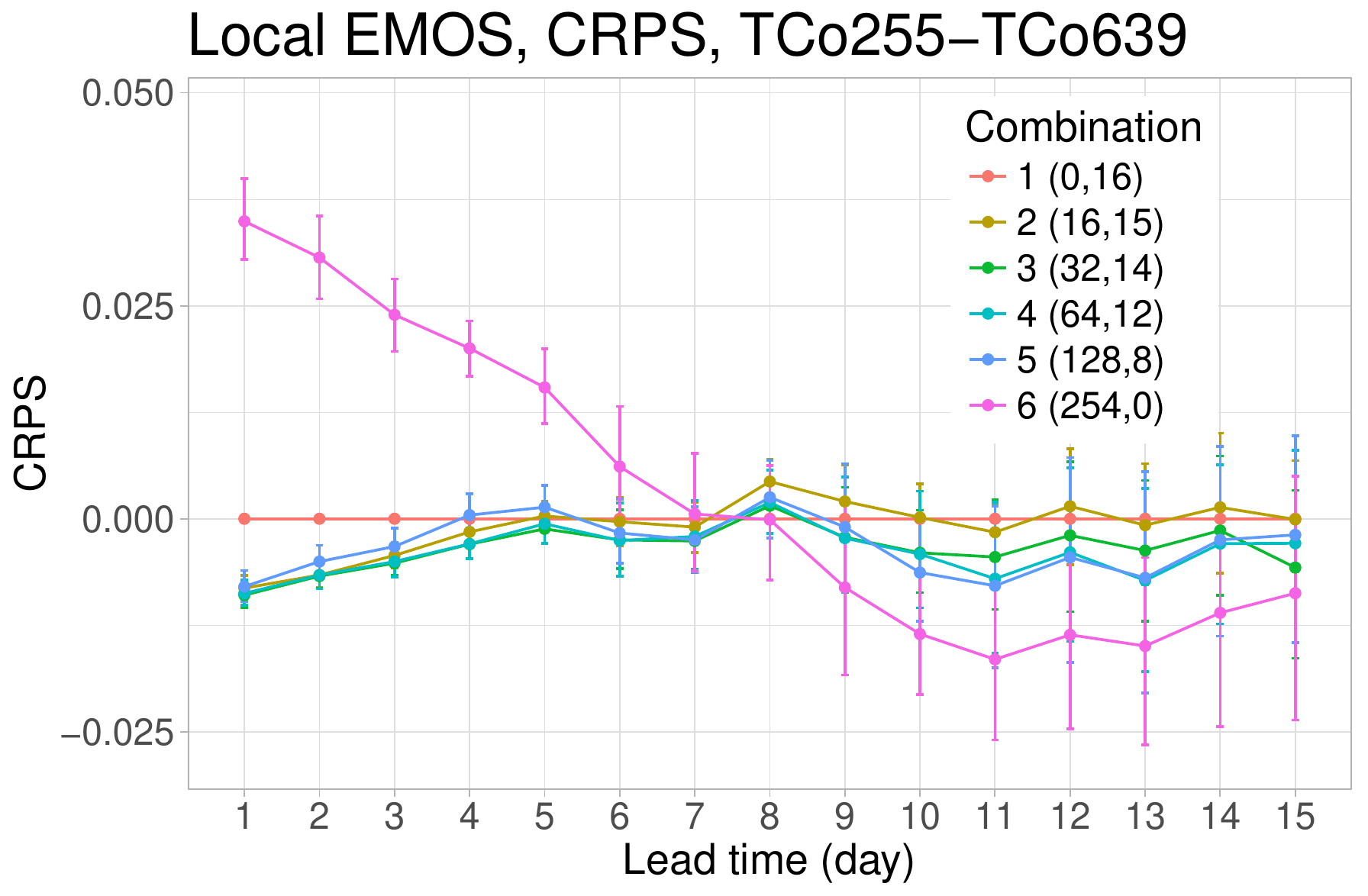, width=.49\textwidth}

\medskip
\epsfig{file=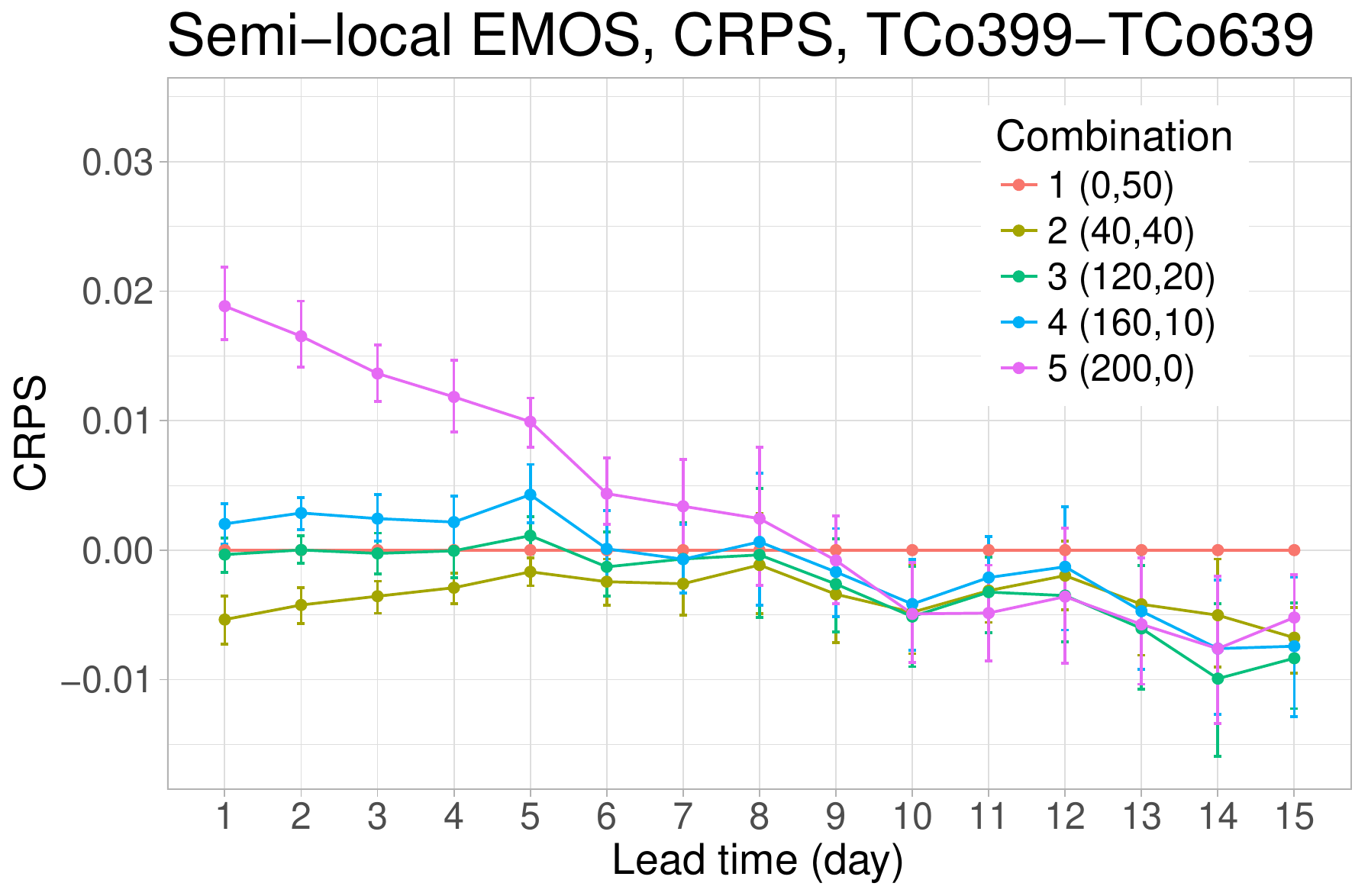, width=.49\textwidth} \
\epsfig{file=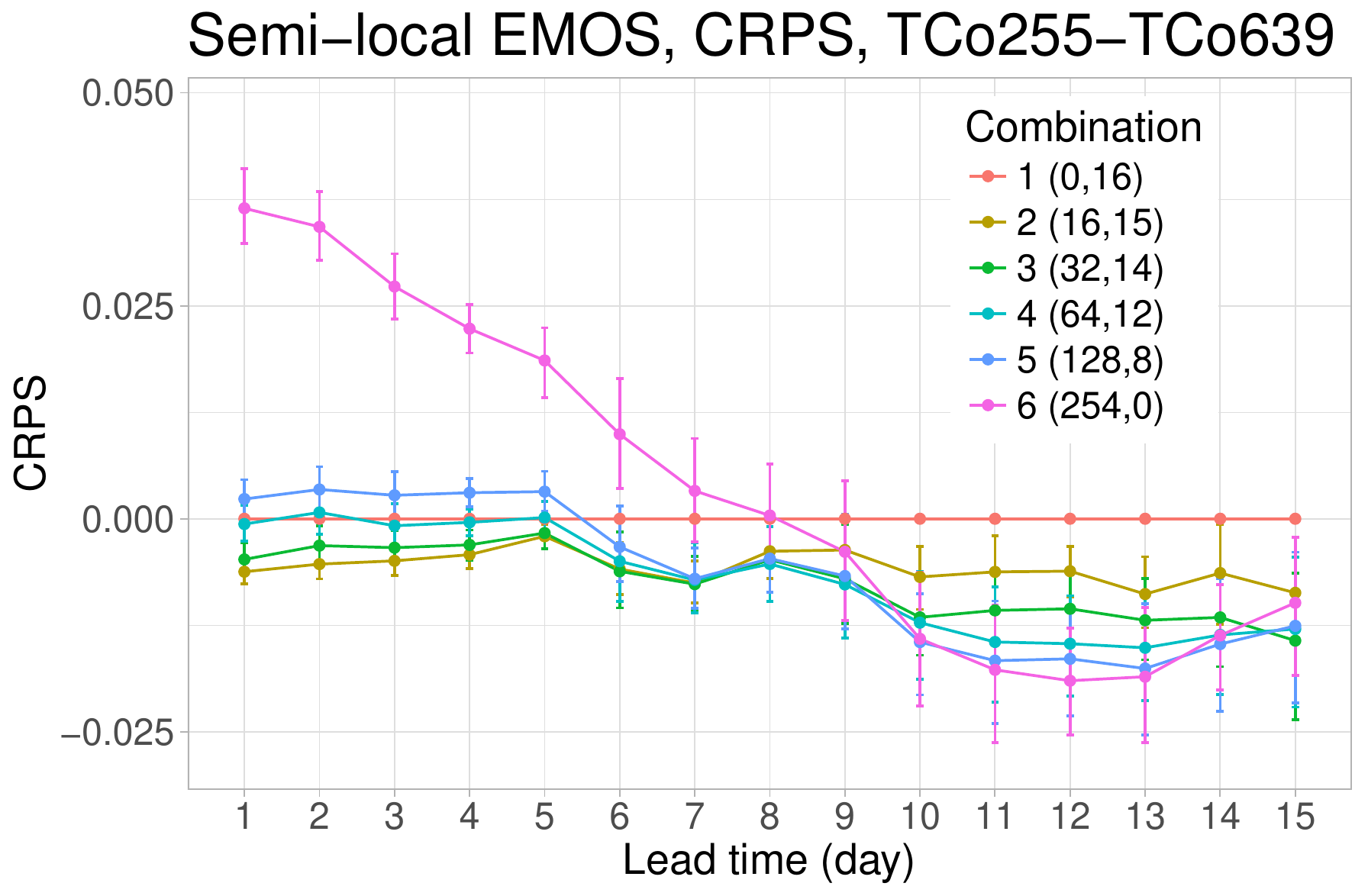, width=.49\textwidth} 
\caption{Difference in mean CRPS (the lower the better) from the reference pure high resolution case with $95\,\%$ confidence intervals of local ({\em top}) and semi-local ({\em bottom}) EMOS post-processed global dual-resolution ensemble forecasts for 2m temperature, LHPC scenario.}
\label{fig:crpsDiffEMOSL}
\end{figure}

For raw ensemble forecasts, both resolution combinations prefer balanced mixtures practically for all lead times, i.e.\ (40,40) for TCo399 - TCo639 and (16,15) for TCo255 - TCo639, which is consistent with the results of \citet{lbb}. This can be clearly observed in Figure~\ref{fig:crpsRawL} showing the mean CRPS values of  dual-resolution ensemble forecasts for 2-metre temperature and the difference in mean CRPS with respect to the pure high resolution case as function of lead time.

\afterpage{\clearpage} 
\begin{figure}[ht]

\epsfig{file=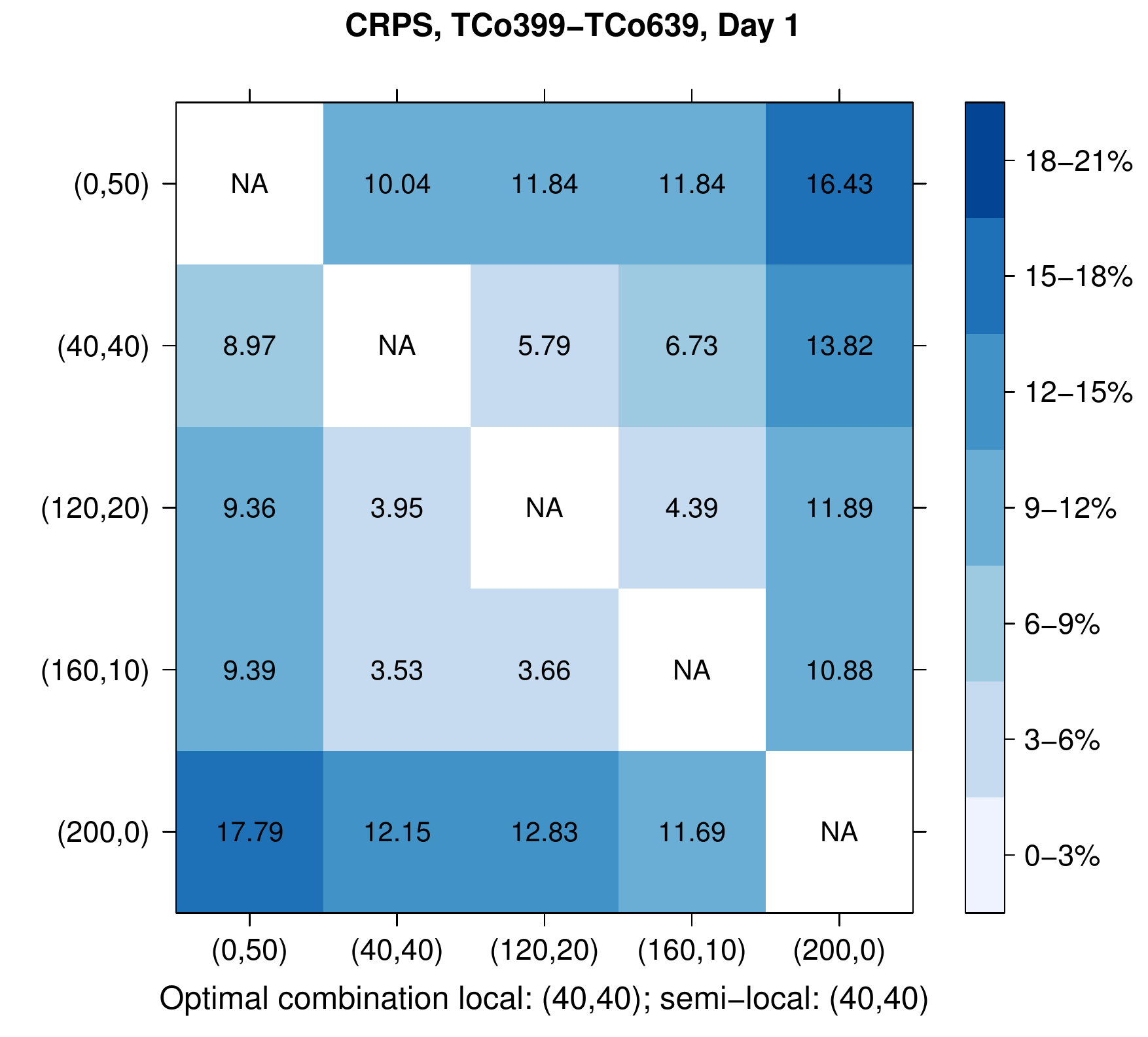, width=.49\textwidth} \
\epsfig{file=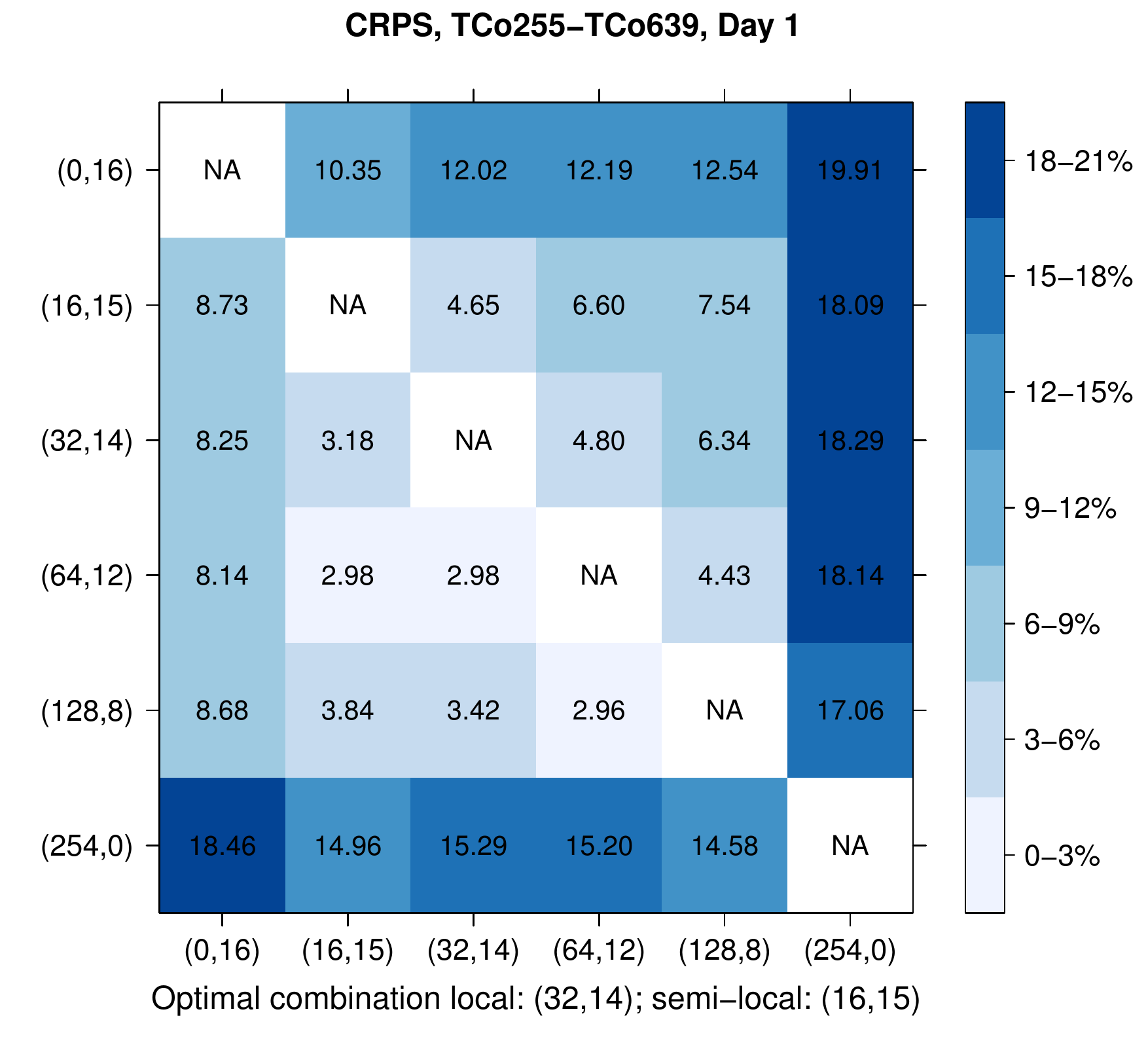, width=.49\textwidth}

\medskip
\epsfig{file=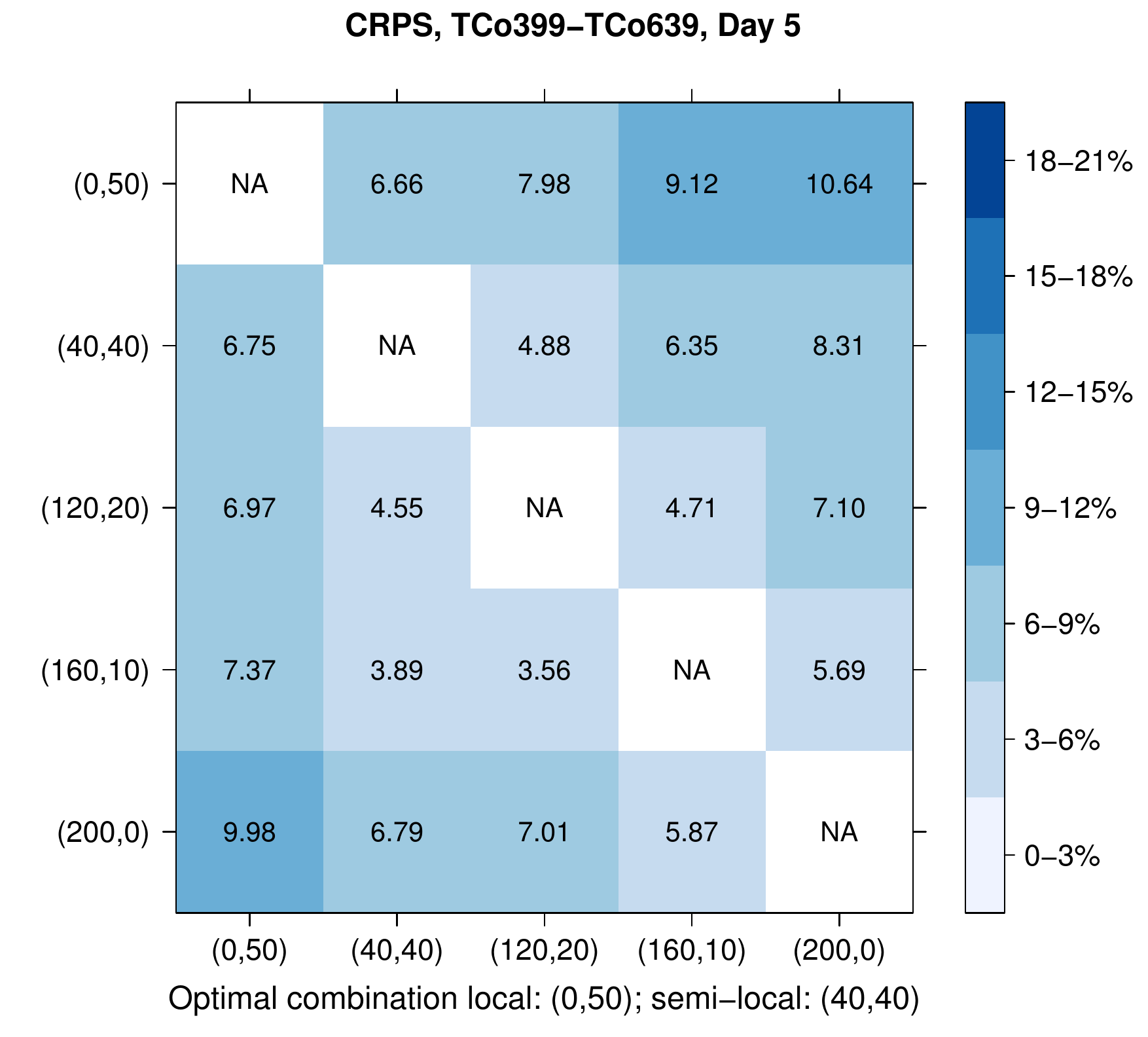, width=.49\textwidth} \
\epsfig{file=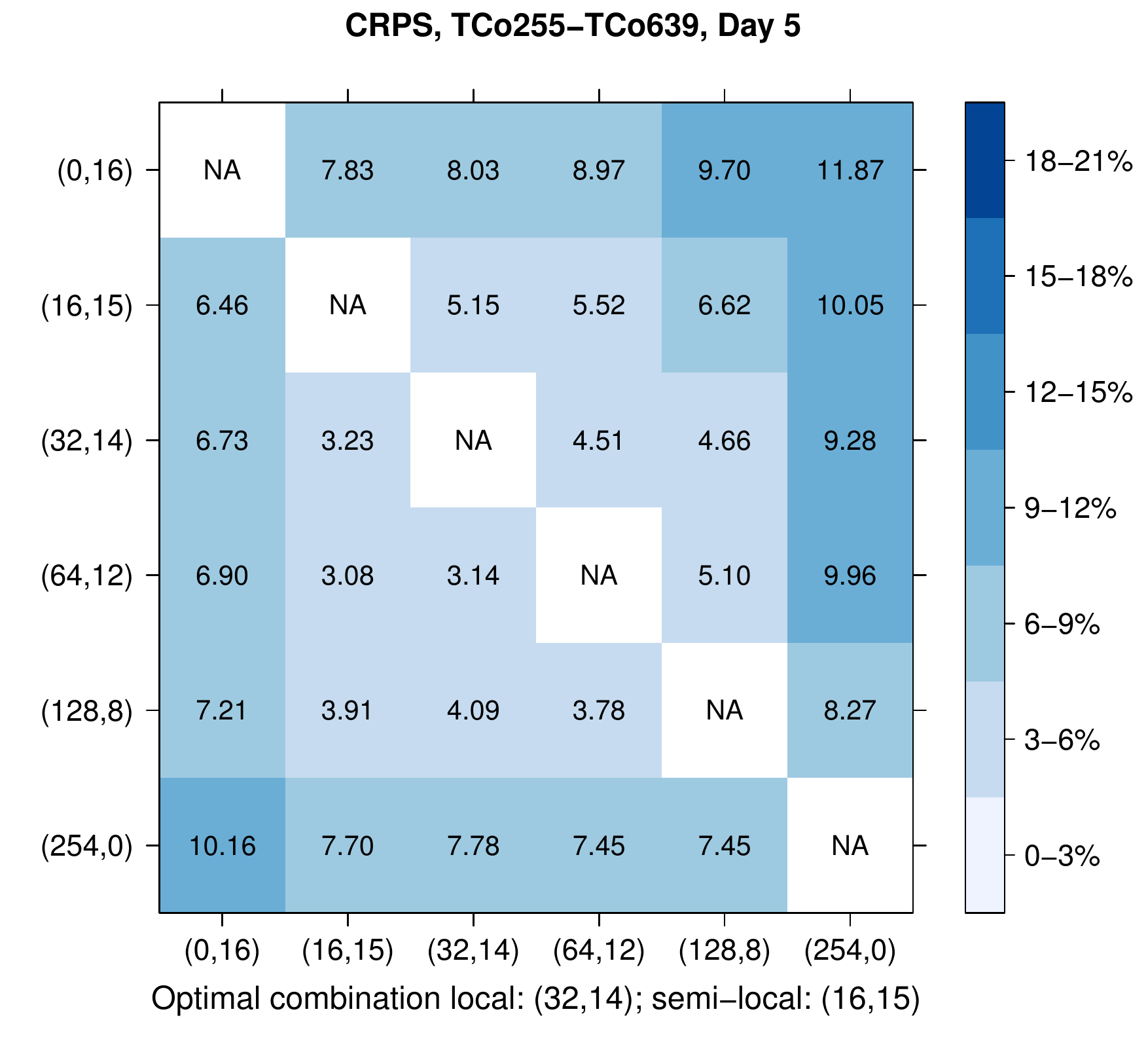, width=.49\textwidth}

\medskip
\epsfig{file=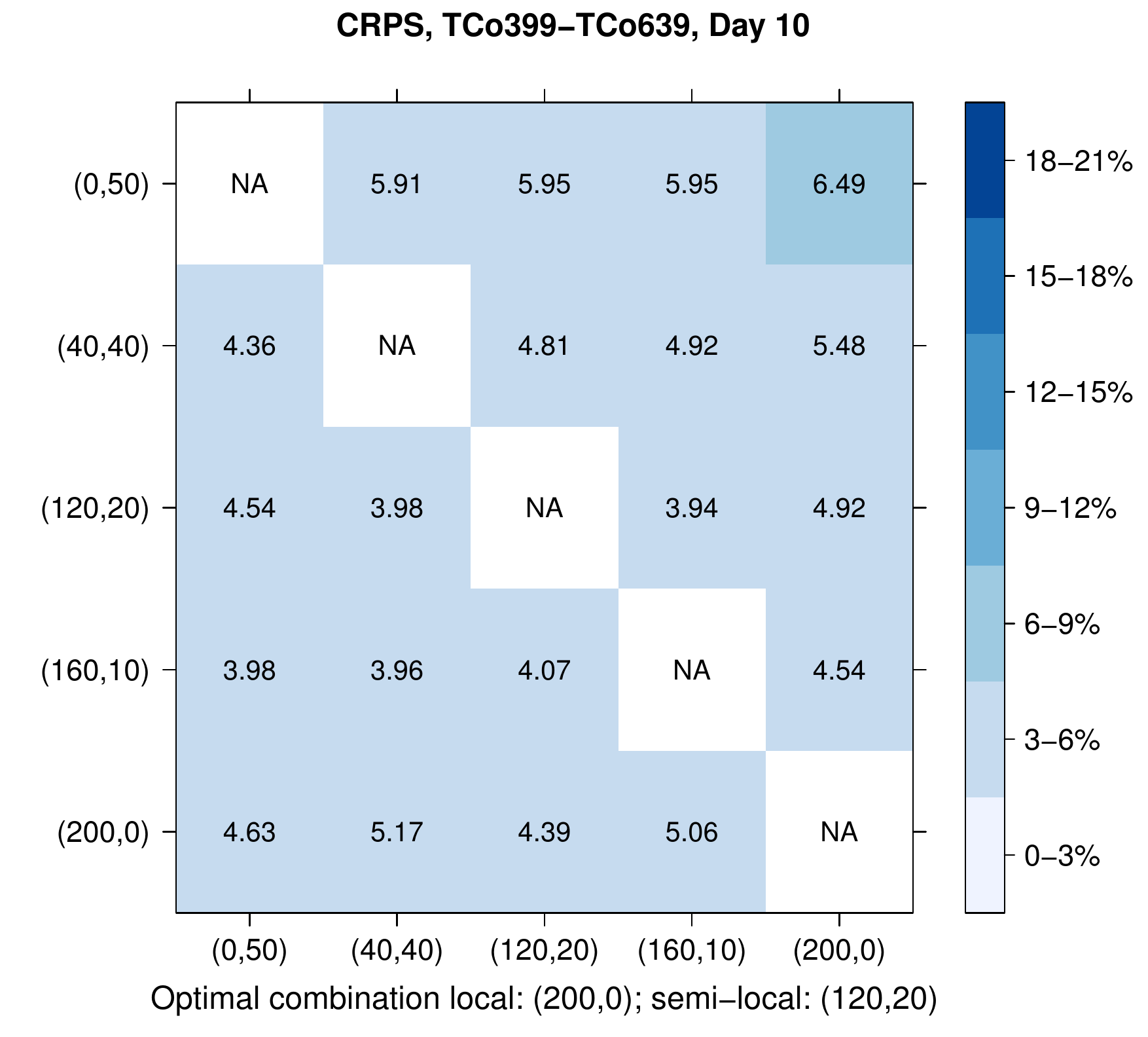, width=.49\textwidth} \
\epsfig{file=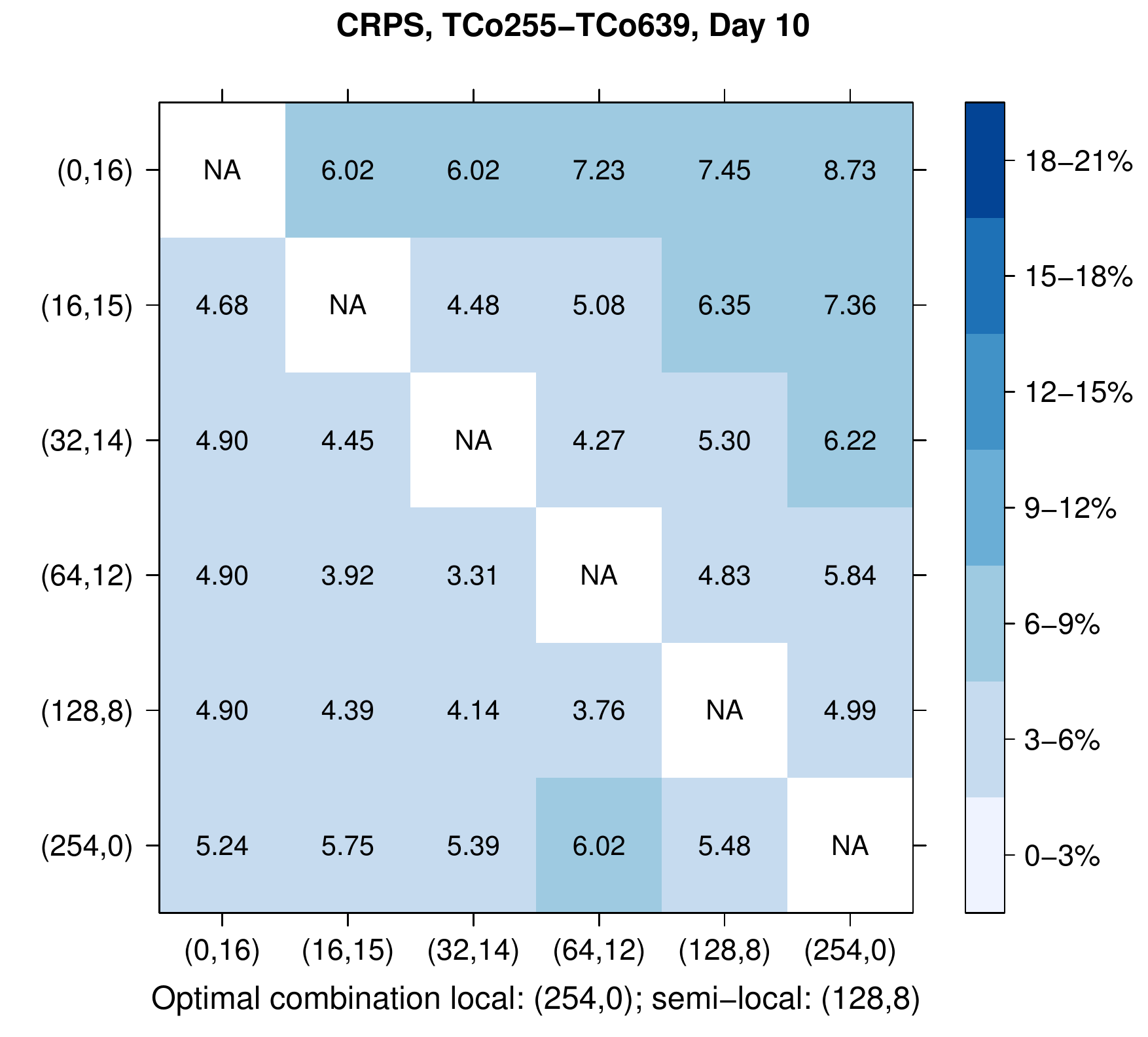, width=.49\textwidth} 
\caption{Proportion of stations with significantly different mean CRPS at a $5\,\%$ level for different lead times for local (lower triangle) and semi-local (upper triangle) parameter estimation approaches, LHPC scenario.}
\label{fig:crpsSigL}
\end{figure}

Statistical post-processing substantially changes this picture. As depicted in Figure~\ref{fig:crpsEMOSL}, semi-local EMOS significantly decreases the mean CRPS for all lead times, and the differences between the predictive performance of the various mixture combinations is strongly reduced: more so for TCo399 - TCo639 than for TCo255 - TCo639. Local EMOS yields very similar results (not shown).

\begin{figure}[!t]
\epsfig{file=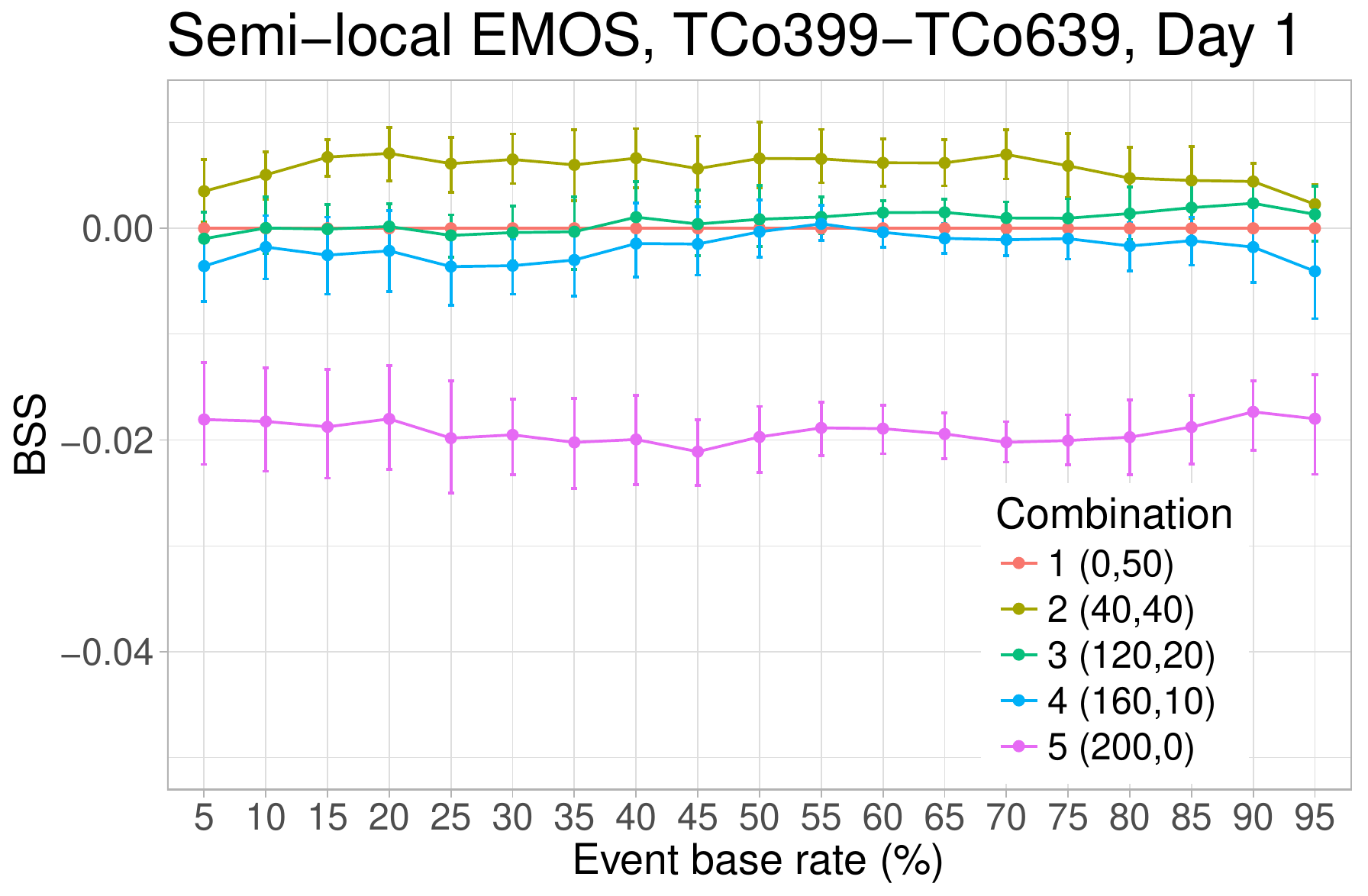, width=.49\textwidth} \ 
\epsfig{file=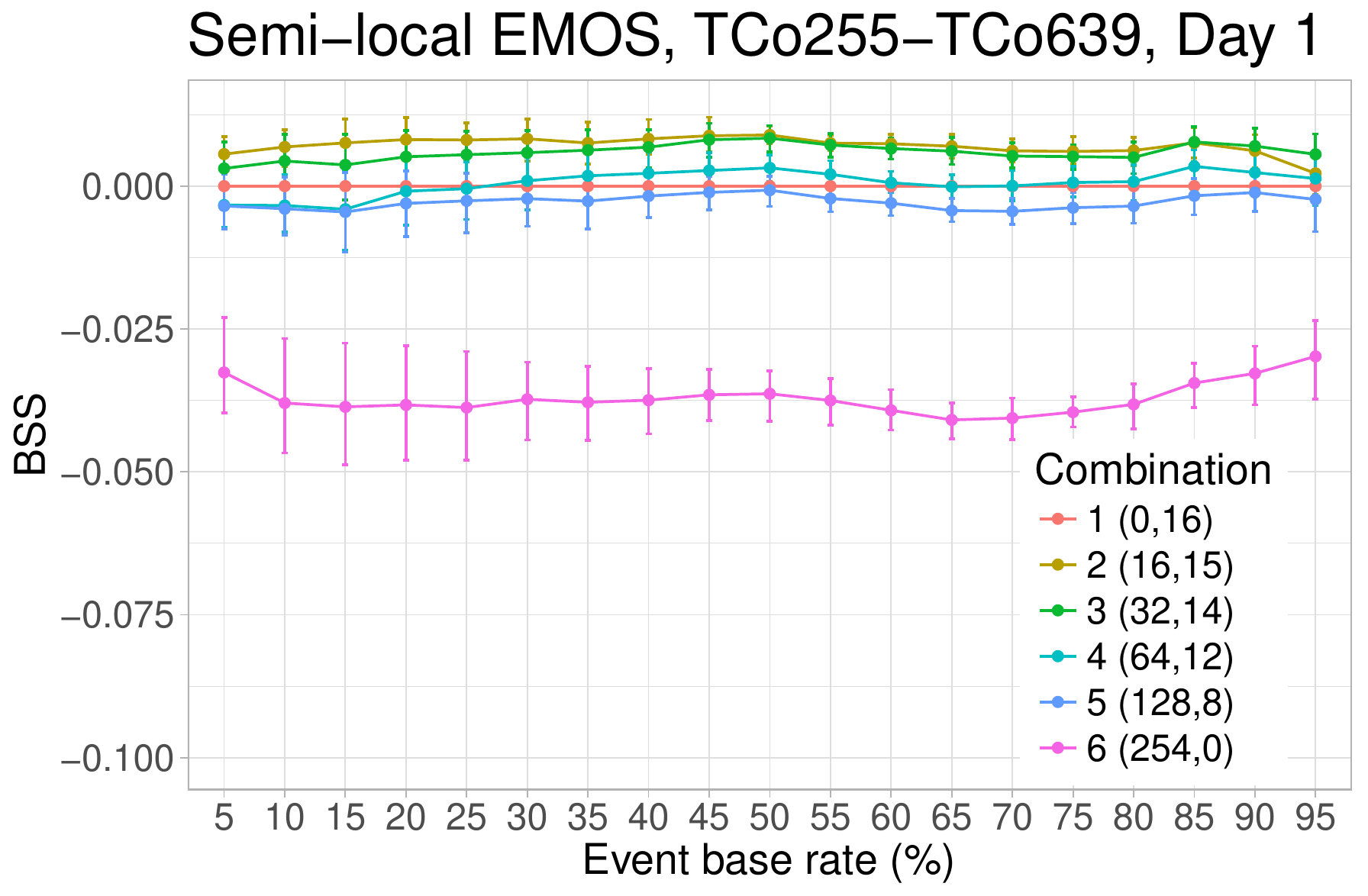, width=.49\textwidth}

\medskip
\epsfig{file=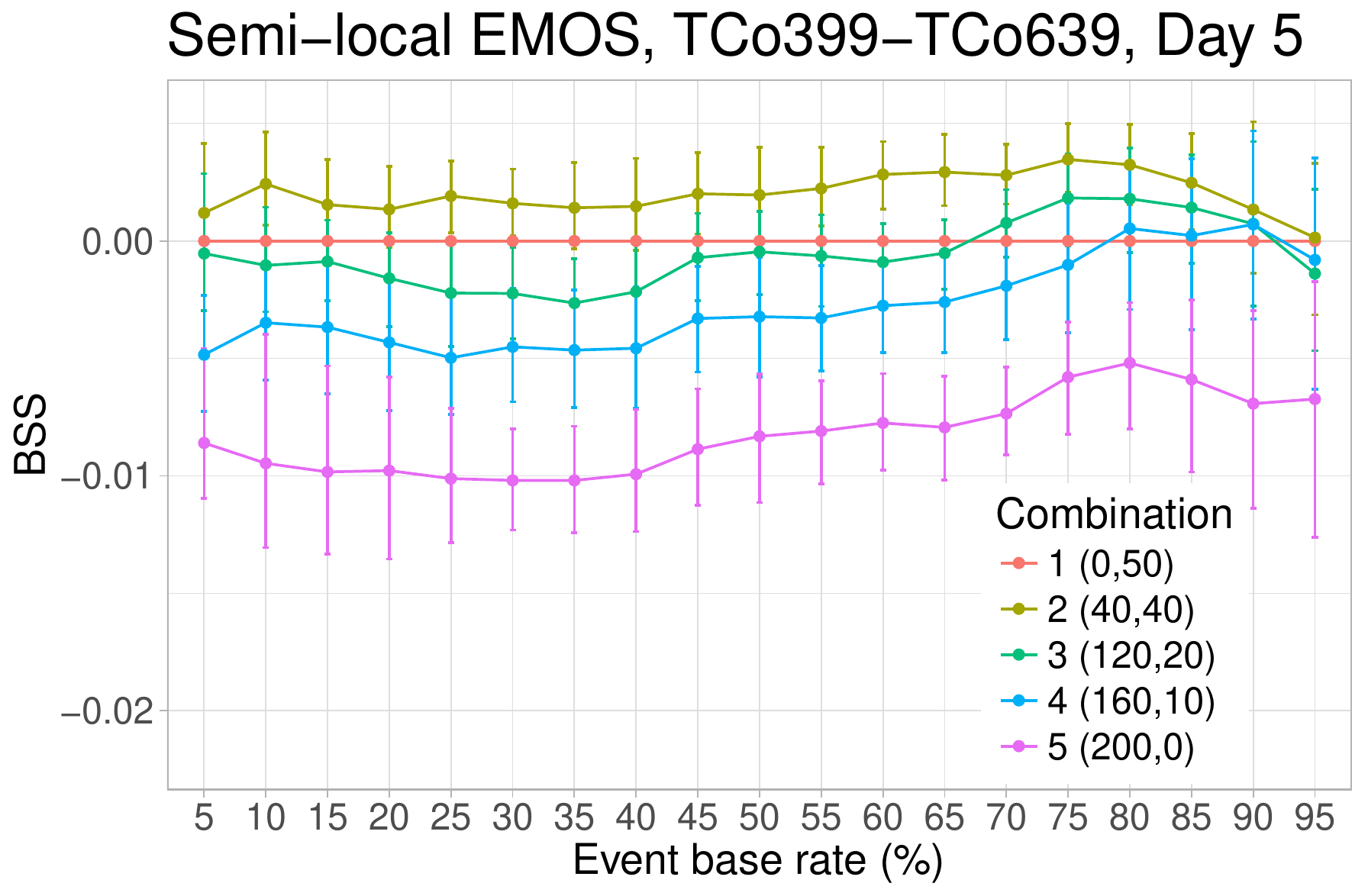, width=.49\textwidth} \ 
\epsfig{file=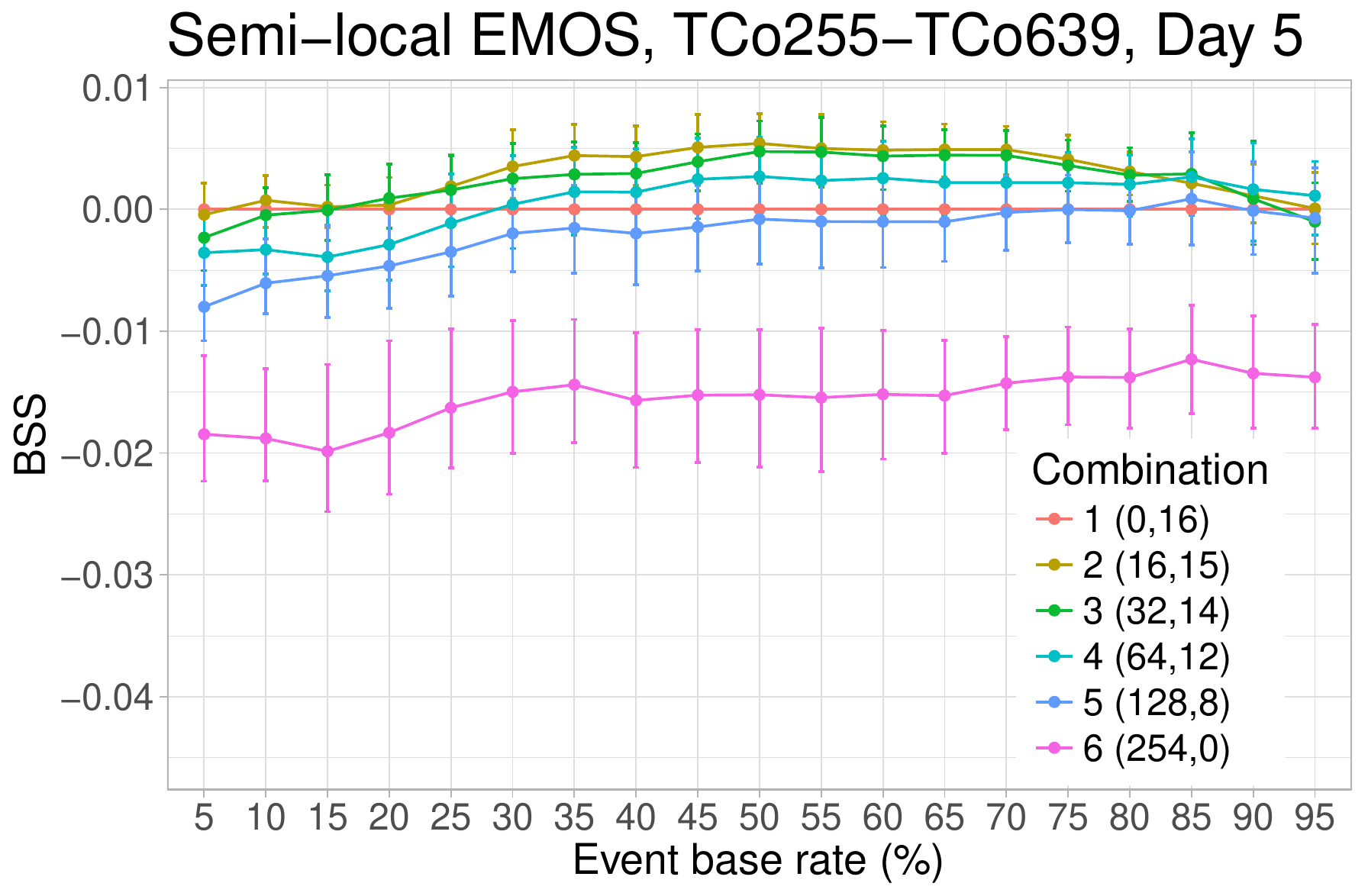, width=.49\textwidth}

\medskip
\epsfig{file=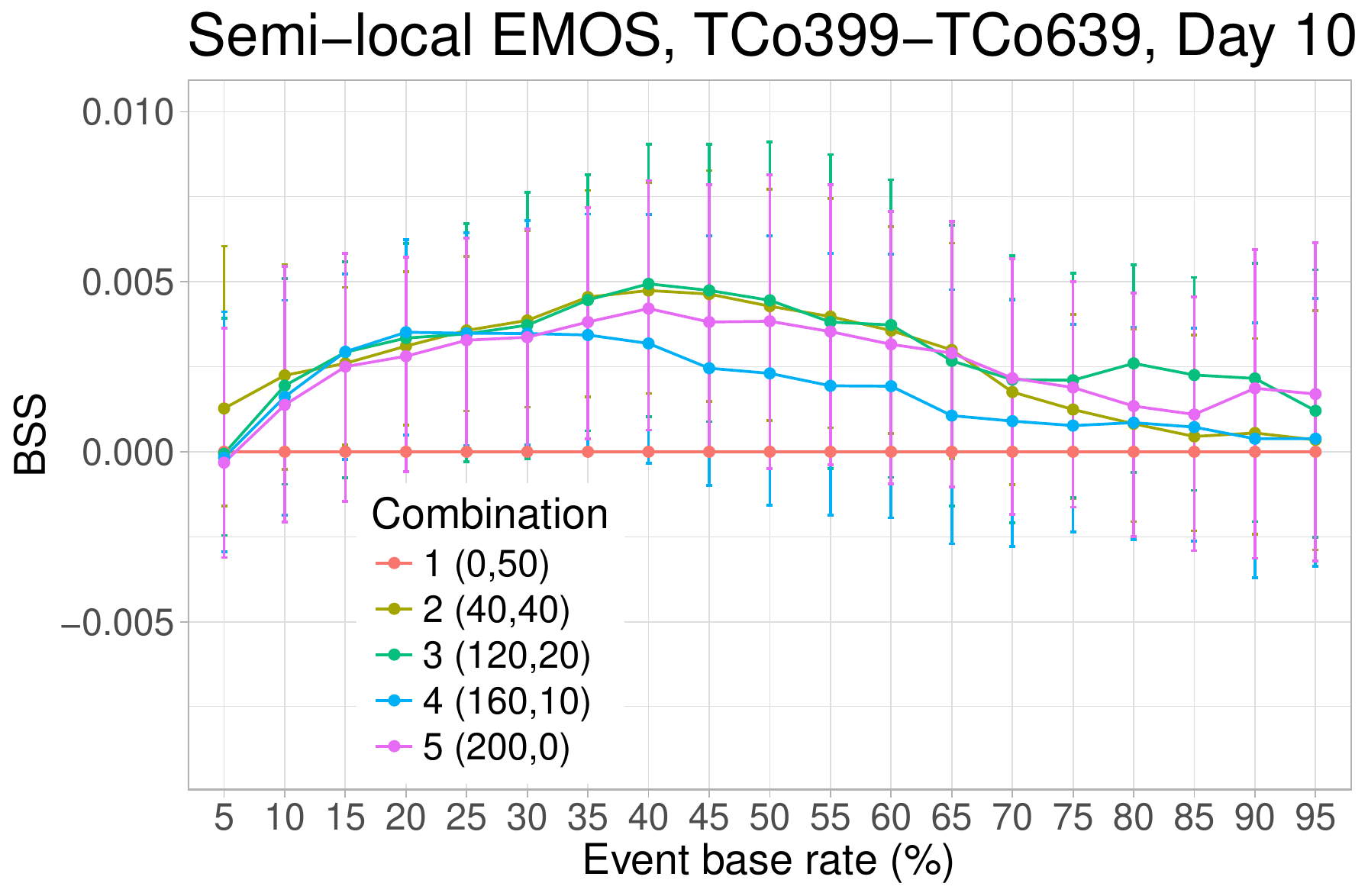, width=.49\textwidth} \ 
\epsfig{file=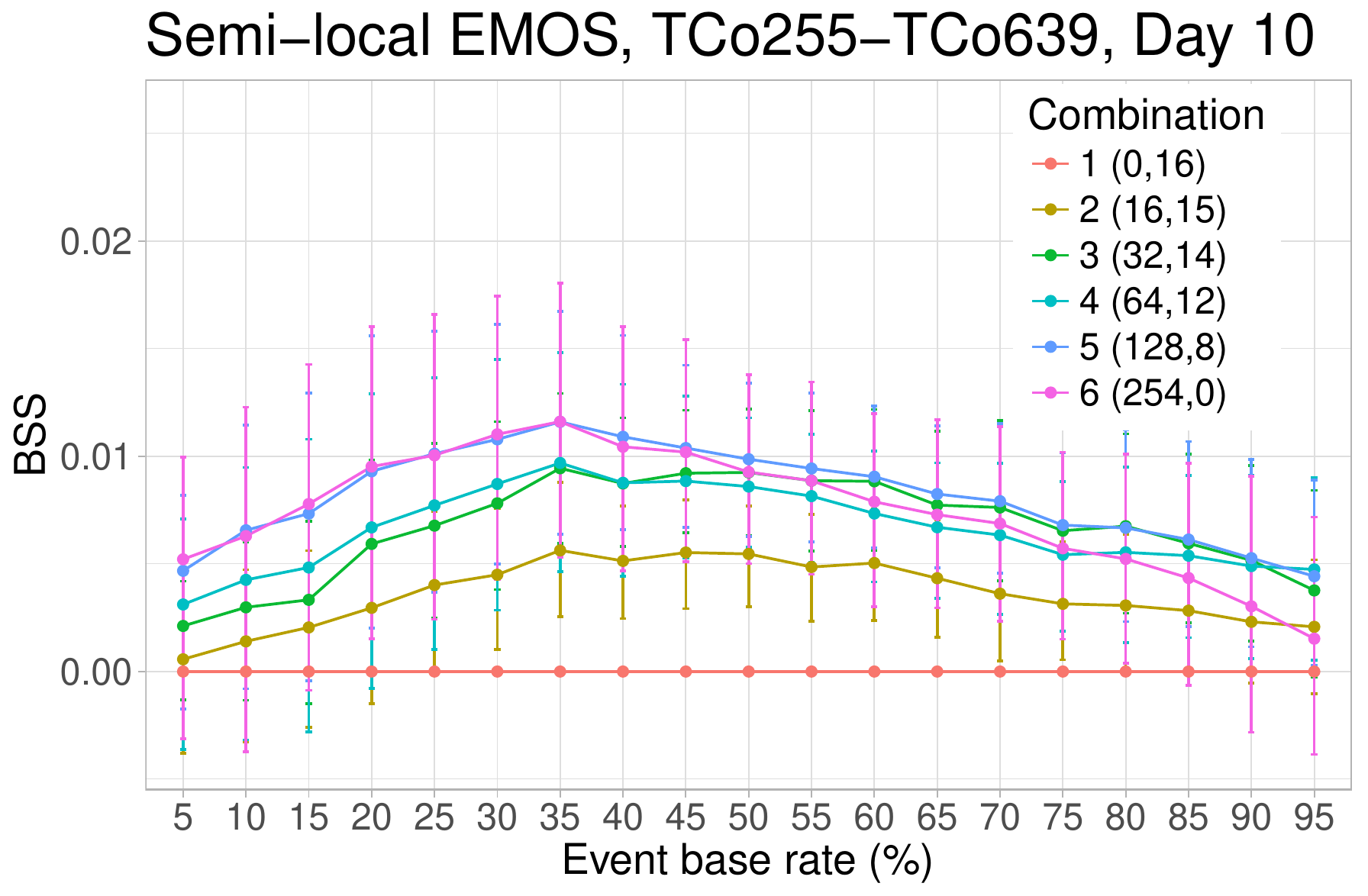, width=.49\textwidth} 
\caption{Brier skill scores (the higher the better) with respect to the reference pure high resolution case with $95\,\%$ confidence intervals of semi-local EMOS post-processed global dual-resolution ensemble forecasts for 2m temperature, LHPC scenario.}
\label{fig:bsDiffEMOSL}
\end{figure}

Differences in mean CRPS with respect to the pure high resolution case are shown in Figure~\ref{fig:crpsDiffEMOSL}. After day 8, the pure low resolution ensemble outperforms the pure high resolution one both for TCo399 - TCo639 and TCo255 - TCo639 resolutions, whereas for short lead times, semi-local EMOS, prefers balanced combinations, which are also optimal in the raw ensemble. Note that for longer lead times score differences based on local EMOS are more variable leading to wider confidence intervals. Hence, in the remaining part of this section we are presenting mainly the results of semi-local EMOS calibration.

\begin{figure}[!t]
\epsfig{file=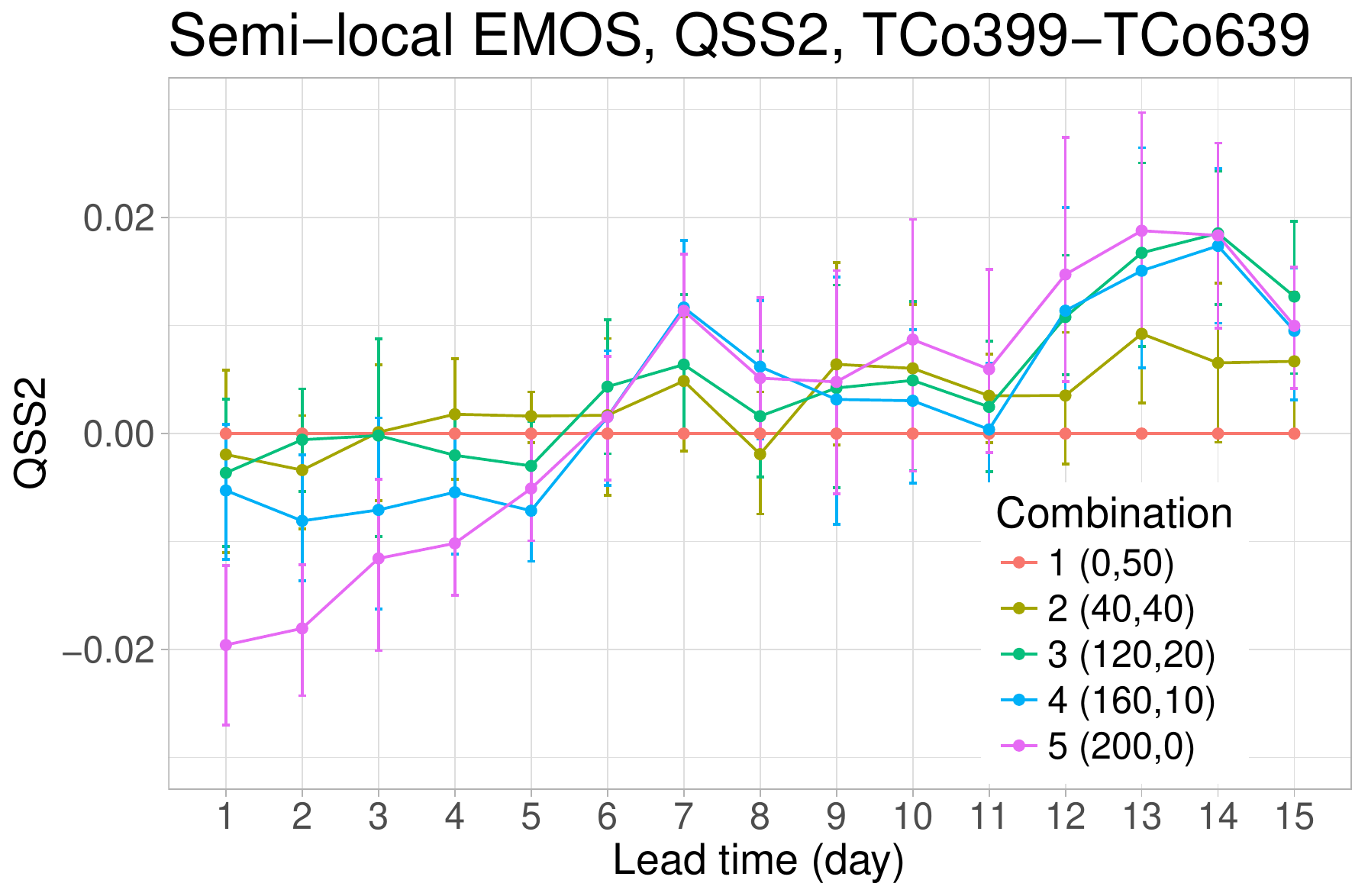, width=.49\textwidth} \ 
\epsfig{file=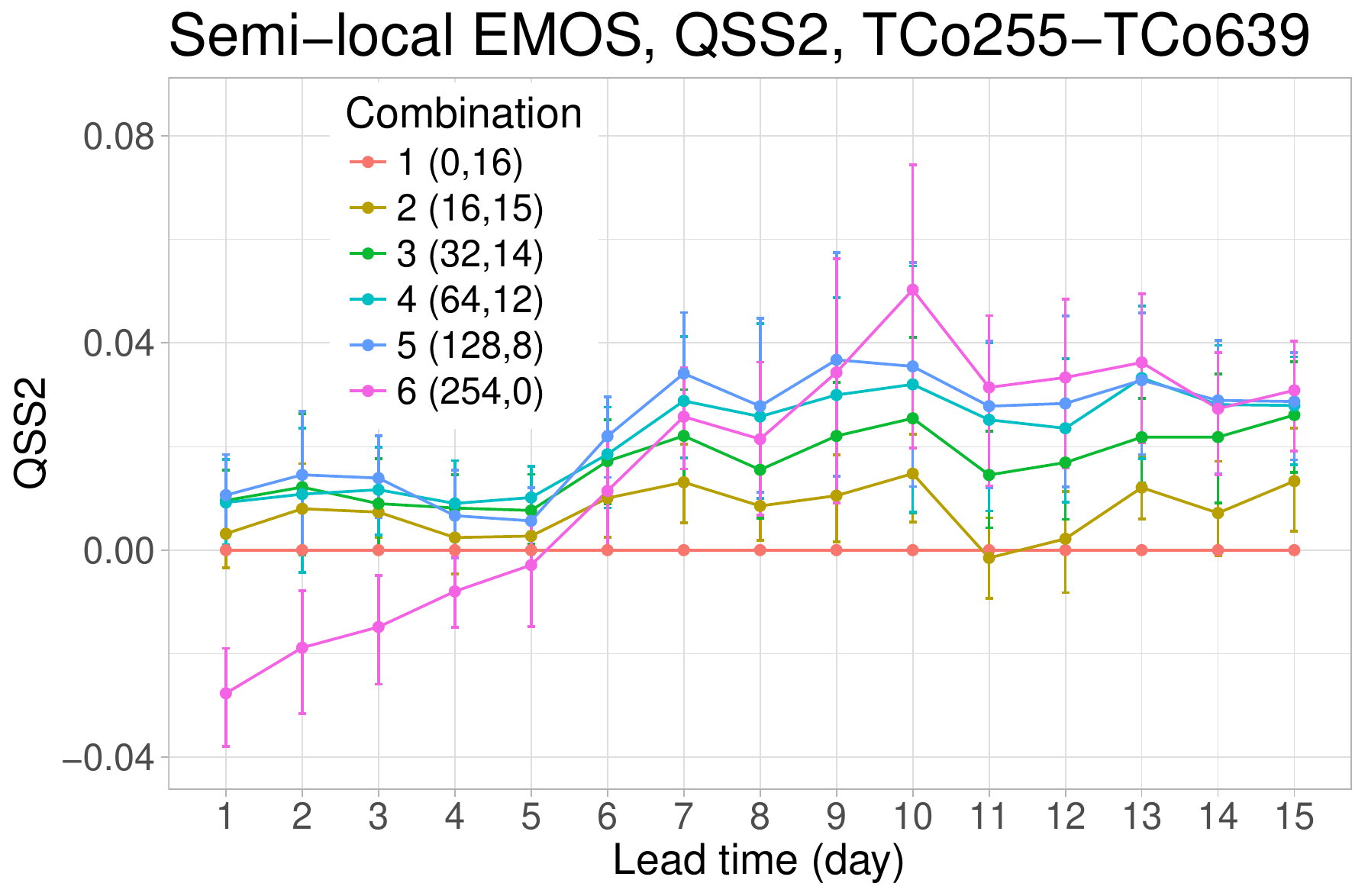, width=.49\textwidth}

\medskip
\epsfig{file=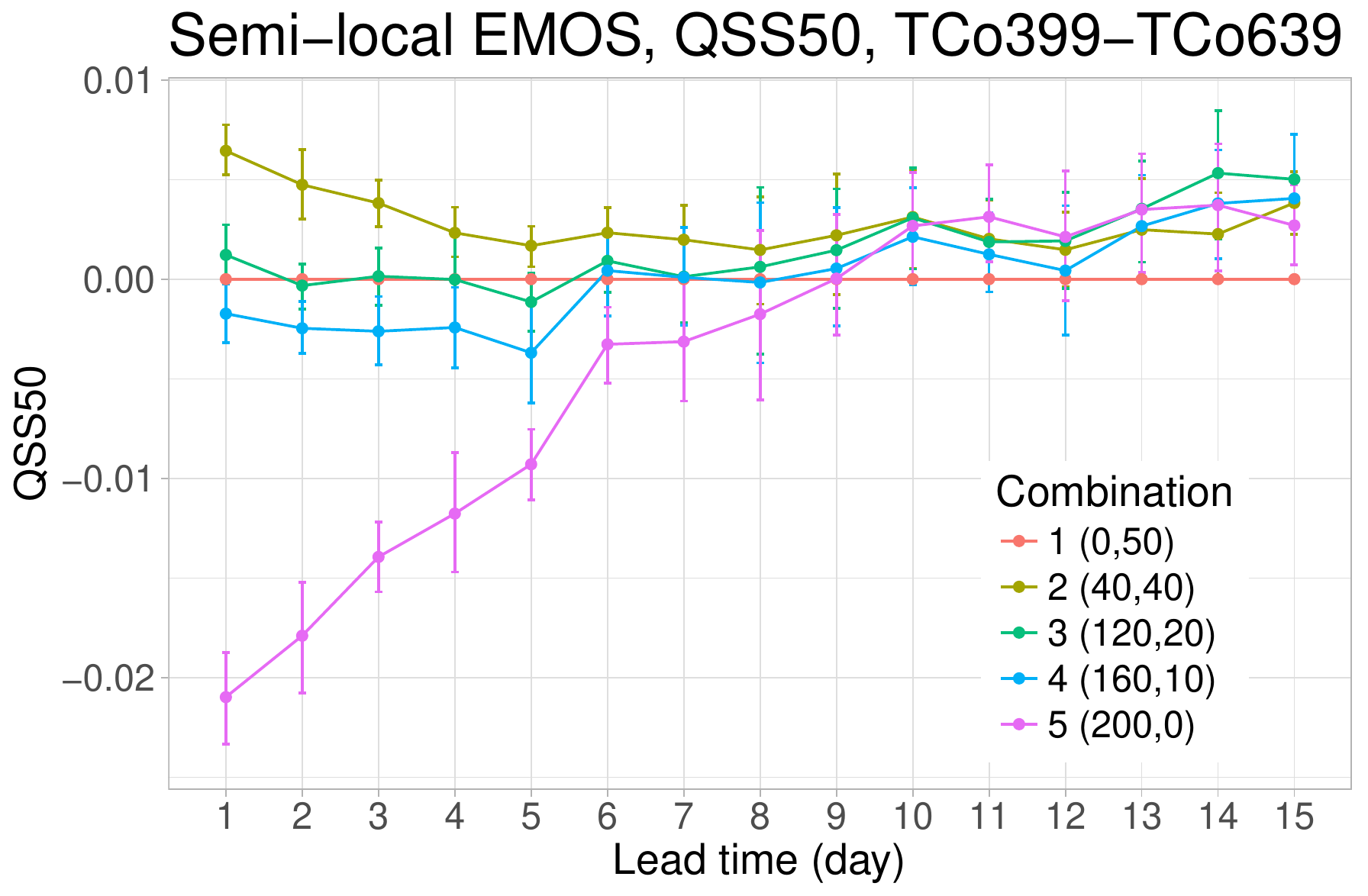, width=.49\textwidth} \ 
\epsfig{file=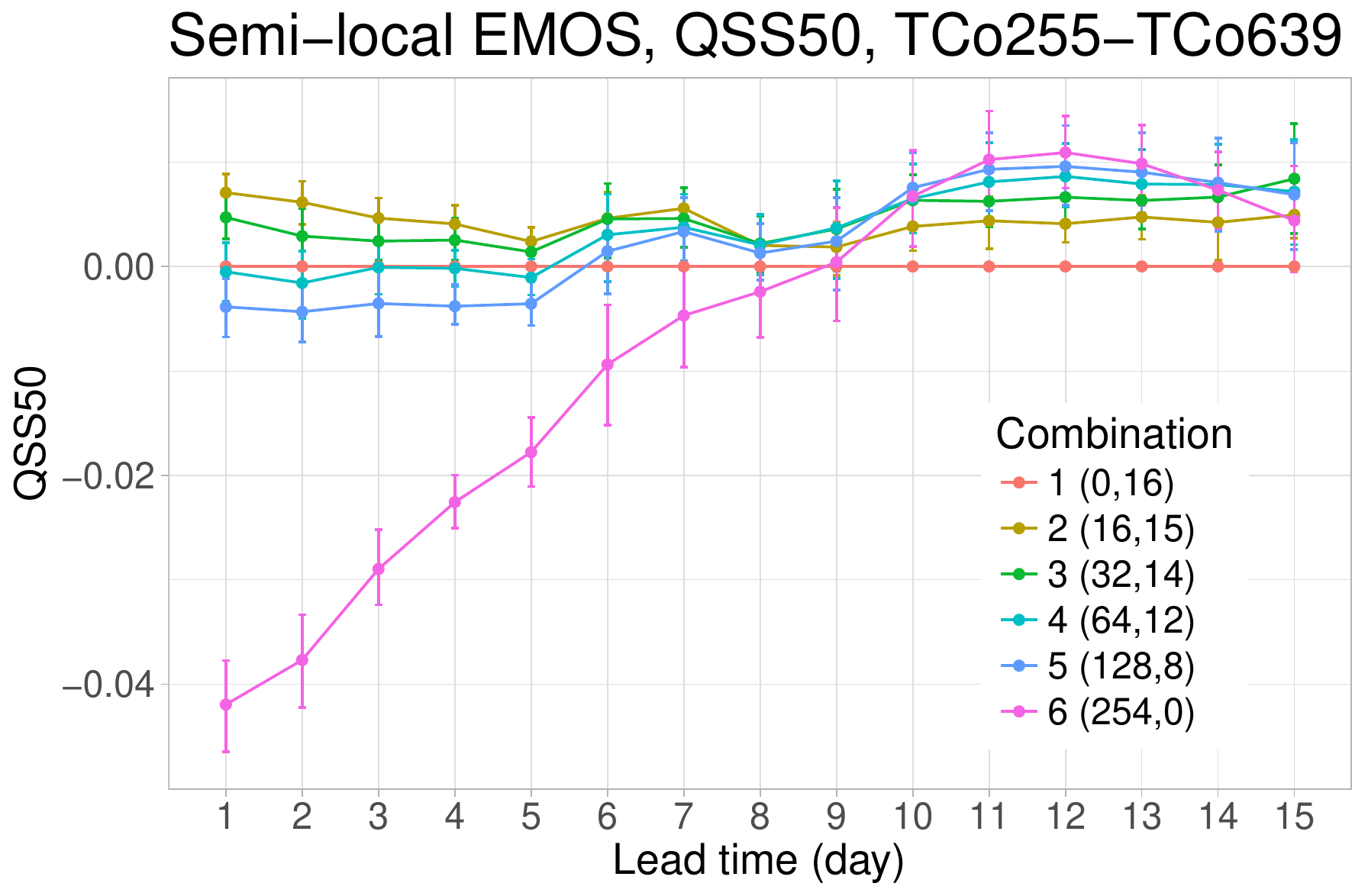, width=.49\textwidth}

\medskip
\epsfig{file=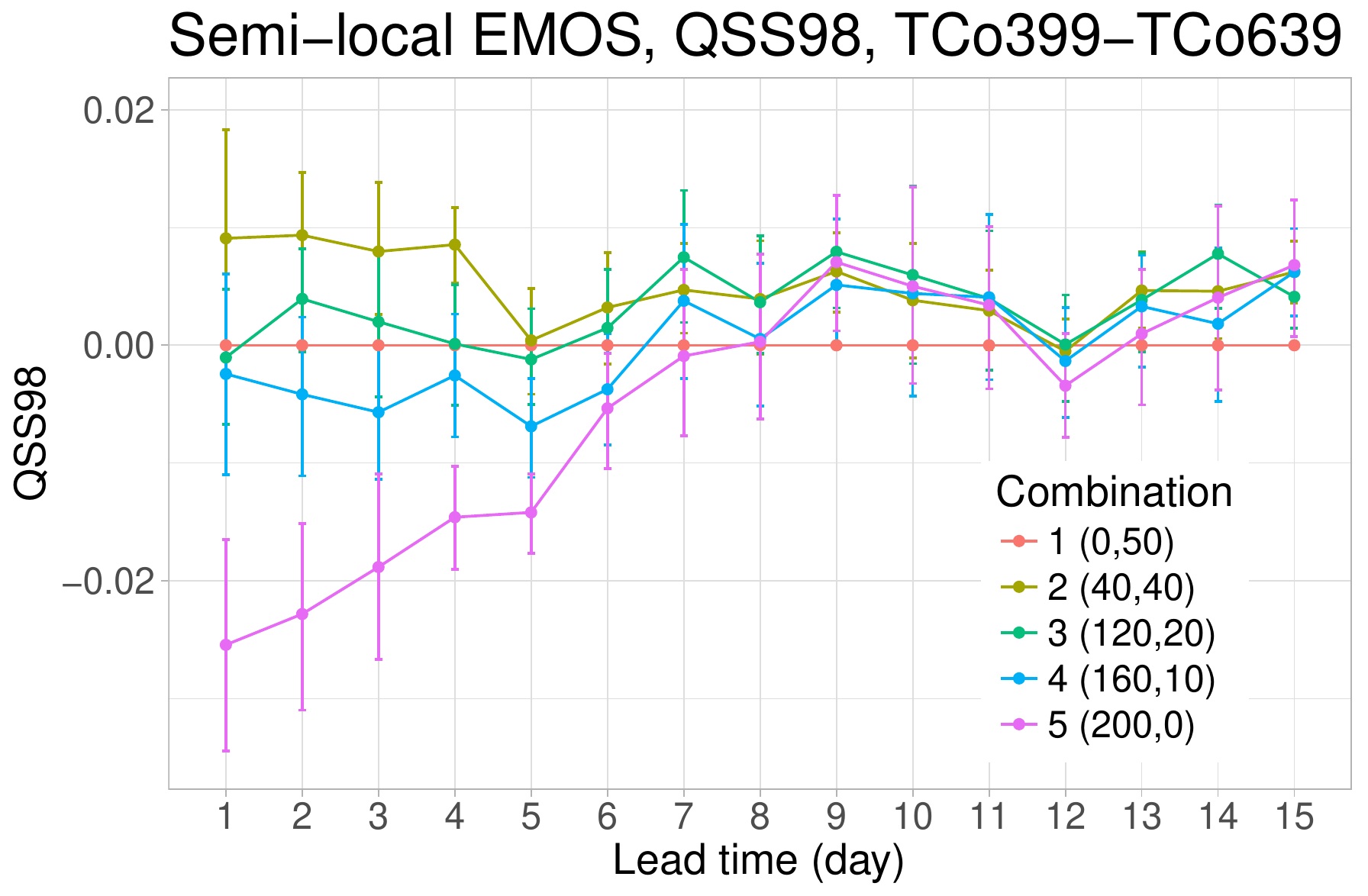, width=.49\textwidth} \ 
\epsfig{file=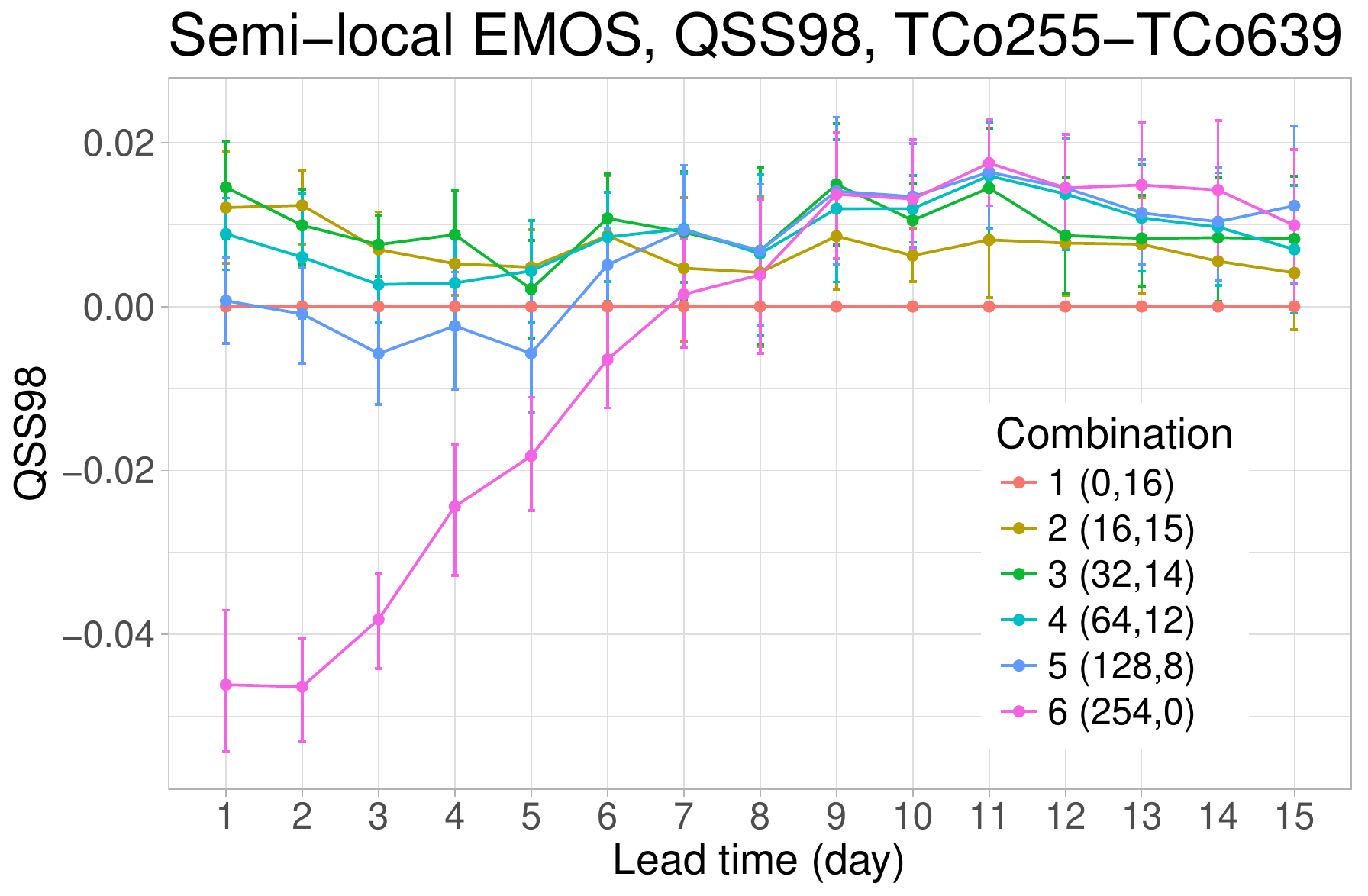, width=.49\textwidth} 
\caption{Quantile skill scores (the higher the better) for percentiles 2 (top), 50 (middle) and 98 (bottom) with respect to the reference pure high resolution case with $95\,\%$ confidence intervals of semi-local EMOS post-processed global dual-resolution ensemble forecasts for 2m temperature, LHPC scenario.}
\label{fig:qsDiffEMOSL}
\end{figure}

To investigate the statistical significance of differences between the mean CRPS values corresponding to various combinations, stationwise DM tests were performed. Figure~\ref{fig:crpsSigL} shows the proportion of stations with significantly different mean CRPS at a $5\,\%$ level for different lead times for local (lower triangle) and semi-local (upper triangle) parameter estimation approaches. In general, longer lead times result in a smaller proportion of stations with significant difference both for local and semi-local EMOS post-processing and the values in the first column and row of each matrix are consistent with Figure~\ref{fig:crpsDiffEMOSL}, respectively.

A natural question to investigate is whether stations with significant difference in mean CRPS between the optimal combination and the reference pure high resolution case exhibit some clear spatial pattern. However, visualization of stations with significant difference on a $5\,\%$ level on maps does not reveal any connection to their location.

The analysis of Brier skill scores with thresholds equal to the \ $5,10,\ldots ,95$ \ percentiles of the corresponding station sample climatology for the verification period leads us to similar conclusions.  BSS values with respect to the pure high resolution case of semi-local EMOS post-processed forecasts displayed in Figure~\ref{fig:bsDiffEMOSL} are fully consistent with the graphs in the bottom row of Figure~\ref{fig:crpsDiffEMOSL}. At days 1 and 5 the balanced combinations have the best forecast skill for all thresholds, whereas at day 10 all other combinations outperform the pure high resolution one.

Figure~\ref{fig:qsDiffEMOSL} shows differences of quantile skill scores for ensemble configurations post-process\-ed with semi-local EMOS. All differences are with respect to the pure higher-resolution configuration. For the median ($50\,\%$ quantile, middle row), results are fully consistent with the CRPS differences.  For the more extreme quantiles ($2\,\%$ and $98\,\%$, top and bottom rows),  score differences are comparable to those for the median but the confidence intervals tend to be larger. In any case, quantile score differences between configurations are statistically not strongly significant for the longer lead times.

\begin{figure}[!t]
\epsfig{file=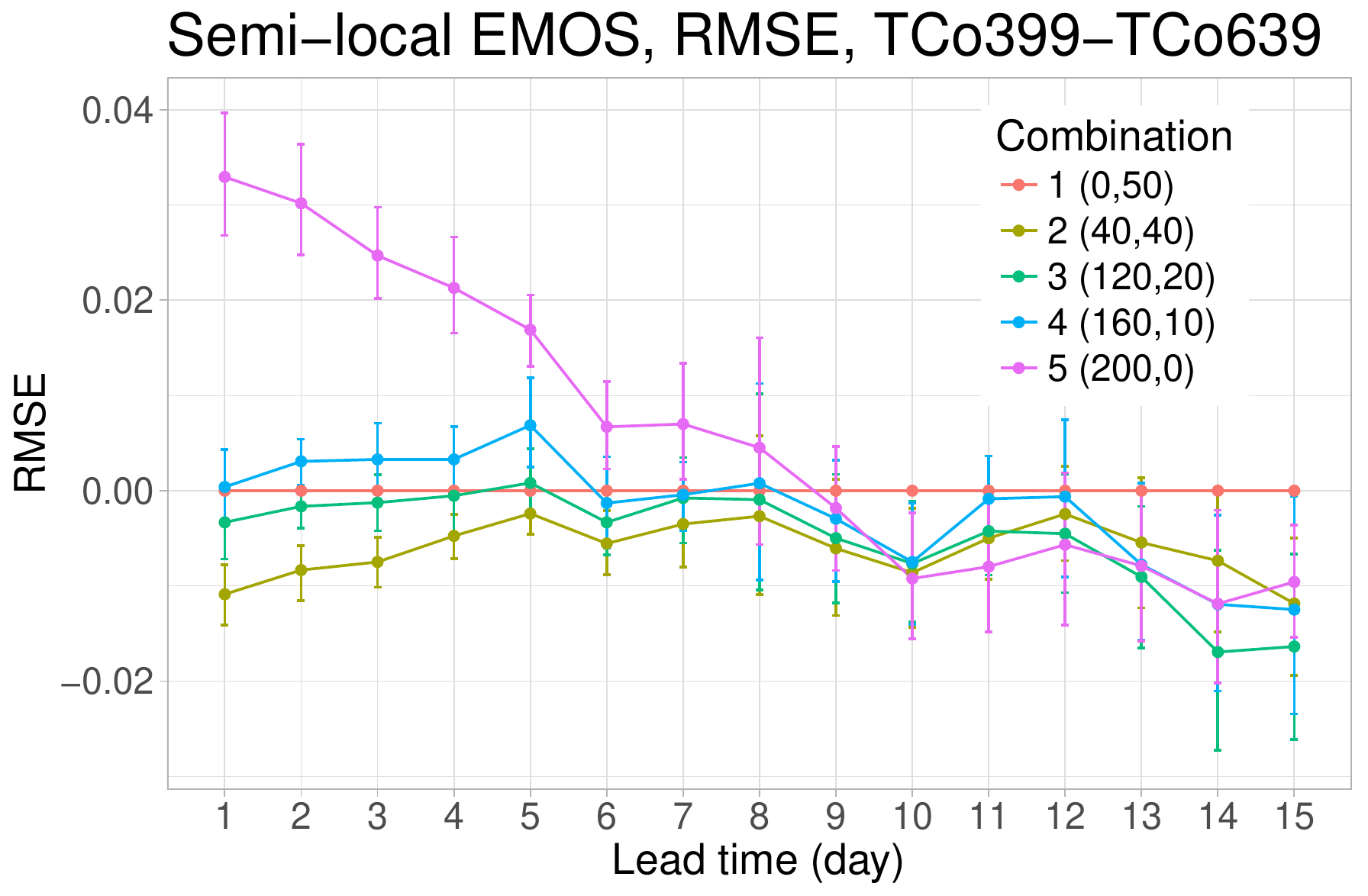, width=.49\textwidth} \
\epsfig{file=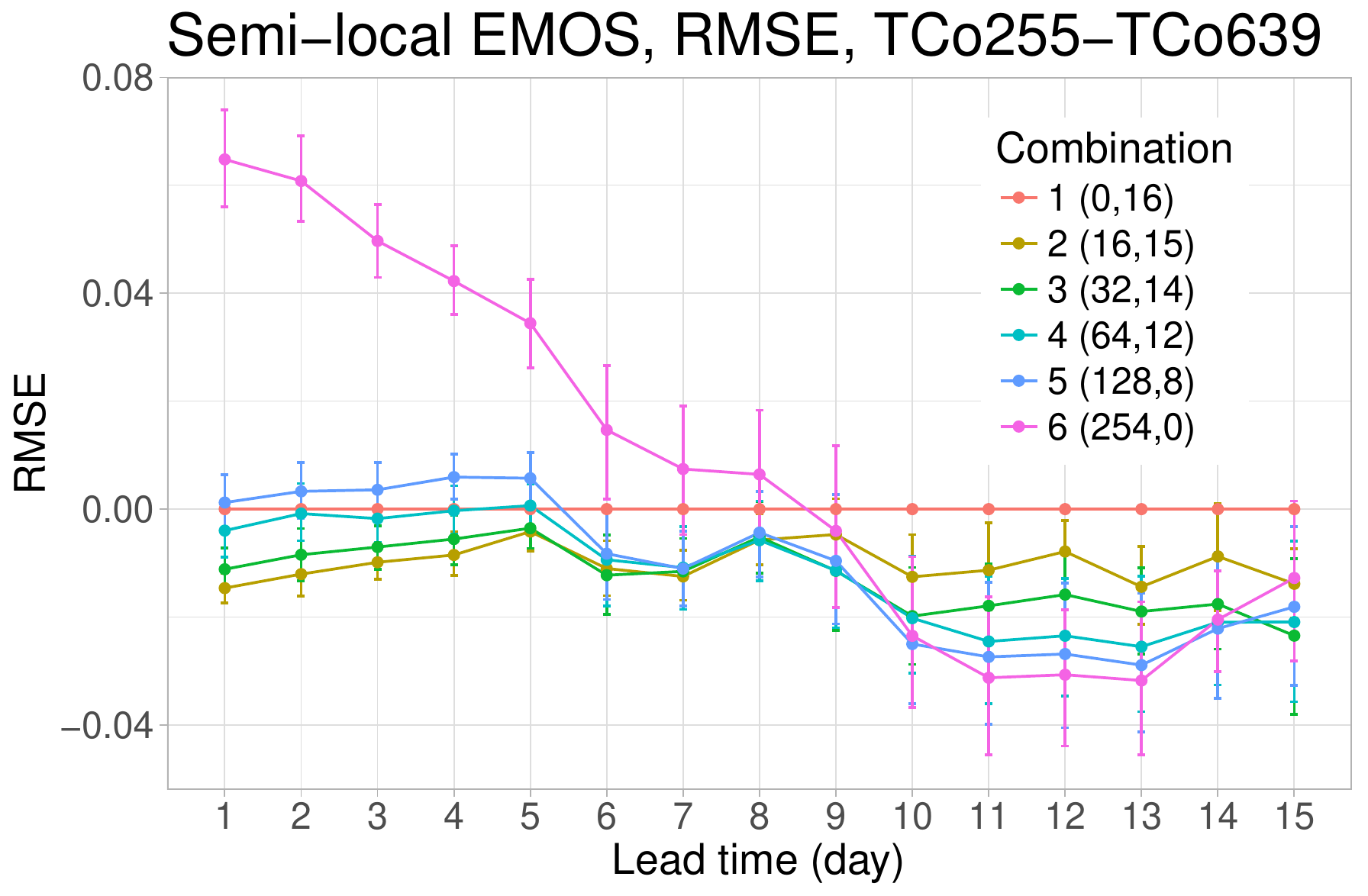, width=.49\textwidth} 
\caption{Difference in RMSE values (the lower the better) from the reference pure high resolution case with $95\,\%$ confidence intervals of semi-local EMOS post-processed global dual-resolution ensemble forecasts for 2m temperature, LHPC scenario.}
\label{fig:maeDiffEMOSL}
\end{figure}

Finally, consider the root mean squared errors of EMOS median forecasts. Figure~\ref{fig:maeDiffEMOSL} displays difference in RMSE values of semi-local EMOS post-processed combinations from the reference pure high resolution case. Note that for both mixtures the graphs are very similar to the ones in the bottom line of Figure~\ref{fig:crpsDiffEMOSL} and the same similarity of differences in mean CRPS (Figure~\ref{fig:crpsDiffEMOSL}, top) and RMSE (not reported) can be observed for local EMOS as well.

\begin{figure}[!t]
\epsfig{file=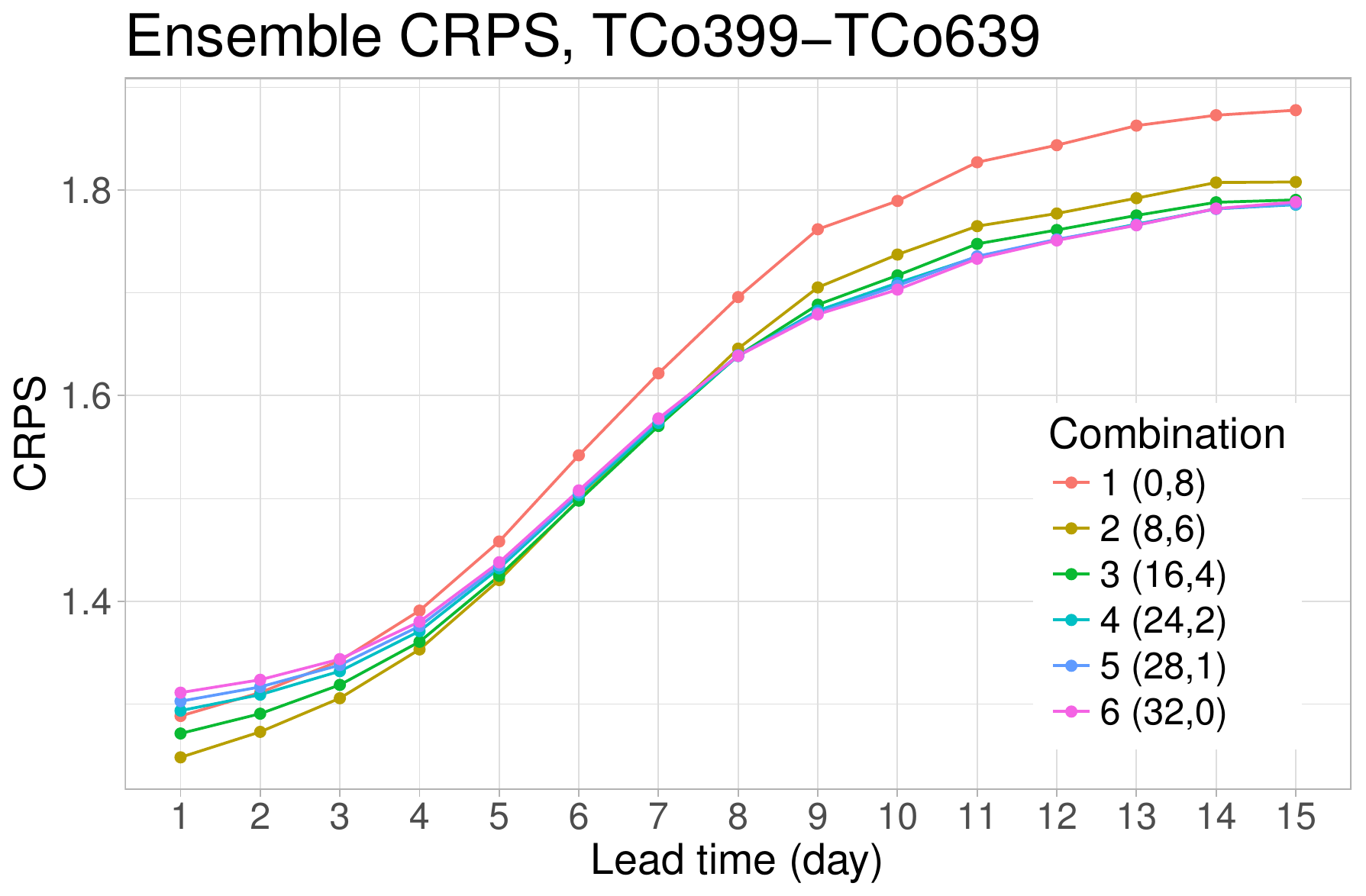, width=.49\textwidth} \
\epsfig{file=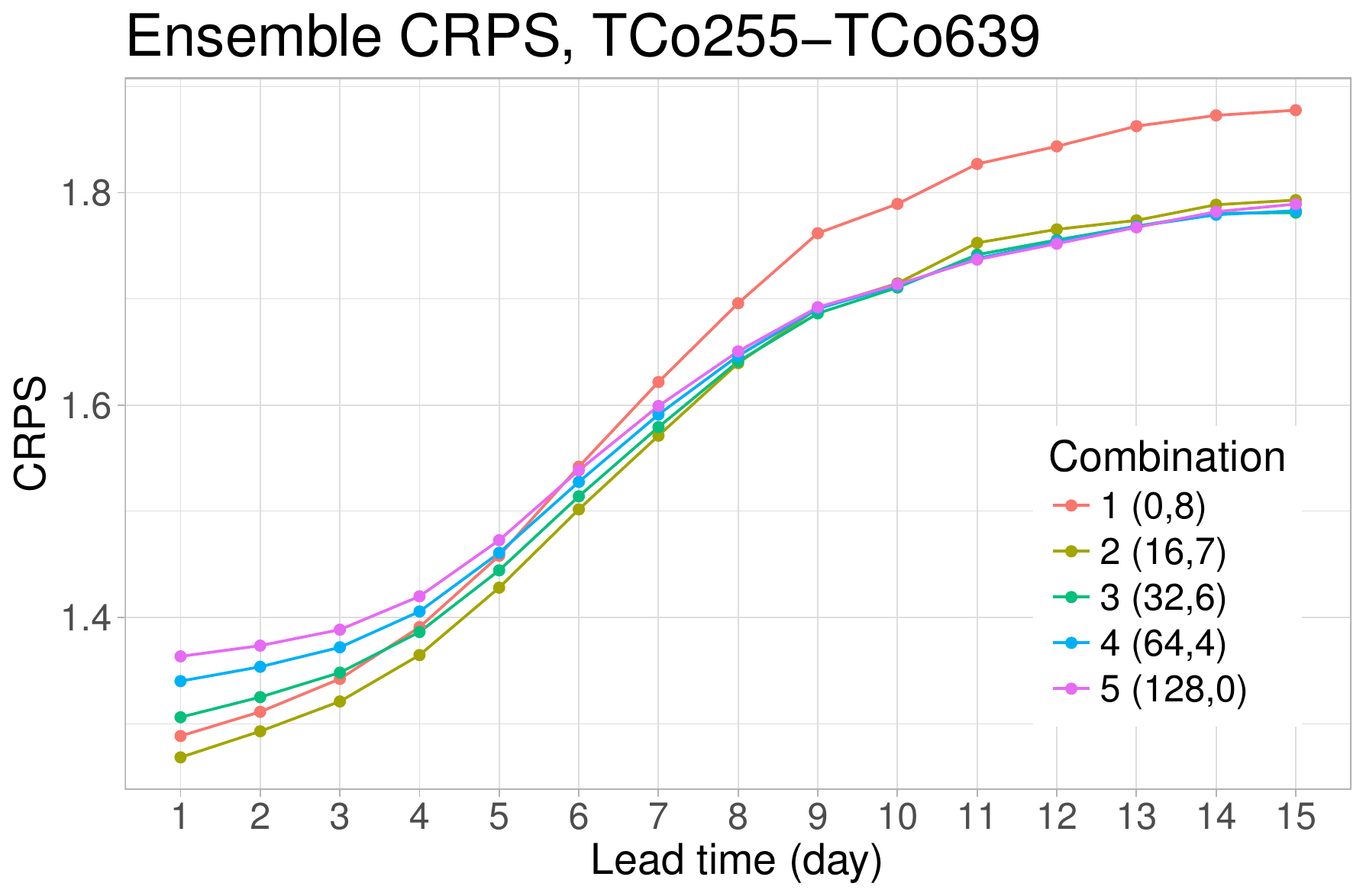, width=.49\textwidth}

\medskip
\epsfig{file=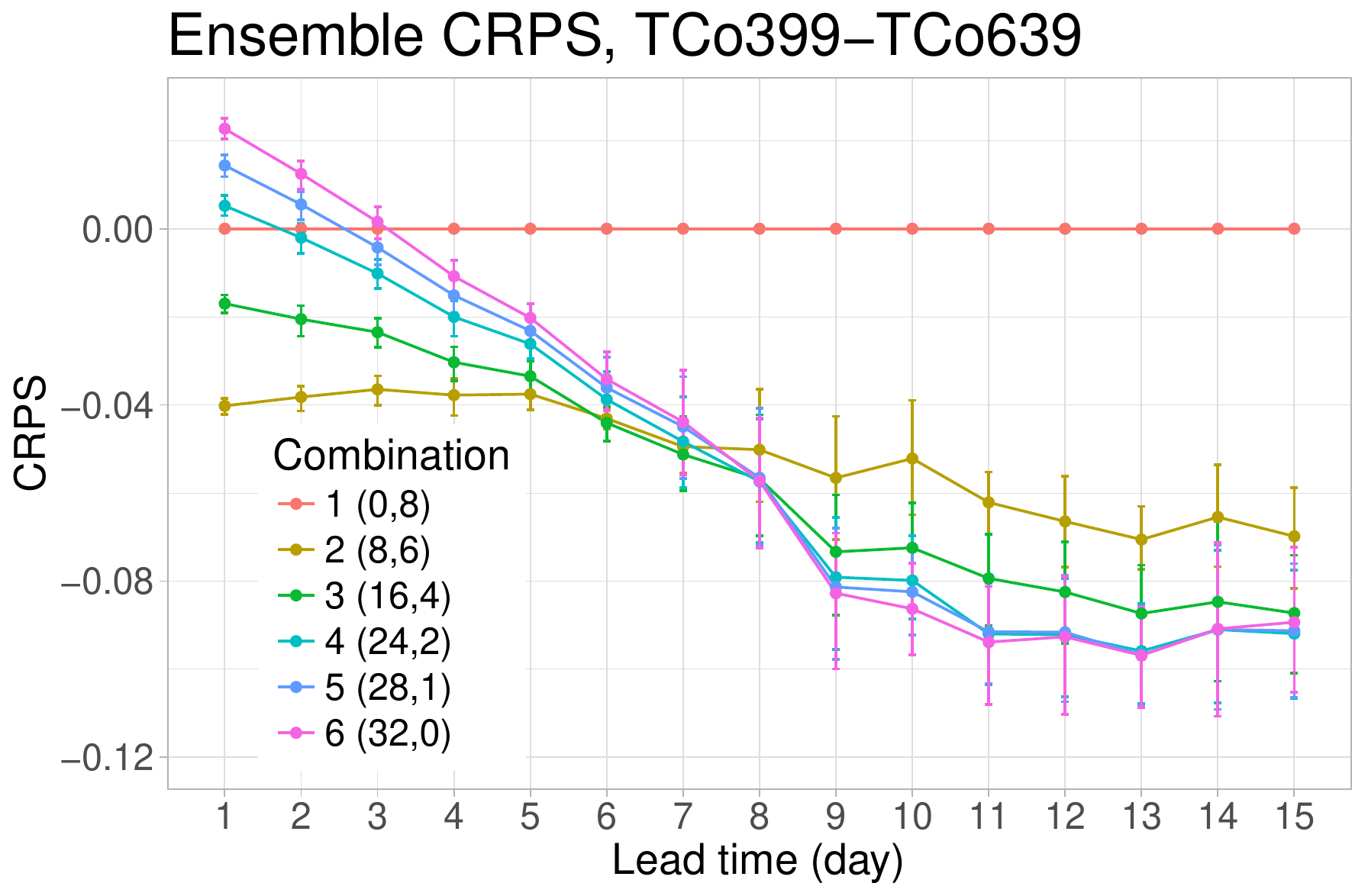, width=.49\textwidth} \
\epsfig{file=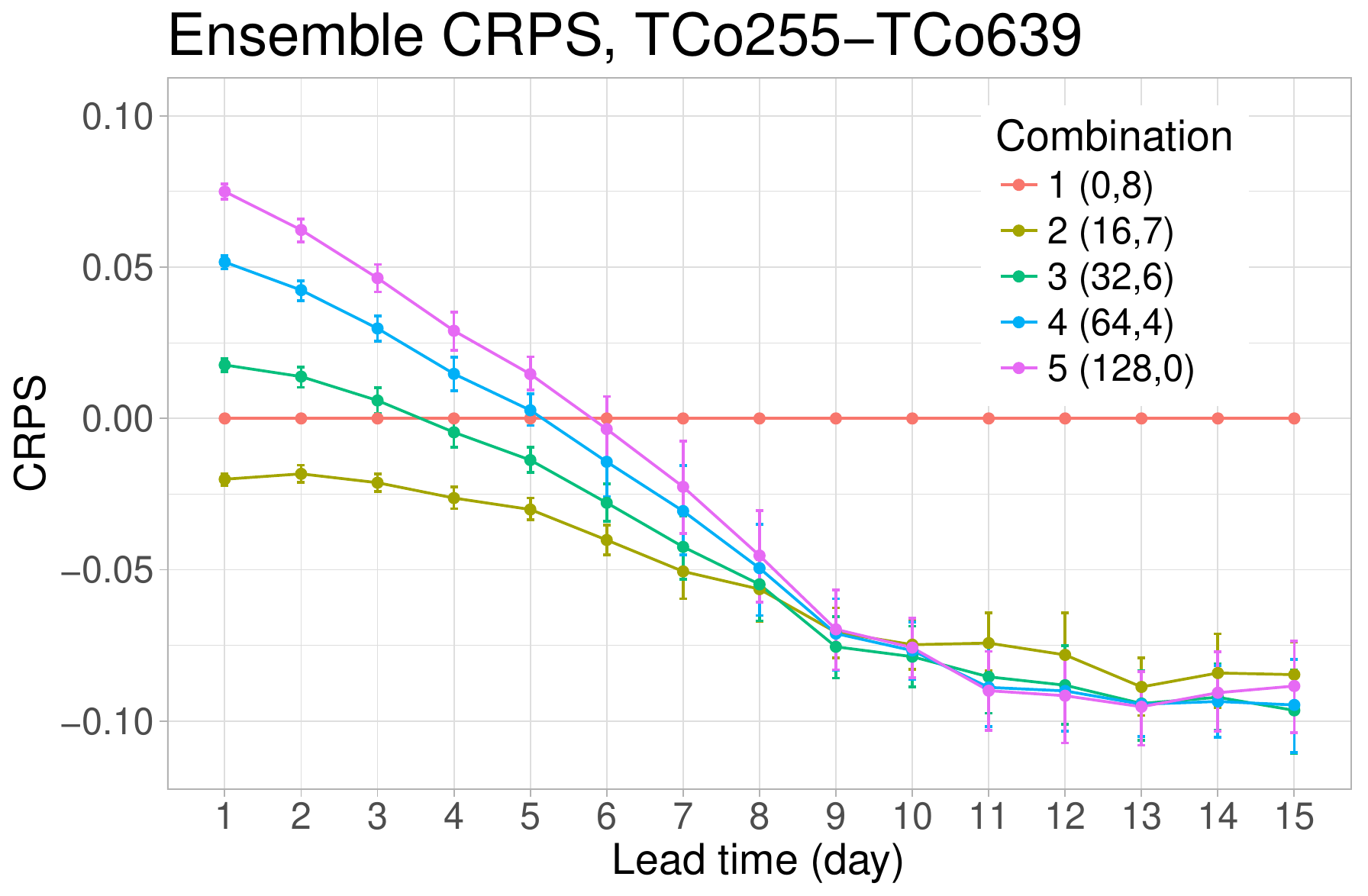, width=.49\textwidth} 
\caption{Mean CRPS values (the lower the better) of global dual-resolution ensemble forecasts for 2m temperature ({\em top}) and the difference in mean CRPS (the lower the better) from the reference pure high resolution ensemble ({\em bottom}) with $95\,\%$ confidence intervals, SHPC scenario.}
\label{fig:crpsRawS}
\end{figure}

One can conclude that for the LHPC scenario all examined verification scores show a consistent picture. In case of the TCo399 - TCo639 mixture up to day 9 the balanced combination (40,40) has the best predictive performance, whereas from day 10 all combinations outperform the reference pure high resolution one, however, the difference between the scores of combinations involving TCo399 forecasts are often very small. This situation differs from the case of the raw ensemble where the balanced combination exhibits the best forecast skills for all lead times (see Figure~\ref{fig:crpsRawL}, left column). For the TCo255 - TCo639 mixture the balanced combination (16,15) is clearly preferred up to day 5, from day 10 again all combinations outperform the reference pure high resolution one, however, with more pronounced differences than for the other mixture. For long lead times, the pure low resolution ensemble performs very well and the ordering of combinations again differs from the one based on the raw ensemble forecasts, which identify a balanced combination as best for all lead times (see Figure~\ref{fig:crpsRawL}, right column).

\subsection{Calibration of mixtures for small supercomputer}
\label{subs:subs4.2}

Now, we examine dual-resolution configurations for the small HPC scenario. In terms of the raw ensemble,  the balanced combinations ((8,6) for TCo399 - TCo639 and (16,7) for TCo255 - TCo639) are the most skillful for short lead times, whereas for longer lead times, the larger the ensemble size, the better, i.e.\ a pure low resolution ensemble has the best forecast skill (Figure~\ref{fig:crpsRawS}).

\begin{figure}[!t]
\epsfig{file=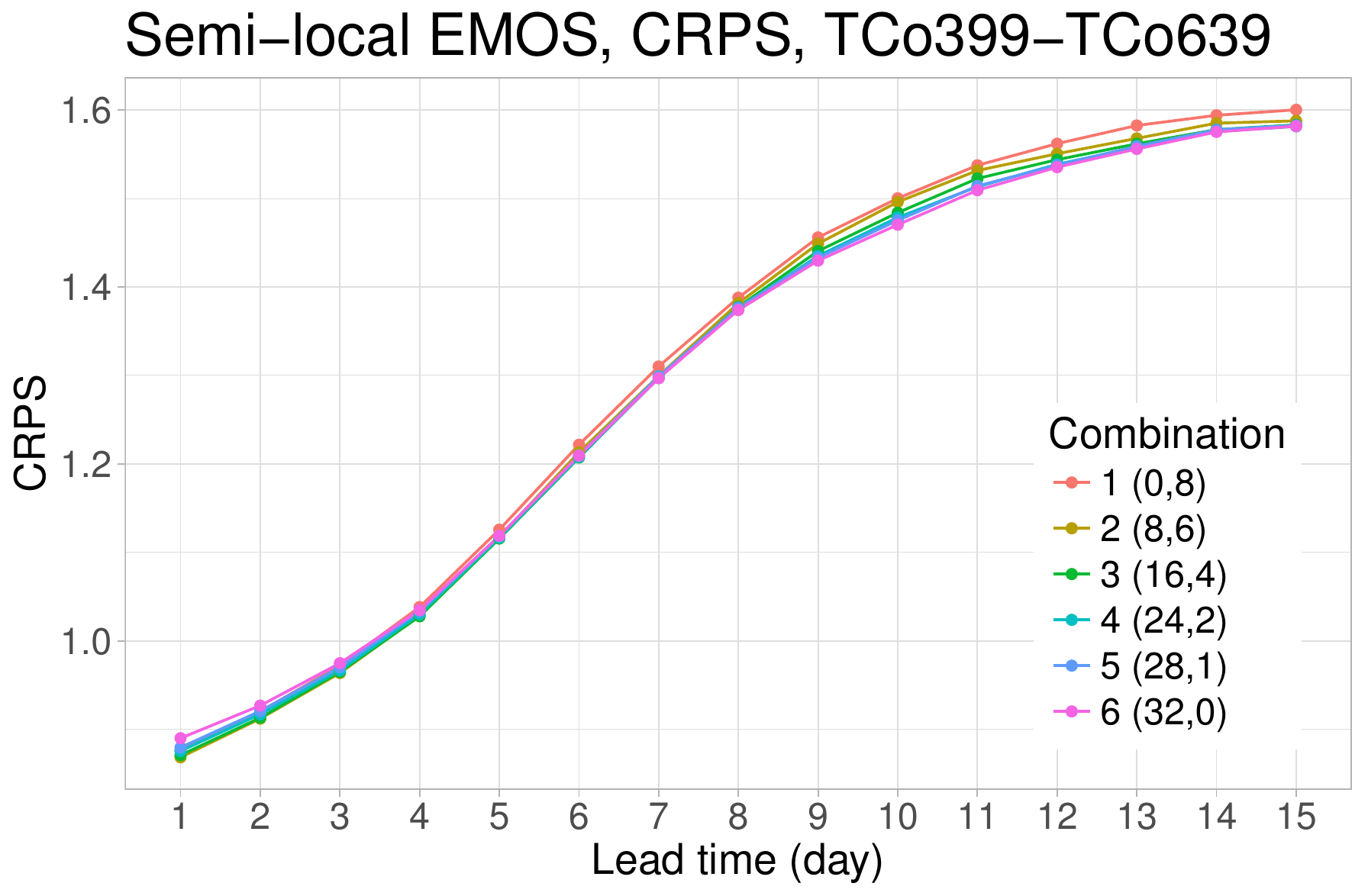, width=.49\textwidth} \
\epsfig{file=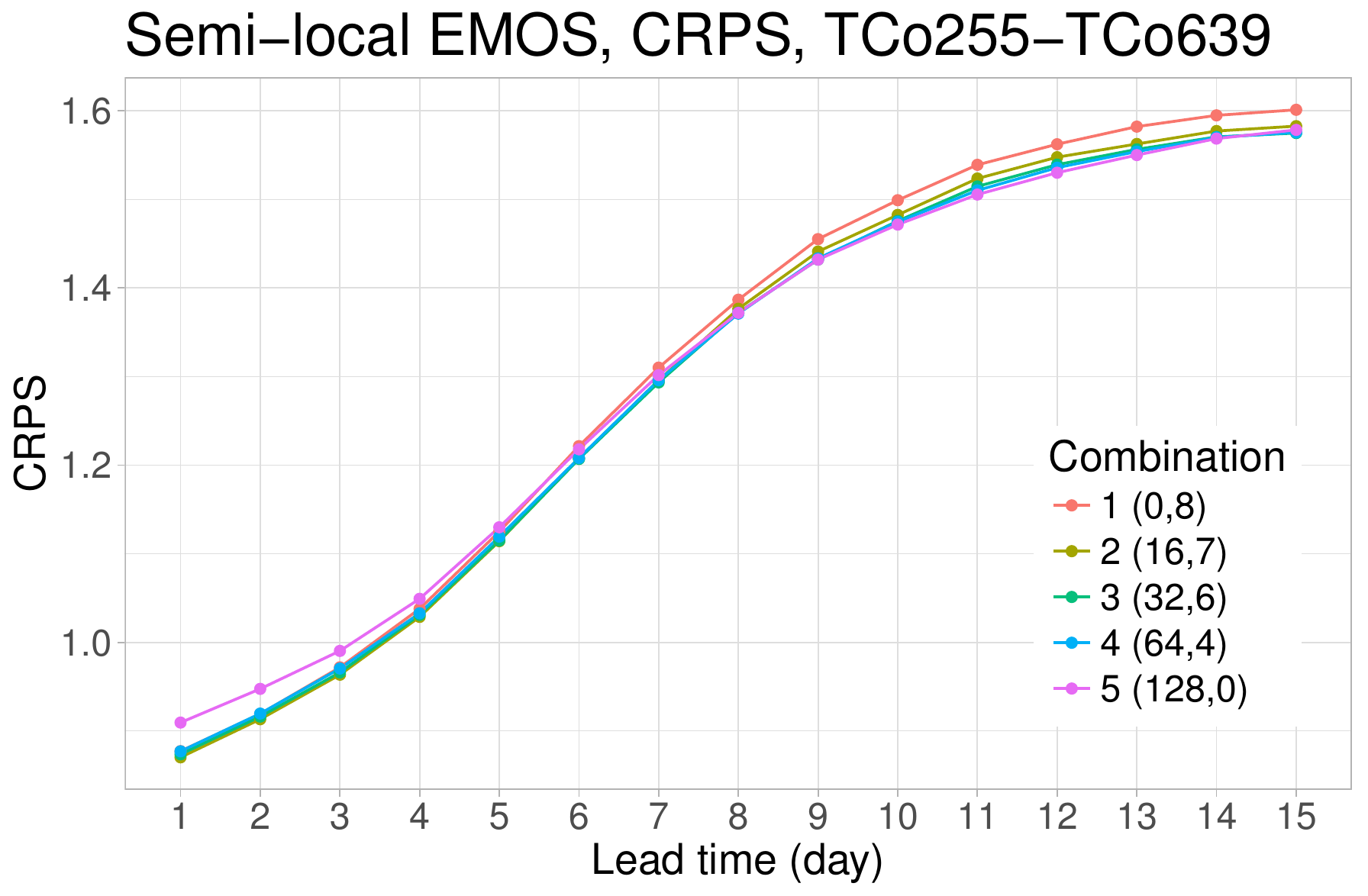, width=.49\textwidth} 
\caption{Mean CRPS values (the lower the better) of semi-local EMOS post-processed global dual-resolution ensemble forecasts for 2m temperature, SHPC scenario.}
\label{fig:crpsEMOSS}
\end{figure}

\begin{figure}[!t]
\epsfig{file=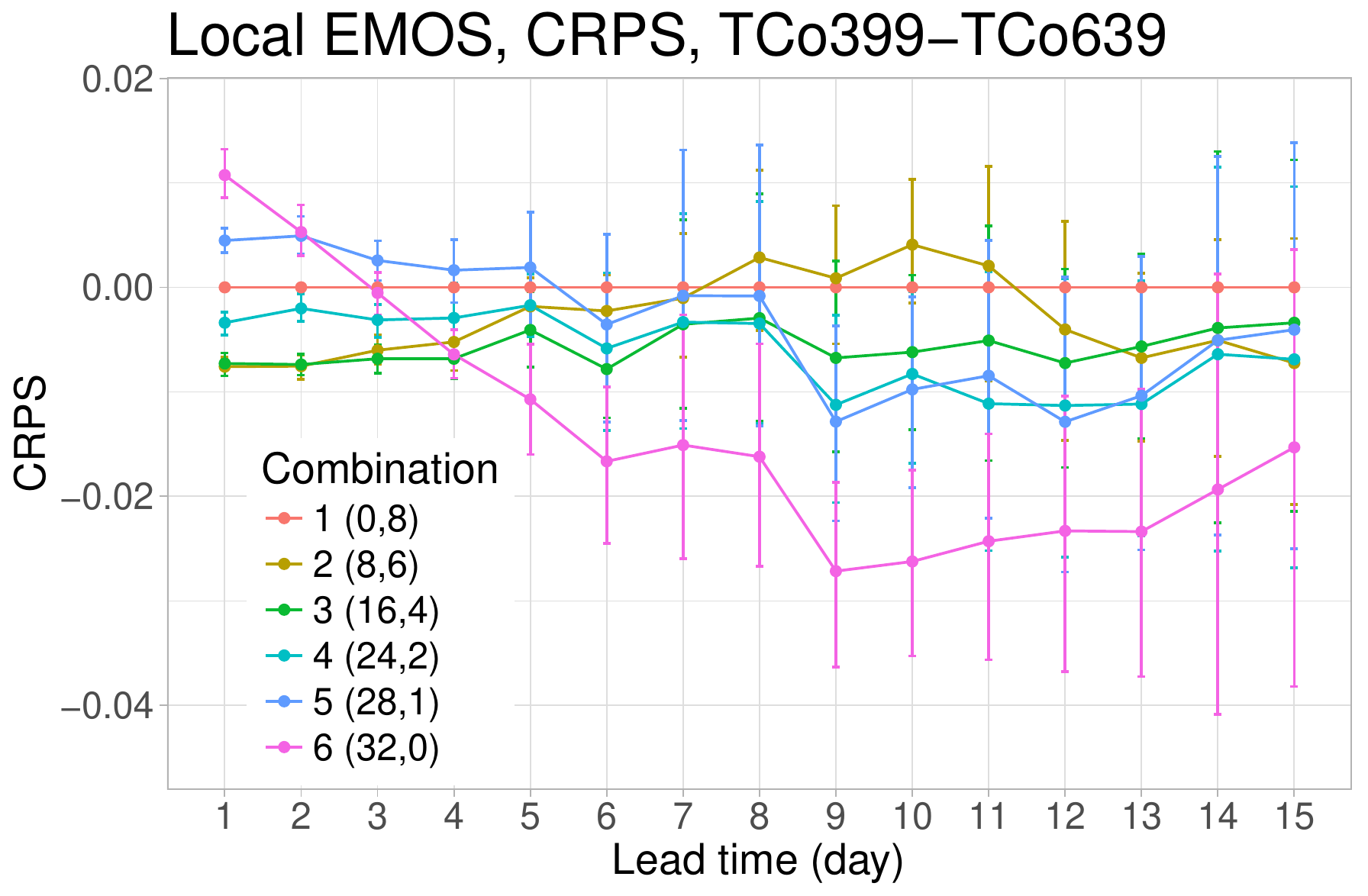, width=.49\textwidth} \
\epsfig{file=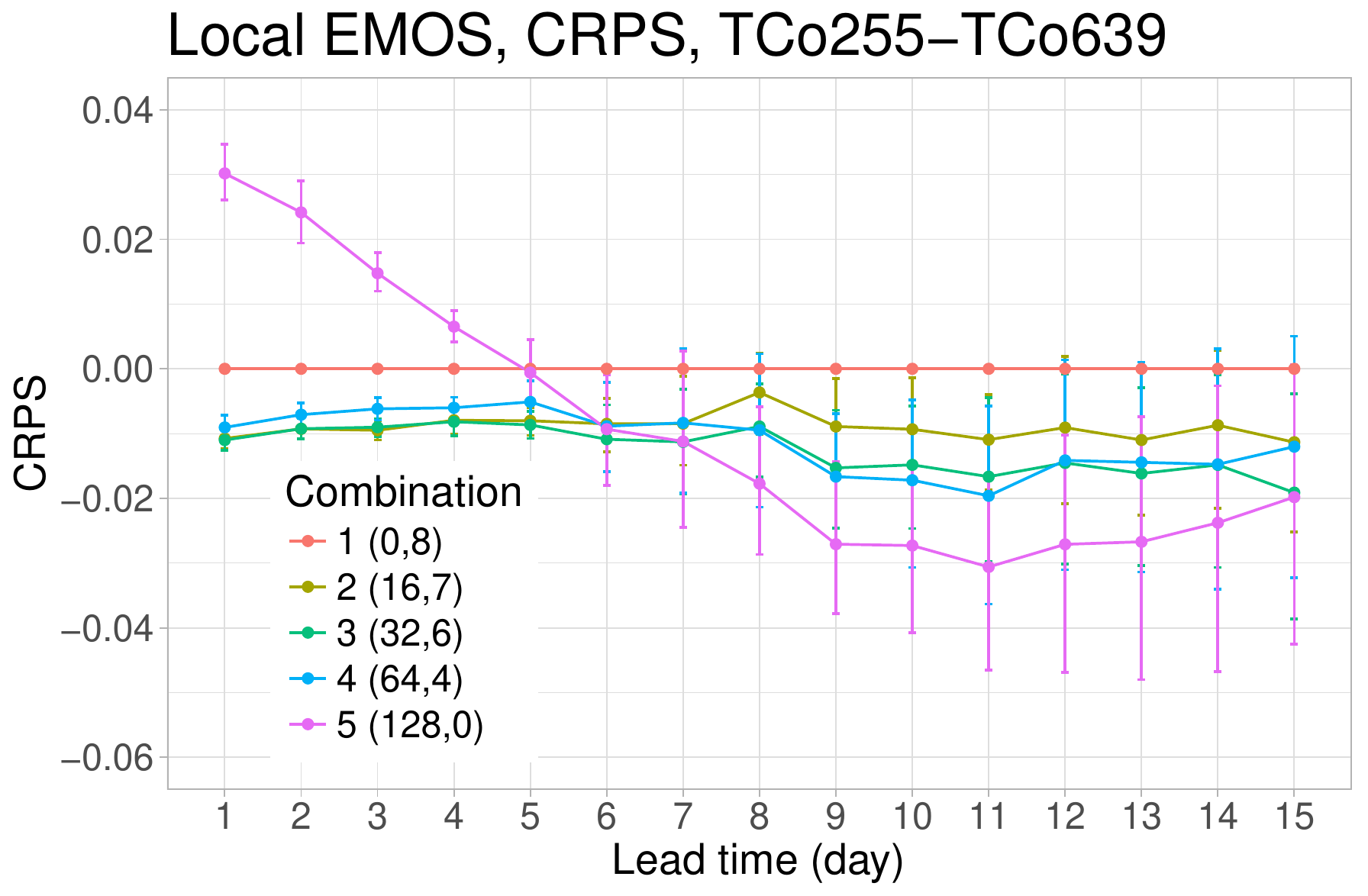, width=.49\textwidth}

\medskip
\epsfig{file=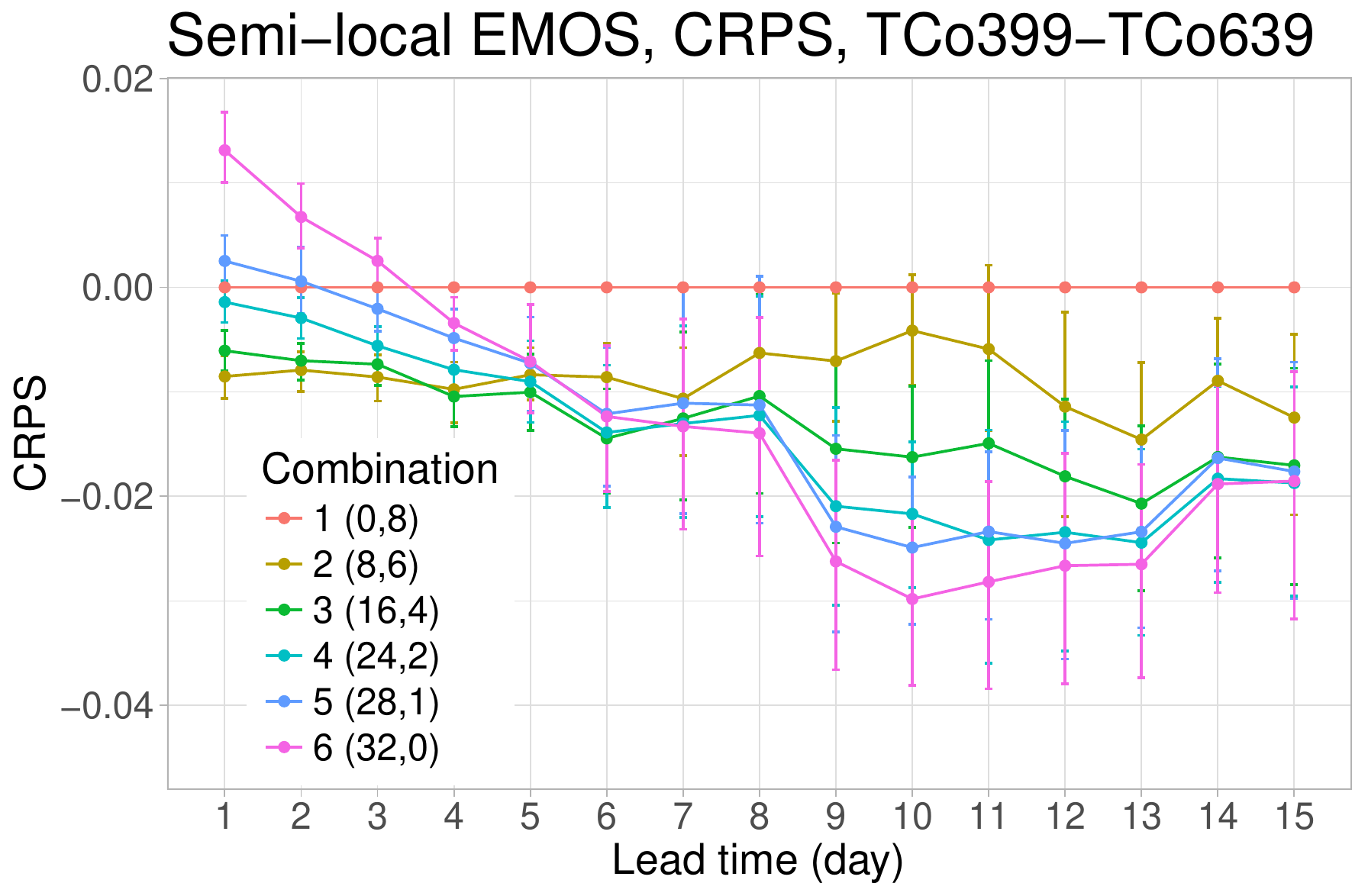, width=.49\textwidth} \
\epsfig{file=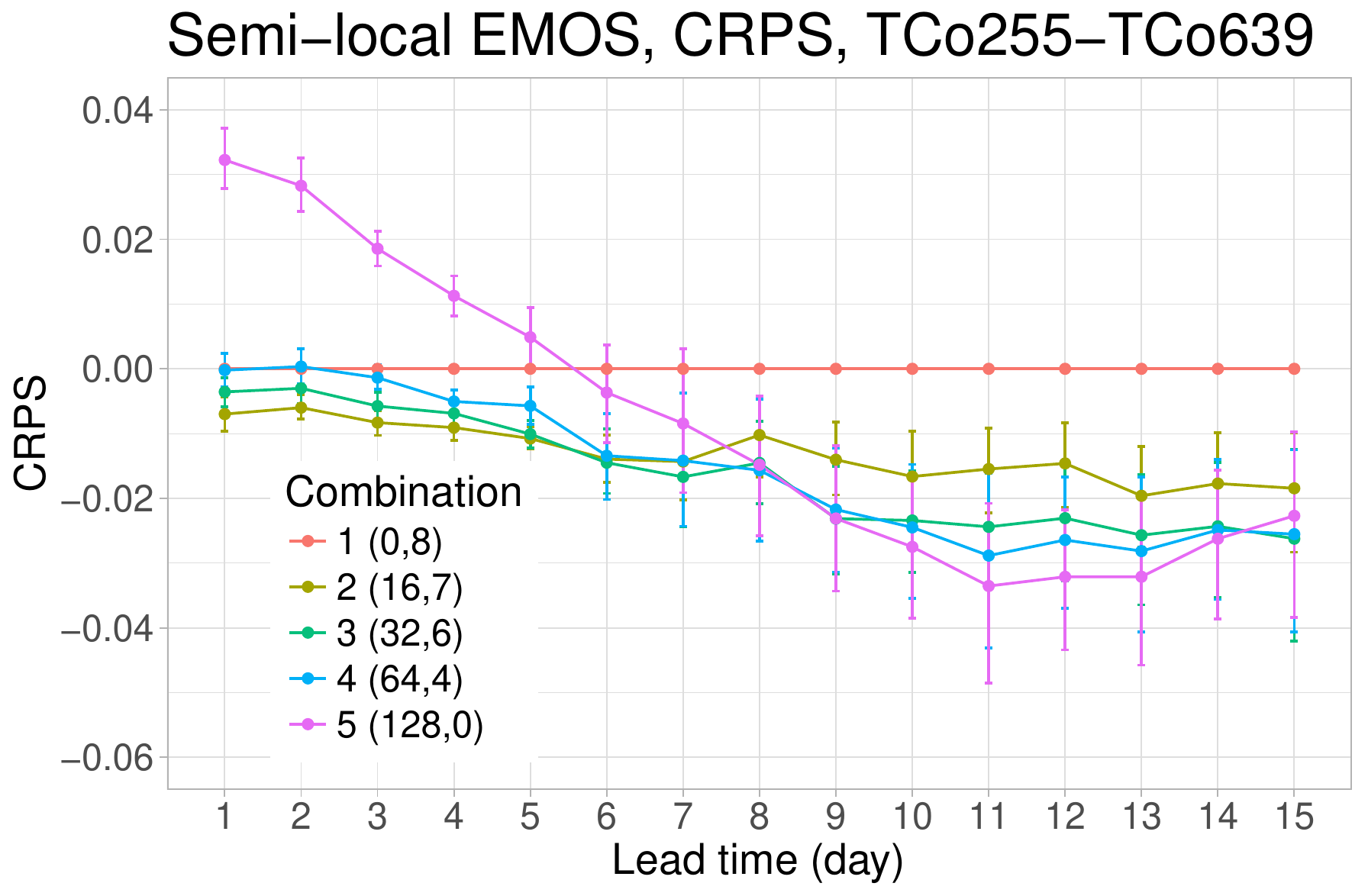, width=.49\textwidth}
\caption{Difference in mean CRPS (the lower the better) from the reference pure high resolution case with $95\,\%$ confidence intervals of local ({\em top}) and semi-local ({\em bottom}) EMOS post-processed global dual-resolution ensemble forecasts for 2m temperature, SHPC scenario.}
\label{fig:crpsDiffEMOSS}
\end{figure}

Statistical post-processing with semi-local EMOS significantly improves the calibration in terms of the mean CRPS and reduces the differences between the various combinations consistent with LHPC results (Figure~\ref{fig:crpsEMOSS}).  
The score differences are shown in Figure~\ref{fig:crpsDiffEMOSS} for local and semi-local EMOS. Both parameter estimation approaches provide an overall consistent ranking: at smaller lead times, the more balanced combinations are best, whereas for longer lead times the larger the ensemble size, the better the performance.

Statistical significance of score differences between configurations has also been computed stationwise (not shown). The longer the lead time, the smaller is the proportion of stations where  differences are significant. Compared with the LHPC scenario, at day 1 the pattern is similar to the top line of Figure~\ref{fig:crpsSigL}. At day 5, the proportions are smaller than for LHPC both for  TCo399 - TCo639 and TCo255 - TCo639 mixtures, whereas at day 10, there are more stations where the difference in mean CRPS is significant in SHPC than in LHPC, which can be explained by the fact that at longer lead times, the ensemble size is particularly important in the SHPC scenario. 

\begin{figure}[!t]
\epsfig{file=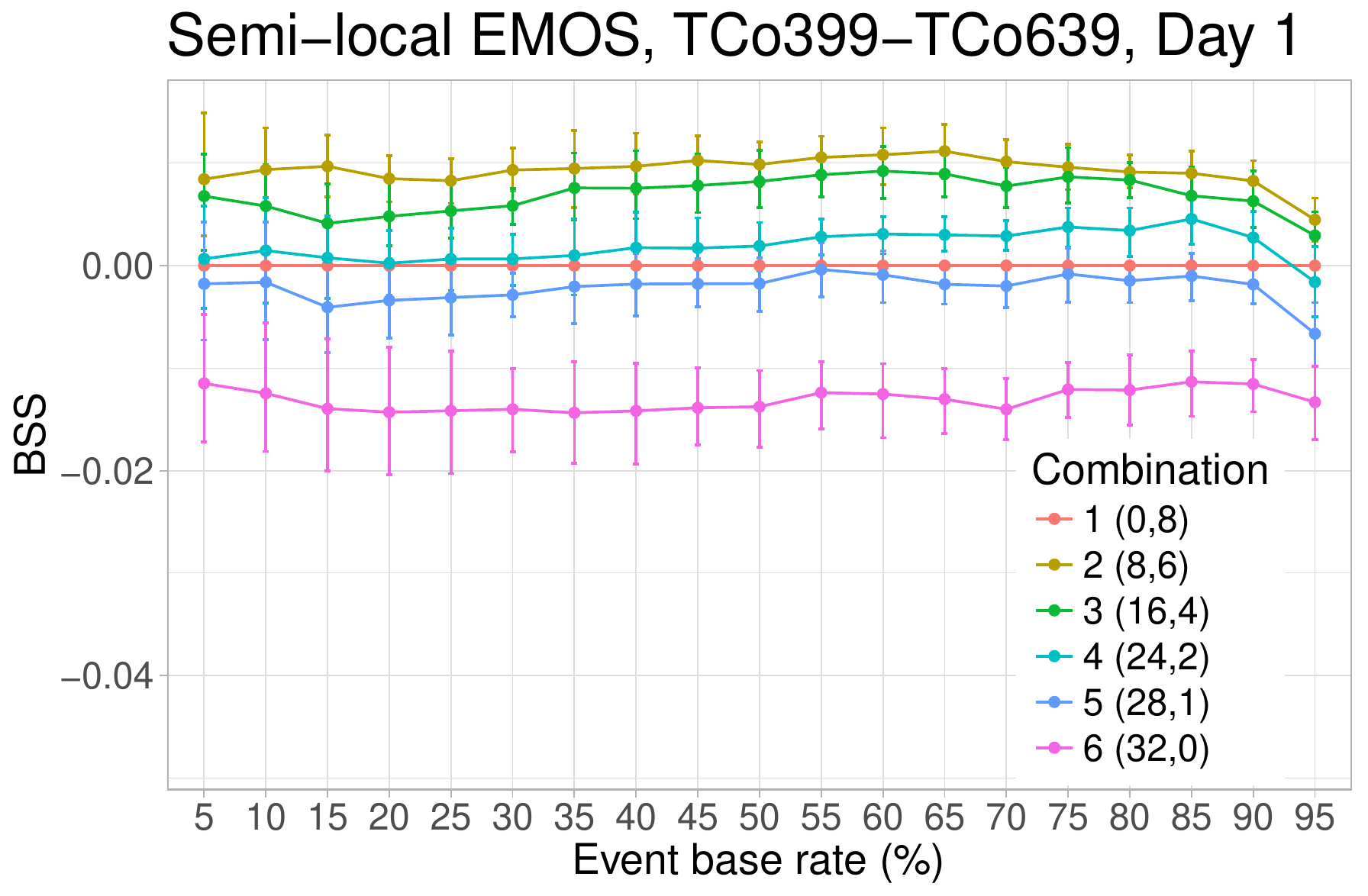, width=.49\textwidth} \
\epsfig{file=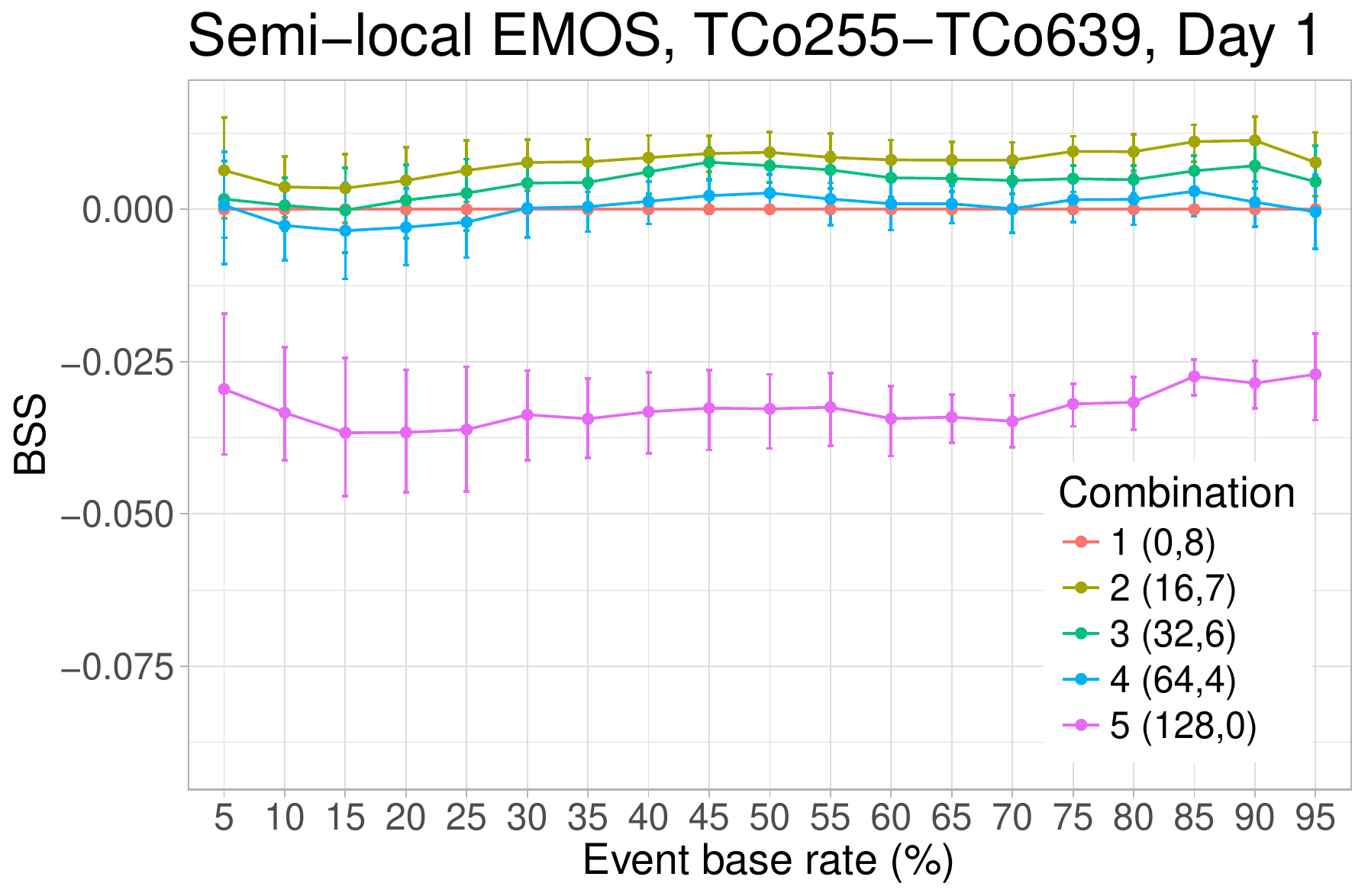, width=.49\textwidth}

\medskip
\epsfig{file=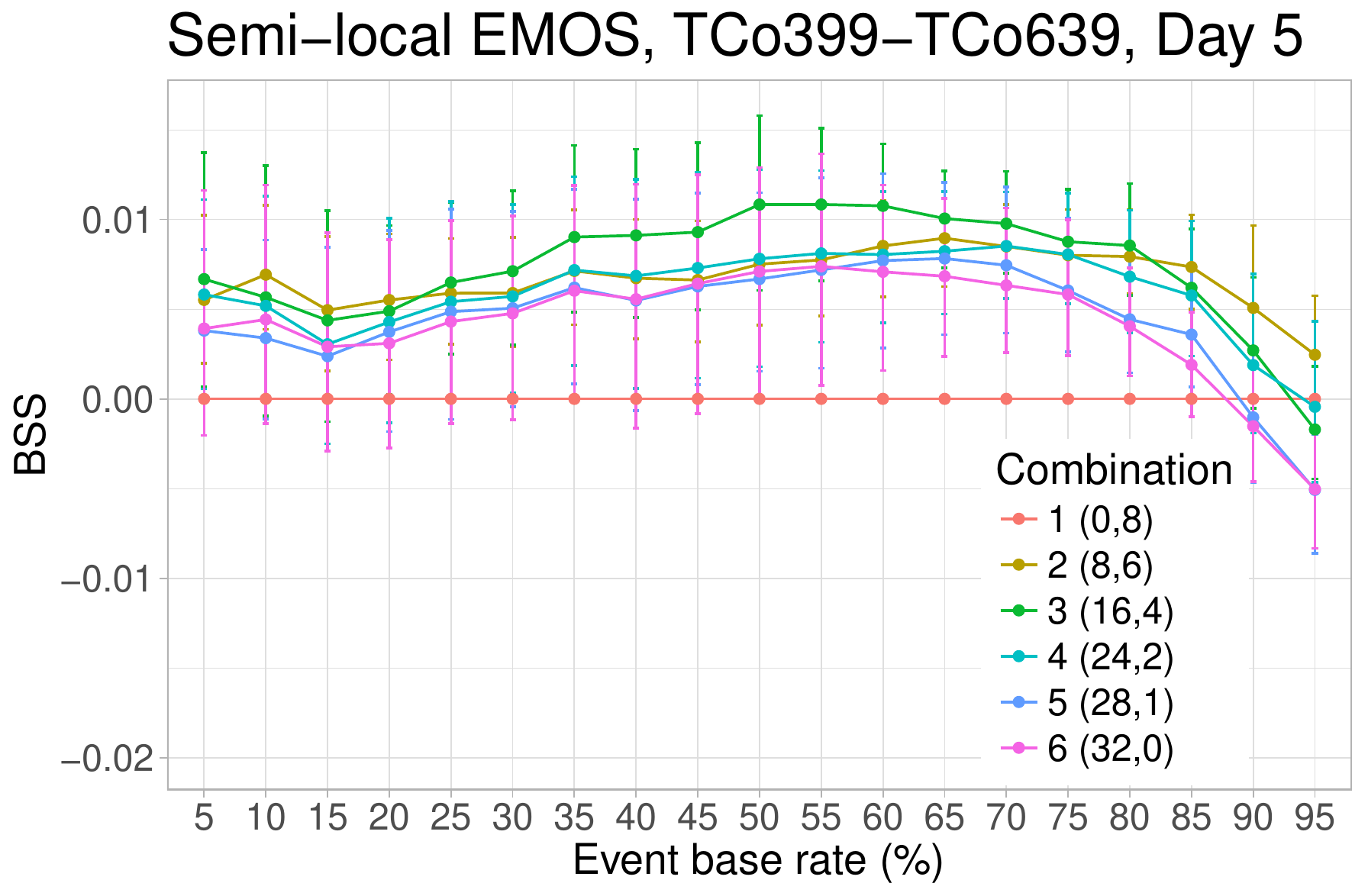, width=.49\textwidth} \
\epsfig{file=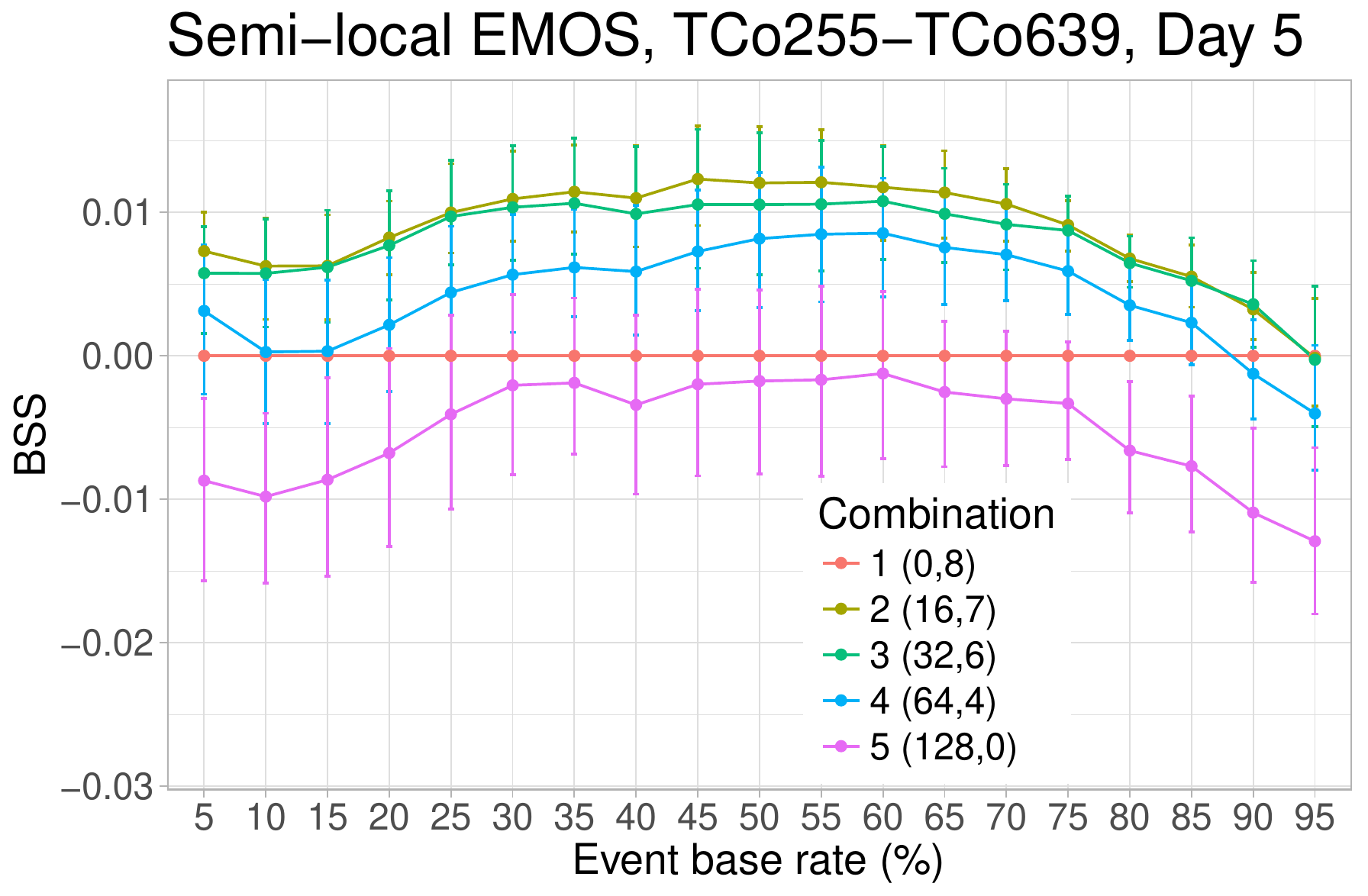, width=.49\textwidth}

\medskip
\epsfig{file=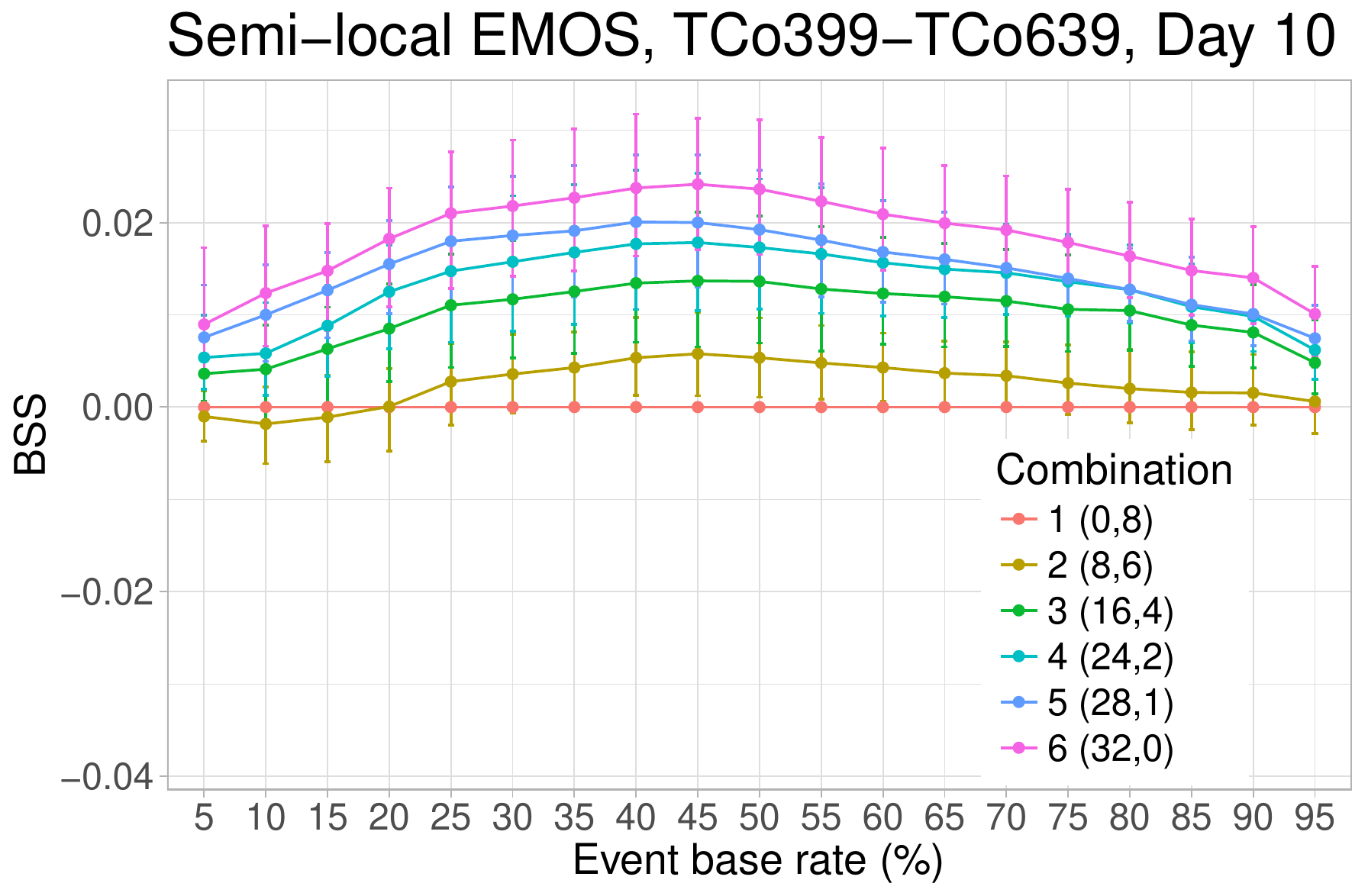, width=.49\textwidth} \
\epsfig{file=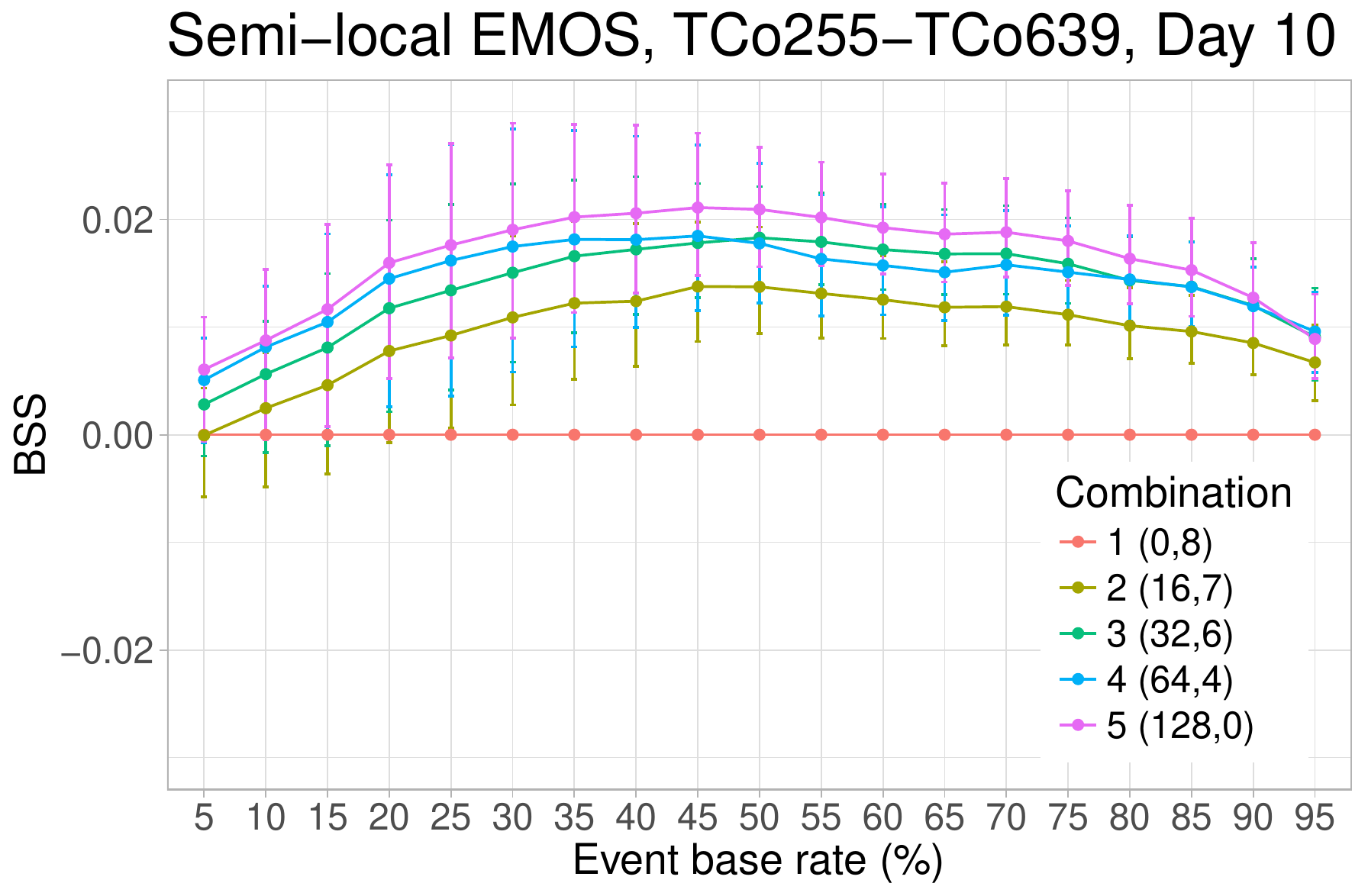, width=.49\textwidth} 
\caption{Brier skill scores (the higher the better) with respect to the reference pure high resolution case with $95\,\%$ confidence intervals of semi-local EMOS post-processed global dual-resolution ensemble forecasts for 2m temperature, SHPC scenario.}
\label{fig:bsDiffEMOSS}
\end{figure}

For TCo399 - TCo639 mixtures, the Brier skill scores with respect to the reference pure high resolution case (Figure~\ref{fig:bsDiffEMOSS}, left column) at day 1 indicate the advantage of the most balanced combination (8,6), at day 5 all combinations involving low resolution members outperform the post-processed pure high resolution ensemble, whereas at day 10 the ordering clearly reflects the ensemble size, that is the larger the better. 
For the TCo255 - TCo639 mixtures  (Figure~\ref{fig:bsDiffEMOSS}, right column), combinations with larger ensemble sizes are considered and both at days 1 and 5 the post-processed 128 member pure low resolution ensemble underperforms the 8 member pure high resolution one and the effect of the ensemble size on the predictive performance dominates only at day 10.  Note that the results for the Brier scores are fully consistent with those for the CRPS (see \ Figures~\ref{fig:bsDiffEMOSS} and \ref{fig:crpsDiffEMOSS}, respectively).

Further, similar to the LHPC scenario, the $50\,\%$ quantile skill score differences with respect to the reference pure high resolution case (not shown) are in line with the results for the CRPS. For the tails, the benefit of the various combinations with respect to the pure high resolution case are of similar amplitude as that for the LHPC scenario, and especially for the TCo399 - TCo639 mixture the differences are often non significant.

Finally, evaluation of the accuracy of the mean of different post-processed dual-resoluti\-on ensemble forecasts in terms of the root mean squared error shows the same picture as the verification scores for probabilistic forecasts. The differences in RMSE values from the reference pure high resolution case for semi-local EMOS (not shown) are very similar to those for the CRPS.

Similar to the LHPC scenario, all investigated verification scores indicate similar conclusions. The balanced combination (8,6) and (16,7) are preferred up to 4 days, although the differences compared to combinations (16,4) and (32,6), respectively, are very small. For  TCo399 - TCo639 after day 7, whereas for TCo255 - TCo639 after day 9, the post-processed pure low resolution ensemble shows the best predictive performance and for long lead times the forecast skill depends on the ensemble size. This behaviour is rather similar to that seen in the raw ensemble (compare Figures~\ref{fig:crpsRawS} and \ref{fig:crpsDiffEMOSS}).

\subsection{Calibration using a very short training period}
\label{subs:subs4.3}

\begin{table}[t]
  {\scriptsize
\begin{center}
\begin{tabular}{c| l |c c c c c | c c c c c}
  \hline
Combi-&Model&\multicolumn{5}{c|}{Day 1}&\multicolumn{5}{c}{Day 5}\\ \cline{3-12}  
nation&&CRPS&RMSE&QS2&QS50&QS98&CRPS&RMSE&QS2&QS50&QS98\\
  \hline
&Raw ensemble&1.258&2.068&0.391&0.754&0.340&1.391&2.425&0.277&0.909&0.243\\ \cline{2-12}
&Local EMOS, 10d&0.925&1.720&0.134&0.626&0.111&1.205&2.226&0.166&0.819&0.140\\ 
 (0,50)&Semi-local EMOS, 10d&0.894&1.683&0.105&0.619&0.086&1.142&2.135&0.129&0.794&0.107\\ \cline{2-12}
 &Local EMOS, 30d&0.851&1.617&0.097&0.591&0.081&1.099&2.070&0.118&0.767&0.102\\
 &Semi-local EMOS, 30d&0.869&1.648&0.093&0.606&0.081&1.102&2.073&0.110&0.773&0.098\\
\hline
&Raw ensemble&1.226&2.053&0.351&0.752&0.292&1.377&2.418&0.253&0.908&0.216\\ \cline{2-12} 
&Local EMOS, 10d&0.935&1.733&0.141&0.630&0.117&1.226&2.266&0.174&0.831&0.147\\  
(40,40) &Semi-local EMOS, 10d&0.889&1.671&0.106&0.615&0.085&1.142&2.138&0.129&0.794&0.108\\  \cline{2-12}
&Local EMOS, 30d&0.846&1.605&0.099&0.587&0.082&1.101&2.073&0.119&0.768&0.103\\
&Semi-local EMOS, 30d&0.864&1.637&0.093&0.602&0.080&1.100&2.070&0.110&0.771&0.098\\      
\hline
&Raw ensemble&1.270&2.093&0.362&0.771&0.300&1.403&2.447&0.255&0.921&0.220\\  \cline{2-12}
&Local EMOS, 10d&0.935&1.732&0.141&0.630&0.117&1.225&2.263&0.174&0.830&0.147\\ 
(120,20)&Semi-local EMOS, 10d&0.894&1.679&0.106&0.619&0.086&1.144&2.140&0.129&0.795&0.108\\ \cline{2-12}
&Local EMOS, 30d&0.847&1.606&0.099&0.588&0.082&1.101&2.073&0.119&0.768&0.103\\
&Semi-local EMOS, 30d&0.869&1.645&0.093&0.606&0.081&1.103&2.073&0.111&0.774&0.098\\     
\hline
&Raw ensemble&1.289&2.109&0.373&0.774&0.311&1.412&2.459&0.260&0.924&0.225\\  \cline{2-12}
&Local EMOS, 10d&0.937&1.736&0.142&0.631&0.118&1.228&2.269&0.176&0.832&0.149\\ 
(160,10)&Semi-local EMOS, 10d&0.896&1.685&0.106&0.621&0.087&1.147&2.148&0.130&0.797&0.108\\  \cline{2-12}
&Local EMOS, 30d&0.848&1.608&0.099&0.588&0.082&1.105&2.079&0.120&0.770&0.103\\
&Semi-local EMOS, 30d&0.871&1.648&0.093&0.607&0.081&1.106&2.080&0.111&0.776&0.098\\      
\hline
&Raw ensemble&1.304&2.121&0.406&0.776&0.345&1.422&2.467&0.270&0.926&0.238\\  \cline{2-12}
&Local EMOS, 10d&0.943&1.754&0.136&0.639&0.113&1.211&2.241&0.166&0.825&0.141\\ 
(200,0)&Semi-local EMOS, 10d&0.913&1.718&0.108&0.632&0.089&1.152&2.152&0.129&0.801&0.109\\  \cline{2-12}
&Local EMOS, 30d&0.869&1.650&0.099&0.604&0.083&1.108&2.087&0.118&0.773&0.103\\
&Semi-local EMOS, 30d&0.888&1.681&0.095&0.619&0.083&1.112&2.090&0.111&0.780&0.099\\
      \hline
\end{tabular}
\end{center}
}
\caption{Verification scores of 2-metre temperature (K) for the raw ensemble,  local and semi-local EMOS post-processed forecasts using 10-day and 30-day training periods, TCo399 - TCo639 mixture, LHPC scenario. Lower values imply more skillful forecasts for all metrics.}
\label{tab:scores10day}
\end{table} 

Additional results have been obtained using a 10-day training period in order to compare improvements from local and semi-local EMOS post-processing. The TCo399 - TCo639 ensemble configurations of the LHPC scenario have been calibrated for lead times up to 5 days. For local EMOS, this implies 10 forecast-observation pairs are used to estimate 4--5 parameters, whereas semi-local EMOS uses on average around 225 forecast-observation pairs  for the same purpose for the price of having a universal set of parameters within each cluster. For verification, we use data of the same 62 calendar days as in Sections \ref{subs:subs4.1} and \ref{subs:subs4.2}, i.e. ensemble forecasts initialized between 1 July and 31 August 2016 and the corresponding validating observations.

Table \ref{tab:scores10day} lists verification scores for lead times 1 and 5 days. Note that all post-processing approaches for all combinations substantially outperform the raw ensemble. For the 30-day training period, there is no clear preference for local EMOS or semi-local EMOS. However, one should mention that the uncertainty in score differences of local EMOS post-processed forecasts is much higher especially for longer lead times. In contrast, for the 10-day training period semi-local EMOS is clearly superior to local EMOS. Thus, clustering based semi-local estimation of EMOS parameters provides a reasonable alternative to the local approach, especially in situations where ensemble data cover only a rather short time period.

\section{Conclusions}
\label{sec:sec5}

The EMOS calibration leads to substantial improvements in skill for all examined single and dual-resolution ensemble forecasts. For instance, the CRPS  decreases from values around 1.3\,K to values just under 1.0\,K at day~3 using the semi-local EMOS. The improvements are large although not as large as those reported by \cite{hemri14}. Their raw ensemble forecasts have significantly larger CRPS values compared to ours. We attribute this difference to having applied an orographic correction to the forecasts. Our 'raw forecasts' have been adjusted with a simple correction that is proportional to the altitude difference between the model orography and the station height while \cite{hemri14} showed results for uncorrected forecasts. 

In terms of calibration methodology, the clustering based semi-local estimation of EMOS parameters provides a reasonable alternative to the local approach, especially in situations where ensemble data cover only a rather short time period. This is fully in line with results reported by \cite{lb17}, where multi-model ensemble forecasts of wind speed over Europe and North-Africa were calibrated.

The EMOS calibration parameters were obtained by optimizing skill in terms of the CRPS. EMOS improves not only the probabilistic forecasts but also the accuracy of point forecasts like the ensemble mean or the median. Moreover, when comparing ensemble configurations after calibration, score differences for the Brier score and the quantile score show fairly consistent rankings of the ensemble configurations across different event thresholds and probability levels, respectively. 

The calibration of the ensemble forecasts with EMOS strongly reduces the differences in skill among the equal cost configurations of single and dual-resolution ensembles. The reduction of the differences is much bigger than the reduction in CRPS seen from raw forecasts to calibrated distribution. This implies that making the correct selection of the 'best resolution/ensemble size configuration' is less important for those users that will use EMOS calibrated forecasts instead of the raw forecasts.

What emerges as the best single- or dual-resolution configuration can change due to EMOS calibration. For instance, the TCo639-TCo399 dual-resolution configuration $(40,40)$ is best at all lead times for raw forecasts in the large supercomputer scenario. After calibration, this configuration is still best until about day~7; however, at longer lead times the other configurations with at least 140 members are equally skillful. Likewise, 50 members at TCo639 are as skillful as 200 members at TCo399 resolution for the raw forecasts while, after calibration, the 200 lower-resolution members are slightly more skillful than the 50 higher-resolution members.

For the small supercomputer scenario, the overall ranking is similar before and after EMOS calibration. At lead times beyond day~7, the skill is mainly determined by the ensemble size and  the pure low resolution ensemble shows the best predictive performance.

\cite{lbb} identified situations where a non-trivial dual-resolution ensemble forecasts of 2-metre temperature is considerably more skillful than a single-resolution configuration with the same computational cost. Here, EMOS calibration was applied to the same data. Results suggest that the benefit of a dual-resolution configurations is more marginal after calibration than before.

\bigskip
\noindent
{\bf Acknowledgments.} \ S\'andor Baran and Marianna Szab\'o were supported by the EFOP-3.6.3-VEKOP-16-2017-00002 project. The project was co-financed by the Hungarian Government and the European Social Fund.
S\'andor Baran also acknowledges the support of the Hungarian
National Research, Development and Innovation Office under Grant No. NN125679.
He is also grateful to the ECMWF for supporting his research stays in Reading and to the Heidelberg Institute for Theoretical Studies for hosting him as visiting scientist.

\end{document}